\documentclass[a4paper,11pt]{article}
\pdfoutput=1 

\usepackage{jheppub} 
\makeatletter
\DeclareRobustCommand*{\bfseries}{%
  \not@math@alphabet\bfseries\mathbf
  \fontseries\bfdefault\selectfont
  \boldmath
}
\makeatother

\usepackage{calc}
\usepackage{rotating}
\usepackage[english]{babel}
\usepackage{graphicx}
\usepackage{subfig}
\usepackage{float}
\usepackage{amsmath}
\usepackage{amssymb}
\usepackage{amsthm}
\usepackage{latexsym}
\usepackage{dcolumn}
\usepackage{hyperref}
\newcommand{\newc}{\newcommand*}

\long\def\begincomment#1\endcomment{%
        \begingroup\sf\baselineskip12pt#1\endgroup}

\newc{\etal}{\textrm{et al.}} 
\newc{\eg}{\textrm{e.g.}} 
\newc{\ie}{\textrm{i.e.}}
\newc{\etc}{\textrm{etc}}
\newc\vs{\textrm{vs.}}
\newc{\cl}{\rm {C.L.}}

\newc{\ev}{\ensuremath{\,\mathrm{eV}}}
\newc{\kev}{\ensuremath{\,\mathrm{keV}}}
\newc{\mev}{\ensuremath{\,\mathrm{MeV}}}
\newc{\gev}{\ensuremath{\,\mathrm{GeV}}}
\newc{\tev}{\ensuremath{\,\mathrm{TeV}}}
\newc{\MeV}{\mev} 
\newc{\TeV}{\tev}
\newc{\invpb}{\ensuremath{/\text{pb}}}
\newc{\invfb}{\ensuremath{\,\text{fb}^{-1}}}
\newc\nb{\ensuremath{\,\mathrm{nb}}} \newc\pb{\ensuremath{\,\mathrm{pb}}} \newc\fb{\ensuremath{\,\mathrm{fb}}}
\newc\pc{\ensuremath{\,\mathrm{pc}}}
\newc\kpc{\ensuremath{\,\mathrm{kpc}}}
\newc\mpc{\ensuremath{\,\mathrm{Mpc}}}
\newc\ps{\ensuremath{\,\mathrm{ps}}} 
\newc\cmeter{\ensuremath{\,\mathrm{cm}}} 
\newc\meter{\ensuremath{\,\mathrm{m}}} 
\newc\kmeter{\ensuremath{\,\mathrm{km}}}
\newc\second{\ensuremath{\,\mathrm{s}}}
\newc\msecond{\ensuremath{\,\mathrm{ms}}}
\newc\nsecond{\ensuremath{\,\mathrm{ns}}}
\newc\psecond{\ensuremath{\,\mathrm{ps}}}

\newc{\chisqmin}{\ensuremath{\chi^2_{\mathrm{min}}}}
\newc{\Delchisq}{\ensuremath{\Delta\chi^2}}
\newc{\chisq}{\ensuremath{\chi^2}}
\newc{\like}{\ensuremath{\mathcal{L}}}

\newc\lsim{\ensuremath{\mathrel{\rlap{\lower4pt\hbox{\hskip1pt$\sim$}}\raise1pt\hbox{$<$}}}}
\newc\gsim{\ensuremath{\mathrel{\rlap{\lower4pt\hbox{\hskip1pt$\sim$}}\raise1pt\hbox{$>$}}}}
\newc{\VEV}[1]{\ensuremath{\langle #1 \rangle}}
\newc{\dl}{\ensuremath{\stackrel{\leftarrow}{D}}}
\newc{\dr}{\ensuremath{\stackrel{\rightarrow}{D}}}

\newc{\scr}[1]{\ensuremath{\mathcal{#1}}}

\newc{\bcenter}{\begin{center}}    \newc{\ecenter}{\end{center}}
\newc{\bfl}{\begin{flushleft}}    \newc{\efl}{\end{flushleft}}
\newc{\bfr}{\begin{flushright}}    \newc{\efr}{\end{flushright}}

\newc{\bi}{\begin{itemize}}
\newc{\ei}{\end{itemize}}
\newc{\bed}{\begin{description}}
\newc{\eed}{\end{description}}
\newc{\ben}{\begin{enumerate}}
\newc{\een}{\end{enumerate}}

\newc{\be}{\begin{equation}}
\newc{\ee}{\end{equation}}
\newc{\bea}{\begin{eqnarray}}
\newc{\eea}{\end{eqnarray}}
\newc{\ra}{\rightarrow}

\newc{\alphas}{\ensuremath{\alpha_s}}
\newc{\alphatwo}{\ensuremath{\alpha_2}}
\newc{\alphaone}{\ensuremath{\alpha_1}}
\newc{\alphai}[1]{\ensuremath{\alpha_{#1}}}
\newc{\alphaem}{\ensuremath{\alpha_{\mathrm{em}}}}
\newc{\alphaeff}{\ensuremath{\alpha_{\mathrm{eff}}}}
\newc{\sineff}{\ensuremath{\sin \theta_{\mathrm{eff}}}}
\newc{\sinsqeff}{\ensuremath{\sin^2 \theta_{\mathrm{eff}}}}
\newc{\dalphahad}{\ensuremath{\Delta \alpha_{\mathrm{had}}}}
\newc{\yt}{\ensuremath{h_t}} \newc{\yb}{\ensuremath{h_b}} \newc{\ytau}{\ensuremath{h_{\tau}}}
\newc\mz{\ensuremath{M_Z}} 
\newc\mw{\ensuremath{m_W}}
\newc\mZ{\mz}        \newc\mW{\mw}
\newc\mhsm{\ensuremath{ m_{H_{\mathrm{SM}}}}}
\newc{\mtop}{\ensuremath{ m_t}}               \newc{\mtpole}{\ensuremath{ M_t}}
\newc{\mbottom}{\ensuremath{ m_b}} 
\newc{\mtau}{\ensuremath{ m_{\tau}}}
\newc{\mt}{\mtpole}
\newc{\mb}{\mbottom} 
\newc{\rgg}{\ensuremath{R_{h}(\gamma\gamma)}}
\newc{\rzz}{\ensuremath{R_{h}(ZZ)}}
\newc{\rtwogg}{\ensuremath{R_{h_2}(\gamma\gamma)}}
\newc{\rtwozz}{\ensuremath{R_{h_2}(ZZ)}}
\newc{\ronegg}{\ensuremath{R_{h_1}(\gamma\gamma)}}
\newc{\ronezz}{\ensuremath{R_{h_1}(ZZ)}}
\newc{\rsiggg}{\ensuremath{R_{h_\textrm{sig}}(\gamma\gamma)}}
\newc{\rsigzz}{\ensuremath{R_{h_\textrm{sig}}(ZZ)}}
\newc{\llbar}{\ensuremath{\ell\bar{\ell}}}
\newc{\tauptaum}{\ensuremath{ \tau^+\tau^-}}
\newc{\qqbar}{\ensuremath{ q\bar{q}}} \newc{\ppbar}{\ensuremath{ p\bar{p}}}
\newc{\bbbar}{\ensuremath{ b\bar{b}}} \newc{\ttbar}{\ensuremath{ t\bar{t}}}
\newc{\ffbar}{\ensuremath{ f\bar{f}}} \newc{\tautaubar}{\ensuremath{ \tau\bar{\tau}}}
\newc{\mchi}{\ensuremath{m_{\chi}}}
\newc{\squark}{\ensuremath{\tilde{q}}}
\newc{\slepton}{\ensuremath{\tilde{l}}}
\newc{\gluino}{\ensuremath{\tilde{g}}} 
\newc{\mgluino}{\ensuremath{{m_{\gluino}}}}
\newc{\tone}{\ensuremath{{\tilde{t}_1}}}

\newc{\sthw}{\ensuremath{ \sin\theta_W}}              \newc{\cthw}{\ensuremath{\cos\theta_W}}
\newc{\tanthw}{\ensuremath{ \tan\theta_W}}              \newc{\cotthw}{\ensuremath{\cot\theta_W}}
\newc{\ssqthw}{\ensuremath{\sin^2 \theta_W}}
\newc{\msbar}{\ensuremath{\overline{MS}}} \newc{\drbar}{\ensuremath{\overline{DR}}}
\newc{\mtmtsmmsbar}{\ensuremath{ m_t(m_t)^{\msbar}_{{\mathrm{SM}}}}}
\newc{\mtmtsmdrbar}{\ensuremath{ m_t(m_t)^{\drbar}_{{\mathrm{SM}}}}}
\newc{\mtmtmssmdrbar}{\ensuremath{ m_t(m_t)^{\drbar}_{{\mathrm{SUSY}}}}}
\newc{\mbmbmsbar}{\ensuremath{ m_b(m_b)^{\msbar} }}
\newc{\mbmbsmmsbar}{\ensuremath{ m_b(m_b)^{\msbar}_{{\mathrm{SM}}}}}
\newc{\mbmzsmmsbar}{\ensuremath{ m_b(\mz)^{\msbar}_{{\mathrm{SM}}}}}
\newc{\mbmzsmdrbar}{\ensuremath{ m_b(\mz)^{\drbar}_{{\mathrm{SM}}}}}
\newc{\mbmzmssmdrbar}{\ensuremath{ m_b(\mz)^{\drbar}_{{\mathrm{SUSY}}}}}
\newc{\mtaumzsmmsbar}{\ensuremath{ m_{\tau}(\mz)^{\msbar}_{{\mathrm{SM}}}}}
\newc{\mtaumzsmdrbar}{\ensuremath{ m_{\tau}(\mz)^{\drbar}_{{\mathrm{SM}}}}}
\newc{\mtaumzmssmdrbar}{\ensuremath{ m_{\tau}(\mz)^{\drbar}_{{\mathrm{SUSY}}}}}
\newc{\alphasmzms}{\ensuremath{\alpha_s(M_Z)^{\overline{MS}}}}
\newc{\alphaimzms}[1]{\ensuremath{\alpha_{#1}(M_Z)^{\overline{MS}}}}
\newc{\alphaemmz}{\ensuremath{\alpha_{\mathrm{em}}(M_Z)^{\overline{MS}}}}

\newc{\mzero}{\ensuremath{{m_0}}}
\newc{\mhalf}{\ensuremath{ m_{1/2}}}
\newc{\tanb}{\ensuremath{\tan\beta}}
\newc{\azero}{\ensuremath{ A_0}}
\newc{\bzero}{\ensuremath{ B_0}}
\newc{\signmu}{\ensuremath{\rm{sgn}\,\mu}}
\newc{\mueff}{\ensuremath{\mu_{\rm{eff}}}}
\newc{\lam}{\ensuremath{{\lambda}}}
\newc{\kap}{\ensuremath{{\kappa}}}
\newc{\alam}{\ensuremath{{A_{\lambda}}}}
\newc{\akap}{\ensuremath{{A_{\kappa}}}}
\newc{\hs}{\ensuremath{ H_s}}      
\newc{\mhs}{\ensuremath{ m_{H_s}}} 
\newc{\mgut}{\ensuremath{ M_{\rm GUT}}}
\newc{\mplanck}{\ensuremath{ M_{\rm P}}}      \newc{\mpl}{\ensuremath{ M_{\rm Pl}}}
\newc{\msusy}{\ensuremath{ M_{\rm SUSY}}}      \newc{\ms}{\ensuremath{ M_{\rm S}}}
 \newc{\mhl}{\ensuremath{m_\hl}} 
 \newc{\mhone}{\ensuremath{m_{h_1}}} 
 \newc{\mhtwo}{\ensuremath{m_{h_2}}} 
 \newc{\mglu}{\ensuremath{m_{\tilde g}}} 
 \newc{\mul}{\ensuremath{m_{\tilde{u}_L}}} 
 \newc{\mtone}{\ensuremath{m_{\tilde{t}_1}}} 
 \newc{\ma}{\ensuremath{m_A}} 
 \newc{\maone}{\ensuremath{m_{a_1}}} 
 \newc{\matwo}{\ensuremath{m_{a_2}}}
 \newc{\hone}{\ensuremath{h_1}}
 \newc{\htwo}{\ensuremath{h_2}}
 \newc{\aone}{\ensuremath{a_1}}
 \newc{\atwo}{\ensuremath{a_2}}
 \newc{\mhu}{\ensuremath{ m_{H_u}}}       
 \newc{\mhd}{\ensuremath{ m_{H_d}}}
 \newc{\mhusq}{\ensuremath{ m_{H_u}^2}}       
 \newc{\mhdsq}{\ensuremath{ m_{H_d}^2}}
 \newc{\mhuew}{\ensuremath{ m^{\ast}_{H_u}}}       
 \newc{\mhdew}{\ensuremath{ m^{\ast}_{H_d}}}
 \newc{\mhuewsq}{\ensuremath{ m^{\ast\, 2}_{H_u}}}       
 \newc{\mhdewsq}{\ensuremath{ m^{\ast\, 2}_{H_d}}}
 \newc{\hu}{\ensuremath{ H_u}}       
 \newc{\hd}{\ensuremath{ H_d}}
 \newc{\barmhu}{\ensuremath{ \bar{m}_{H_u}}}
 \newc{\barmhd}{\ensuremath{ \bar{m}_{H_d}}}

 \newc{\mqthree}{\ensuremath{m_{\widetilde{Q}_3}^2}}
 \newc{\muthree}{\ensuremath{m_{\tilde{u}_3}^2}}
 \newc{\mdthree}{\ensuremath{m_{\tilde{d}_3}^2}}
 \newc{\mlthree}{\ensuremath{m_{\widetilde{L}_3}^2}}
 \newc{\methree}{\ensuremath{m_{\tilde{e}_3}^2}}
 \newc{\mqtwo}{\ensuremath{m_{\widetilde{Q}_2}^2}}
 \newc{\mutwo}{\ensuremath{m_{\tilde{u}_2}^2}}
 \newc{\mdtwo}{\ensuremath{m_{\tilde{d}_2}^2}}
 \newc{\mltwo}{\ensuremath{m_{\widetilde{L}_2}^2}}
 \newc{\metwo}{\ensuremath{m_{\tilde{e}_2}^2}}
 \newc{\mqone}{\ensuremath{m_{\widetilde{Q}_1}^2}}
 \newc{\muone}{\ensuremath{m_{\tilde{u}_1}^2}}
 \newc{\mdone}{\ensuremath{m_{\tilde{d}_1}^2}}
 \newc{\mlone}{\ensuremath{m_{\widetilde{L}_1}^2}}
 \newc{\meone}{\ensuremath{m_{\tilde{e}_1}^2}}
 \newc{\msmul}{\ensuremath{m_{\tilde{\mu}_L}}}
 \newc{\msmur}{\ensuremath{m_{\tilde{\mu}_R}}}
 \newc{\msneumu}{\ensuremath{m_{\tilde{\nu}_{\mu}}}}
 \newc{\mone}{\ensuremath{M_1}}
 \newc{\monesq}{\ensuremath{M_1^2}}
 \newc{\mtwo}{\ensuremath{M_2}}
 \newc{\mtwosq}{\ensuremath{M_2^2}}
 \newc{\mthree}{\ensuremath{M_3}}
 \newc{\mthreesq}{\ensuremath{M_3^2}}
 \newc{\atau}{\ensuremath{{A_{\tau}}}}
 \newc{\at}{\ensuremath{{A_{t}}}}
 \newc{\ab}{\ensuremath{{A_{b}}}}
 \newc{\atausq}{\ensuremath{{A_{\tau}^2}}}
 \newc{\atsq}{\ensuremath{{A_{t}^2}}}
 \newc{\absq}{\ensuremath{{A_{b}^2}}}

 \newc{\dmzero}{\ensuremath{\Delta{_{m_0}}}}
 \newc{\dmhalf}{\ensuremath{\Delta{_{m_{1/2}}}}}
 \newc{\dmu}{\ensuremath{\Delta{_{\mu}}}}

 \newc{\pten}{\ensuremath{\psi_{10}}}
 \newc{\ffive}{\ensuremath{\phi_{5}}}
 \newc{\hfive}{\ensuremath{h_{5}}}
 \newc{\hbfive}{\ensuremath{h_{\bar{5}}}}
 \newc{\thet}{\ensuremath{\theta_{50}}}
 \newc{\thetb}{\ensuremath{\theta_{\,\overline{50}}}}
 \newc{\ptenhat}{\ensuremath{\hat{\psi}_{10}}}
 \newc{\ffivehat}{\ensuremath{\hat{\phi}_{5}}}
 \newc{\hfivehat}{\ensuremath{\hat{h}_{5}}}
 \newc{\hbfivehat}{\ensuremath{\hat{h}_{\bar{5}}}}
 \newc{\thethat}{\ensuremath{\hat{\theta}_{50}}}
 \newc{\thetbhat}{\ensuremath{\hat{\theta}_{\,\overline{50}}}}
 \newc{\si}{\ensuremath{\Sigma}}
 \newc{\mfive}{\ensuremath{m_5^2}}
 \newc{\mten}{\ensuremath{m_{10}^2}}
 \newc{\dfive}{\ensuremath{\Delta^2_5}}
 \newc{\dbfive}{\ensuremath{\Delta^2_{\bar{5}}}}
 \newc{\dfifty}{\ensuremath{\Delta^2_{50}}}
 \newc{\dfiftyb}{\ensuremath{\Delta^2_{\,\overline{50}}}}
 \newc{\msi}{\ensuremath{m_{\Sigma}^2}}
 \newc{\lamh}{\ensuremath{\lambda_{H}}}
 \newc{\lamhb}{\ensuremath{\lambda_{\bar{H}}}}
 \newc{\ah}{\ensuremath{A_{H}}}
 \newc{\ahb}{\ensuremath{A_{\bar{H}}}}
 \newc{\lams}{\ensuremath{\lambda_{S}}}
 \newc{\as}{\ensuremath{A_{S}}}
 \newc{\lamsig}{\ensuremath{\lambda_{\si}}}
 \newc{\asig}{\ensuremath{A_{\si}}}

 \newc{\msten}{\ensuremath{m_{16}^2}}
 \newc{\mhun}{\ensuremath{m_{126}^2}}
 \newc{\mhunb}{\ensuremath{m_{\bar{126}}^2}}
 \newc{\mthun}{\ensuremath{m_{210}^2}}
 \newc{\ahun}{\ensuremath{A_{\bar{126}}}}
 \newc{\yhun}{\ensuremath{Y_{\bar{126}}}}
 \newc{\aten}{\ensuremath{A_{10}}}
 \newc{\yten}{\ensuremath{Y_{10}}}
 \newc{\alone}{\ensuremath{A_{\lambda_1}}}
 \newc{\altwo}{\ensuremath{A_{\lambda_2}}}
 \newc{\althree}{\ensuremath{A_{\lambda_3}}}
 \newc{\althreeb}{\ensuremath{A_{\bar{\lambda_3}}}}
 \newc{\lone}{\ensuremath{\lambda_1}}
 \newc{\ltwo}{\ensuremath{\lambda_2}}
 \newc{\lthree}{\ensuremath{\lambda_3}}
 \newc{\lthreeb}{\ensuremath{\bar{\lambda_3}}}

\newc{\sigsip}{\ensuremath{\sigma^{\rm SI}_{p}}}	\newc{\sigsin}{\ensuremath{\sigma^{\rm SI}_{n}}}
\newc{\sigsdp}{\ensuremath{\sigma^{\rm SD}_{p}}}	\newc{\sigsdn}{\ensuremath{\sigma^{\rm SD}_{n}}}
\newc{\sigsi}{\ensuremath{\sigma^{\rm SI}}}	\newc{\sigsd}{\ensuremath{\sigma^{\rm SD}}}
\newc{\sigv}{\ensuremath{\sigma v}}
\newc{\abund}{\ensuremath{ \Omega h^2}}
\newc{\omegadm}{\ensuremath{ \Omega_{{\rm DM}}}}     \newc{\abunddm}{\ensuremath{ \Omega_{{\rm DM}} h^2}} 
\newc{\omegam}{\ensuremath{ \Omega_{{\rm m}}}}       \newc{\abundm}{\ensuremath{ \Omega_{{\rm m}} h^2}}
\newc{\omegab}{\ensuremath{ \Omega_{{\rm b}}}}	\newc{\abundb}{\ensuremath{ \Omega_{{\rm b}} h^2}}
\newc{\omegatot}{\ensuremath{ \Omega_{{\rm TOT}}}}
\newc{\omegacdm}{\ensuremath{ \Omega_{{\rm CDM}}}}   \newc{\abundcdm}{\ensuremath{ \Omega_{{\rm CDM}} h^2}}
\newc{\omegalambda}{\ensuremath{ \Omega_{\Lambda}}} \newc{\abundlambda}{\ensuremath{ \Omega_{\Lambda} h^2}}
\newc{\omegarad}{\ensuremath{ \Omega_{{\rm rad}}}}  \newc{\abundrad}{\ensuremath{ \Omega_{{\rm rad}} h^2}}
\newc{\rhocrit}{\ensuremath{ \rho_{\rm crit}}}
\newc{\rhochi}{\ensuremath{ \rho_{\chi}}}
\newc{\abunchi}{\ensuremath{\Omega_\chi h^2}}
\newc{\abundlsp}{\ensuremath{\Omega_{\rm LSP}h^2}}

\newc{\amu}{\ensuremath{ a_{\mu}}}        \newc{\amususy}{\ensuremath{ a_{\mu}^{\mathrm{SUSY}}}}
\newc{\amuexpt}{\ensuremath{ a_{\mu}^{\mathrm{expt}}}}        \newc{\amusm}{\ensuremath{ a_{\mu}^{\mathrm{SM}}}}
\newc\deltaamu{\ensuremath{\Delta a_{\mu}}} \newc{\deltaamususy}{\ensuremath{\delta a_{\mu}^{\mathrm{SUSY}}}}
\newc\gmtwo{\ensuremath{ (g-2)_{\mu}}} 
\newc{\deltagmtwomususy}{\ensuremath{\delta\left(g-2\right)_{\mu}^{\mathrm{SUSY}}}}
\newc{\deltagmtwomu}{\ensuremath{\delta\left(g-2\right)_{\mu}}}
\newc\BR{\ensuremath{\textrm{BR}}}
\newc\bsgamma{\ensuremath{ b\rightarrow s \gamma }}
\newc\bxsgamma{\ensuremath{\overline{B}\rightarrow X_{s}\gamma}}
\newc\brbsgamma{\ensuremath{\BR\left(\bsgamma\right)}}
\newc\brbxsgamma{\ensuremath{\BR\left(\bxsgamma\right)}}
\newc\bsmumu{\ensuremath{B_s\to\mu^+\mu^-}}
\newc\brbsmumu{\ensuremath{\BR\left(B_s\to\mu^+\mu^-\right)}}
\newc\bdmmumu{\ensuremath{\overline{B}_d\to\mu^+\mu^-}}
\newc\bbbarmix{\ensuremath{\overline{B}_s\mbox{-}B_s}}      
\newc\delmbs{\ensuremath{\Delta M_{B_s}}}
\newc{\butaunu}{\ensuremath{B_u \rightarrow \tau \nu}}
\newc{\brbutaunu}{\ensuremath{\BR\left(B_u \rightarrow \tau \nu\right)}}

\newc{\brmuegamma}{\ensuremath{\BR\left(\mu^{\pm}\rightarrow e^{\pm}\gamma\right)}}
\newc{\brtauegamma}{\ensuremath{\BR\left(\tau^{\pm}\rightarrow e^{\pm}\gamma\right)}}
\newc{\brtaumugamma}{\ensuremath{\BR\left(\tau^{\pm}\rightarrow \mu^{\pm}\gamma\right)}}
\newc{\brmuthreee}{\ensuremath{\BR\left(\mu^{\pm}\rightarrow e^{\pm}e^+e^-\right)}}
\newc{\brtauthreee}{\ensuremath{\BR\left(\tau^{\pm}\rightarrow e^{\pm}e^+e^-\right)}}
\newc{\brtauthreemu}{\ensuremath{\BR\left(\tau^{\pm}\rightarrow \mu^{\pm}\mu^+\mu^-\right)}}

\newcommand*{\reffig}[1]{Fig.~\ref{#1}}
 
        \newcommand*{\refeq}[1]{Eq.~(\ref{#1})}
     \newcommand*{\refsec}[1]{Sec.~\ref{#1}}

\newcommand*{\sarah}{{\tt SARAH}}
\newcommand*{\spheno}{{\tt SPheno}}
\newcommand*{\fkit}{{\tt FlavorKit}}

\newcommand*{\chep}{{\tt CalcHEP}}
\newcommand*{\mad}{{\tt MadGraph5$\_$aMC@NLO}}

\newcommand*{\micromegas}{{\tt MicrOMEGAs}}
\newcommand*{\multinest}{{\tt MultiNest}}

\newcommand*{\pythia}{{\tt PYTHIA}}

\newcommand*{\delphes}{{\tt DELPHES 3}}

\let\oldcite\cite
\renewcommand*{\cite}{~\oldcite}

\newcommand*{\hl}{\ensuremath{h}}

\newc{\glzmu}{\ensuremath{{g^{Z\mu \mu}_{L}}}}
\newc{\grzmu}{\ensuremath{{g^{Z\mu \mu}_{R}}}}
\newc{\glwmu}{\ensuremath{{g^{W\mu \nu_\mu}_{L}}}}
\newc{\grwmu}{\ensuremath{{g^{W\mu \nu_\mu}_{R}}}}
\newc{\glzmuSM}{\ensuremath{{g^{Z\mu \mu}_{L,\textrm{SM}}}}}
\newc{\grzmuSM}{\ensuremath{{g^{Z\mu \mu}_{R,\textrm{SM}}}}}
\newc{\glwmuSM}{\ensuremath{{g^{W\mu \nu_\mu}_{L,\textrm{SM}}}}}
\newc{\grwmuSM}{\ensuremath{{g^{W\mu \nu_\mu}_{R,\textrm{SM}}}}}

\restylefloat{figure}

\title{Expectations for the muon $g-2$ in simplified models with dark matter}

\author[a]{Kamila Kowalska}
\author[a,b]{and Enrico Maria Sessolo}

\affiliation{$^a$ Institut f\"ur Physik, Technische Universit\"at Dortmund,\\ Otto-Hahn Str. 4, D-44221 Dortmund, Germany\\
$^b$ National Centre for Nuclear Research,\\
Ho{\.z}a 69, 00-681 Warsaw, Poland} 

\emailAdd{kamila.kowalska@tu-dortmund.de}
\emailAdd{enrico.sessolo@ncbj.gov.pl}

\abstract{We investigate simplified models of new physics that can accommodate the measured value of the anomalous magnetic moment 
of the muon and the relic density of dark matter. 
We define a set of renormalizable, SU(2)$\times$U(1) invariant extensions of the Standard Model, 
each comprising an inert $\mathbb{Z}_2$-odd scalar field and one or more vector-like pairs of colorless fermions 
that communicate to the muons through Yukawa-type interactions. The new sectors are classified according to their transformation 
properties under the Standard Model gauge group and all models are systematically confronted with a variety of 
experimental constraints: LEP mass bounds, direct LHC searches, electroweak precision observables, and direct searches 
for dark matter. We show that scenarios featuring only one type of new fermions 
become very predictive once the relic density and collider constraints are taken into account, 
as in this case \gmtwo\ is not enhanced by chirality flip. Conversely, for models where an additional source of chiral-symmetry violation is generated via fermion mixing, the constraints are much looser and 
new precision experiments with highly suppressed systematic uncertainties may be required to test the parameter space.}

\dedicated{DO-TH 17/13,\\ QFET-2017-11}

\begin{document}
\maketitle
\section{Introduction\label{sec:intro}}

When the anomalous magnetic moment of the muon, \gmtwo, was measured at BNL several years ago\cite{Bennett:2006fi},
it showed a discrepancy with the Standard Model (SM) expectation that has since been widely interpreted as 
a hint for new physics not far from the 
electroweak symmetry-breaking (EWSB) scale. After taking into account a recent update\cite{Davier:2016iru} of the lowest-order hadronic contributions to the SM calculation, the discrepancy, \deltagmtwomu, 
is estimated to be at the level of $\sim3.5\,\sigma$:
$\deltagmtwomu = (27.4 \pm 7.6 ) \times 10^{-10}$.
(A second recent review of the hadronic vacuum polarization 
contributions and uncertainties\cite{Jegerlehner:2017lbd} yields an even more convincing $\deltagmtwomu = (31.3 \pm 7.7 ) \times 10^{-10}$.)

While the interest in this anomaly never really went away, it is bound to receive a boost by the start of the new Muon g-2 
experiment at Fermilab\cite{Grange:2015fou,Chapelain:2017syu},
which will improve the statistical precision of the measurement by a factor of four or so with respect to BNL.
Additionally, just a few years after the Fermilab experiment, \gmtwo\ will also be measured 
at J-PARC\cite{Ishida:2009zz,Mibe:2010zz,Iinuma:2011zz,Saito:2012zz,Otani:2015jra}, which is expected to reach a comparable sensitivity even if the experimental setup is different.

From the theory standpoint, the anomaly can be accommodated in many scenarios beyond the Standard Model (BSM) 
(see, e.g.,\cite{Jegerlehner:2009ry} for a comprehensive review). Early on, the impact of new physics 1-loop contributions to \deltagmtwomu\cite{Leveille:1977rc,Moore:1984eg} was investigated predominantly in the framework of supersymmetry\cite{Fayet,Grifols:1982vx,Ellis:1982by,Barbieri:1982aj,Romao:1984pn,
Kosower:1983yw,Yuan:1984ww,Vendramin:1988rd,Moroi:1995yh,Cho:2000sf,Martin:2001st}, but
the consequences of a positive measurement for more generic models were also explored\cite{Czarnecki:2001pv,Lynch:2001zs}.  
Recently, among the large number of studies appearing every year on the topic, 
Ref.\cite{Freitas:2014pua} and Refs.\cite{Queiroz:2014zfa,Lindner:2016bgg}
have adopted a systematic approach based on simplified models instead of focusing on specific constructions.
Simplified models are characterized by a limited number of free parameters and classified according to 
the gauge quantum numbers of the particles introduced and 
the Lorentz structure of their interactions, and can be confronted with
a variety of experimental information, like LEP/LHC constraints in\cite{Freitas:2014pua} or flavor observables in\cite{Lindner:2016bgg}.

In this paper, extending the approach of\cite{Freitas:2014pua,Queiroz:2014zfa,Lindner:2016bgg}, 
we try to answer the following question: in case a positive measurement of \deltagmtwomu\ is obtained with
large significance at Fermilab, what information can we infer on the couplings, masses, and quantum numbers of the 
new particles involved in the process, 
provided we require that the same physics also yields the 
relic density of dark matter in the Universe.
As the nature of dark matter constitutes one of the greatest mysteries in contemporary particle physics, 
we think it is enticing to entertain the idea that a positive measurement at Fermilab and J-PARC could open a 
window into the nature of the dark sector, possibly in conjunction with other experimental signatures.  
In fact, we will show that requiring the same physics to be responsible for the \gmtwo\ anomaly and dark matter
leads to strong bounds on the allowed parameter space and introduces a series of complementary signatures, 
in particular at the high-luminosity LHC, in future electroweak precision experiments and, to a lesser extent, 
in dark matter direct detection searches.

Obviously, this complementary approach is not original. It is invoked for instance in supersymmetry, 
where neutralinos, sleptons, and charginos often provide at the same time a good fit to \deltagmtwomu\ and the correct relic abundance.
It has also been recently adopted for other BSM models\cite{Belanger:2015nma}, and there exists at least one 
previous study of possible complementary signatures for \gmtwo\ and dark matter in
simplified models based on a minimal set of assumptions\cite{Agrawal:2014ufa}.    

Like Ref.\cite{Agrawal:2014ufa}, we define in this paper a set of minimal extensions of the SM that 
provide a viable weakly interactive massive particle (WIMP) dark matter candidate
and have the potential to give a positive signal in the upcoming \gmtwo\ experiments.
Unlike that work, however, we will not employ the effective field theory approach, 
nor limit ourselves to SM singlets in the interactions with the muons.
Rather, inspired by Ref.\cite{Freitas:2014pua}, we consider a set of scenarios in which both the dark matter and the 
BSM lepton mediator can transform non-trivially under the SU(2) gauge group. We will also 
always adopt the relic abundance as an important constraint on the parameter space. 

The models we construct are based on the following requirements:
\begin{itemize}
\item The dark matter interacts with the muons through renormalizable couplings
\item Interactions are CP conserving and invariant under the SM gauge group, SU(2)$\times$U(1)
\item Each model satisfies the constraints from perturbativity and unitarity
\item The measurement of the relic abundance is an active constraint on the parameter space.
\end{itemize}

We do not consider in this paper dark matter lighter than the mass of the muon so that, to make it stable, 
we introduce an additional discrete symmetry, $\mathbb{Z}_2$, under which the dark matter is odd and the SM is even. 
Also note that the dark matter must be \textit{leptophilic} to evade the stringent current bounds from direct detection experiments, 
and for this reason we neglect dark-matter interactions with quarks.
 
The first of the requirements listed above limits the allowed interactions to fermion--(pseudo)scalar--fermion and  fermion--(axial)vector--fermion types. As one of the participating fermions is necessarily the muon, the discrete symmetry forces us to additionally introduce
$\mathbb{Z}_2$-odd colorless fermions, which must be \textit{vector-like} (VL) to evade the bounds from electroweak precision observables (EWPOs) and to not introduce gauge anomalies.

Note that all our assumptions are trivially satisfied by Yukawa-type interactions fermion--(pseudo)scalar--fermion, 
once the appropriate scalar potential is spelled out. Conversely, interactions involving (axial)vector particles 
require additional assumptions, namely the definition of extra dark gauge groups and charges, as well as a careful treatment of lepton-flavor violating processes. 
For this reason, in this paper we limit ourselves to discussing fermion--(pseudo)scalar--fermion interactions only. 

For the sake of simplicity, and unlike Refs.\cite{Freitas:2014pua,Agrawal:2014ufa}, we also do not require
universal Yukawa interactions of the BSM sector with the SM leptons, which could give extra constraints from LEP measurements. We rather assume that the new scalars and fermions couple exclusively to the SM muons. Lepton-flavor violating processes are obviously absent in such a setup.
This assumption may seem somewhat \textit{ad hoc}, but it allows us to focus on effects arising in the muon sector only, without imposing additional model-dependent symmetries aimed at enforcing Minimal Flavor Violation.
It can also perhaps be seen as realistic in light of the recent flavor anomalies at LHCb\cite{Aaij:2017vbb}, 
which seem to point to the existence of lepton-flavor non-universality. 

The paper is organized as follows. In \refsec{sec:gm2} we briefly review the expressions of the 1-loop new physics contributions 
to \gmtwo\ in the case of scalar couplings. In \refsec{sec:models} we introduce the Lagrangians 
of our simplified models and describe their couplings to the muons. 
In \refsec{sec:constraints} we review the constraints we apply, and  in \refsec{sec:results} we present and extensively discuss 
the results of our numerical analyses. We finally summarize our findings and conclude in \refsec{sec:summary}.

\section{Muon $\boldsymbol{g-2}$ contributions from scalar interactions\label{sec:gm2}}

\begin{figure}[t]
\centering
\includegraphics[width=0.50\textwidth]{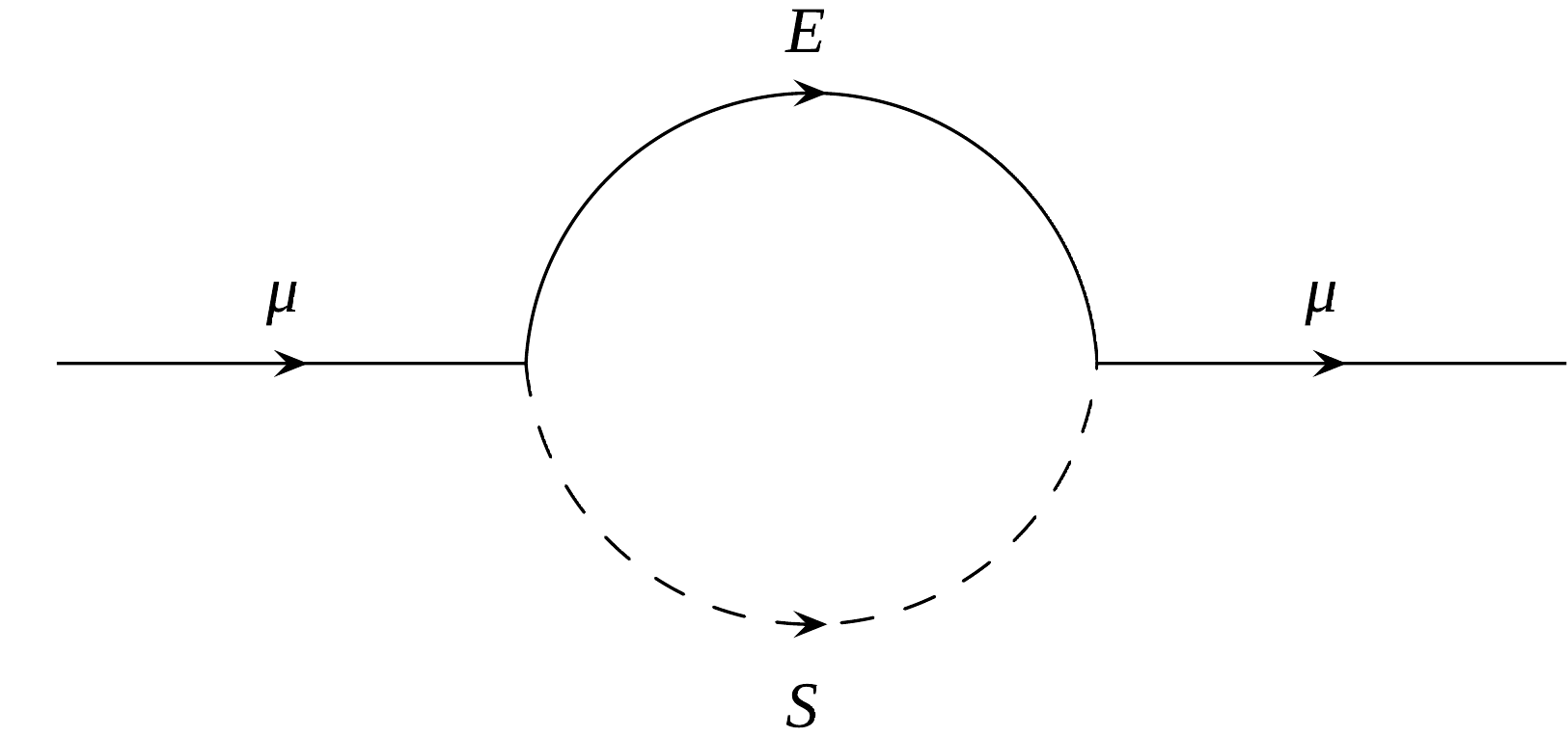}
\caption{The 1-loop contribution to \deltagmtwomu\ in the presence of a new scalar field $S$ and a new lepton $E$. 
A photon line attached to whichever particle is electrically charged is implied.}
\label{fig:g2_gen}
\end{figure}

The generic 1-loop contribution to the muon anomalous magnetic moment involving the $\mathbb{Z}_2$-odd sector 
is schematically shown in \reffig{fig:g2_gen}, 
where at least one of the particles in the loop must be electrically charged,
and there is an implied photon line attached to the charged propagator.
Interaction terms in the Lagrangian of the models we consider are given by
\be\label{scalarint}
\mathcal{L}\supset (g_s\,\bar{\psi}_{E}\psi_{\mu}\phi_S+ig_p\,\bar{\psi}_{E}\gamma^5\psi_{\mu}\phi_S+\textrm{h.c.})
-m_E\,\bar{\psi}_E\psi_E-V(\phi_S), 
\ee
in terms of a Yukawa coupling $g_s$ ($g_p$) of the muon to a (pseudo)scalar field $S$ -- whose dynamics is described by an
appropriate scalar potential $V(\phi_S)$ -- and a generic heavy fermion $E$ of mass $m_E$.

The specific value of \gmtwo\ depends on the electric charge and spin quantum numbers of the particles running in the loop.
Considering, for example, a charged fermion and a neutral scalar, one gets 
(see, e.g.,\cite{Jegerlehner:2009ry,Lindner:2016bgg} for a review of the calculation)
\begin{multline}
\deltagmtwomu=\frac{m_{\mu}^2}{8 \pi^2 m_S^2}\left[\left(|g_s|^2+|g_p|^2\right)\int_0^1dx\,\frac{x^2(1-x)}{(1-x)(1-\lambda^2 x)+\epsilon^2\lambda^2 x}\right.\\
+\left.\epsilon\left(|g_s|^2-|g_p|^2\right)\int_0^1dx\,\frac{x^2}{(1-x)(1-\lambda^2 x)+\epsilon^2\lambda^2 x}\right],\label{mastergm2}
\end{multline}
in terms of the mass ratios of the new particles to the muon, 
$\epsilon=m_E/m_{\mu}$, $\lambda=m_{\mu}/m_S$\,.

Note that the chiral structure of the underlying model plays a crucial role in determining the size of \gmtwo.
If one sets $g_p=0$, the dominant contribution in \refeq{mastergm2} arises from the second term, which is enhanced by a large factor $\epsilon$.
On the other hand, the presence of a non-zero pseudoscalar coupling will reduce the size of this term, and in the case when
$|g_s|=|g_p|$ the only remaining contribution is the one in the first line of \refeq{mastergm2}.

This well-known fact is often used to obtain, by simple inspection of the Lagrangian, back-of-the-envelope estimates
of how well a specific model can perform with respect to \gmtwo.
The prescription is best recast in terms of $c_L$ and $c_R$, the couplings of the new physics to the left- and right-chiral 
Weyl components of the muon.
One can explicitly write the scalar and pseudoscalar coupling of \refeq{scalarint} as $g_s=(c_R+c_L)/2$ 
and $i g_p=(c_R-c_L)/2$, and the integrals can be easily calculated for $\epsilon\lambda$ fixed in the approximation $\lambda\ll 1$. 
By defining 
$\epsilon^2\lambda^2=m_E^2/m_S^2\equiv r$, one gets
\be
\int_0^1dx\,\frac{x^2(1-x)}{1-x+r x}=\frac{2+3r-6r^2+r^3+6r\ln r}{6\,(r-1)^4}\equiv\mathcal{F}_1(r)
\ee
\be
\int_0^1dx\,\frac{x^2}{1-x+r x}=\frac{3-4r+r^2+2\ln r}{2\,(r-1)^3}\equiv\mathcal{F}_2(r)\,,
\ee
which lead to the well known formula
\be\label{chiralint}
\deltagmtwomu=\frac{1}{16\pi^2}\sum_{S^0,E^{\pm}}
\left[\frac{m_{\mu}^2}{m_S^2}\left(|c_L|^2+|c_R|^2\right)\mathcal{F}_1(r)
+2\,\frac{m_{\mu} m_E}{m_S^2}\,\Re(c_L c_R^{\ast})\,\mathcal{F}_2(r)\right]\,,
\ee
where we have generalized \refeq{mastergm2} to include all possible charged fermion and neutral scalar states coupling to the muon. 

Equation~(\ref{chiralint}) expresses the fact that much larger values of \gmtwo\ can be expected if the new physics 
cross-couples to \textit{both} chiral states of the muon, providing a chirality-flip term that is enhanced proportionally 
to the mass of the new fermions.
Note, incidentally, that if $\Re(c_L c_R^{\ast})$ is positive, \deltagmtwomu\ is positive. 

One can follow a similar procedure to derive the \gmtwo\ formula for the case of charged scalars and 
neutral fermions in the loop, obtaining
\be\label{chargescal}
\deltagmtwomu=\frac{1}{16\pi^2}\sum_{S^{\pm},E^0}\left[-\frac{m_{\mu}^2}{m_S^2}\left(|c_L|^2+|c_R|^2\right)\mathcal{G}_1(r)
+2\,\frac{m_{\mu} m_E}{m_S^2}\,\Re(c_L c_R^{\ast})\,\mathcal{G}_2(r)\right]\,,
\ee
where 
\be
\mathcal{G}_1(r)\equiv\frac{1-6 r+3 r^2+2 r^3-6 r^2\ln r}{6(r-1)^4}
\ee
\be
\mathcal{G}_2(r)\equiv\frac{-1+r^2-2 r\ln r}{(r-1)^3}\,.
\ee
This contribution is negative when either $c_L$ or $c_R$ are equal zero, but can become positive for 
$\Re(c_L c_R^{\ast})\neq 0$.

Finally we will also need the contribution to \gmtwo\ from a doubly charged fermion and a charged scalar in the loop.
In this work we will just use the one for the $c_L$ coupling, which reads\cite{Freitas:2014pua} 
\be\label{doubl_char}
\deltagmtwomu=\frac{1}{16\pi^2}\sum_{S^{\mp},E^{\pm\pm}}
\left(\frac{m_{\mu}^2}{m_S^2}\,|c_L|^2\,\mathcal{H}_1(r)\right),
\ee
where 
$\mathcal{H}_1(r)=2\mathcal{F}_1(r)+\mathcal{G}_1(r)$.
 
\section{Muophilic portal models\label{sec:models}}

Before we introduce our simplified models for \gmtwo\ and dark matter, we start by defining the notation we adopt throughout the paper.

We indicate the SM Weyl spinor fields with lower-case letters, and the new $\mathbb{Z}_2$-odd fields with capital letters.
The quantum numbers of the leptons and Higgs boson of the SM are
\begin{equation}
l:(\mathbf{1},\mathbf{2},-1/2)\,\,\,\,\,\,\, e_R:(\mathbf{1},\mathbf{1},-1)\,\,\,\,\,\,\, \phi:(\mathbf{1},\mathbf{2},1/2),
\end{equation}
with respect to  SU(3)$\times$SU(2)$\times$U(1). 
When written explicitly, the lepton doublet reads $l=(\nu_L,e_L)^T$ and the Higgs field is $\phi=[0,(v+h)/\sqrt{2}]^T$ after EWSB. 
The Dirac fermion of charge $-1$ is constructed as $\psi_e=(e_L,e_R)^T$, as usual.

For the SM muon Yukawa coupling we use the convention, in Weyl notation, 
\be
\mathcal{L}\supset -y_{\mu}\,\phi^{\dag} l\,e_R^{\ast}+\textrm{h.c.}\label{LagrSM}
\ee

As was discussed in \refsec{sec:intro}, we limit ourselves to considering CP conserving, renormalizable Yukawa interactions.
We extend the SM particle content with new heavy scalar fields and VL leptons, 
which provide generally safe extensions of the SM because they do not introduce gauge anomalies and they 
avoid the stringent bounds on chiral fermions from LEP and SLC precision measurements.
In the context of non-supersymmetric models, VL leptons for the \gmtwo\ anomaly have also been considered 
in\cite{Czarnecki:2001pv,Kannike:2011ng,Kanemitsu:2012dc,Dermisek:2013gta,Freitas:2014pua,Belanger:2015nma}.

We will next classify our models according to the transformation properties of the new scalar and fermion fields under the SM gauge group.

\subsection{Models with a real neutral scalar singlet\label{sec:realsca}} 

We begin with the simplest case, extending the SM by a singlet real scalar field,
\begin{equation}\label{scalarqn}
s:(\mathbf{1},\mathbf{1},0).
\end{equation}
For simplicity, we assume in this work that any new introduced scalar is \textit{inert}, 
in the sense that it does not develop a vacuum expectation value (vev). 

The most general $\mathbb{Z}_2$-symmetric renormalizable scalar potential that includes 
mass terms and quartic interactions for both the Higgs and the BSM scalar, as well as a portal coupling between the two, takes the form
\begin{equation}
\label{scalpot_real}
V=-\mu^2 \phi^{\dag}\phi+\frac{\lambda}{2}  (\phi^{\dag}\phi)^2+\frac{\mu_s^2}{2} s^2+\frac{\lambda_s}{2} s^4+\lambda_{12}s^2\phi^{\dag}\phi\,,
\end{equation}
in terms of 5 free parameters: $\mu$ and $\lambda$,  the SM mass and quartic coupling, 
$\mu_s$, $\lambda_s$, and the portal coupling $\lambda_{12}$.
The tree level mass of the BSM scalar is in this case given by 
\begin{equation}
\label{real_mass}
m_s^2=\mu^2_s+\lambda_{12}v^2.
\end{equation}

The parameters of the scalar potential are constrained by theoretical requirements. 
In order to guarantee that the electroweak vacuum is a global minimum, 
it is required that $\mu_s^2+\lambda_{12}\,v^2>0$. One also needs  $\mu_s^2>0$, so that the direction $v=0,v_s\neq 0$ is not a minimum  and the scalar $s$ remains inert. The latter also guarantees that the $\mathbb{Z}_2$ symmetry is preserved in the electroweak broken phase. As a result, there is no mixing between the new scalar and the SM Higgs.
Vacuum stability, i.e. requiring that the potential is bounded from below, requires $\lambda>0$, $\lambda_s>0$, and 
$\lambda_{12}>-\sqrt{\lambda\lambda_s}$. Moreover, perturbative unitarity bounds give $\lambda<4.2$, 
$\lambda_s<4.2$, and $\lambda_{12}<25$\cite{Lewis:2017dme}.

\begin{figure}[t]
\centering
\includegraphics[width=1.0\textwidth]{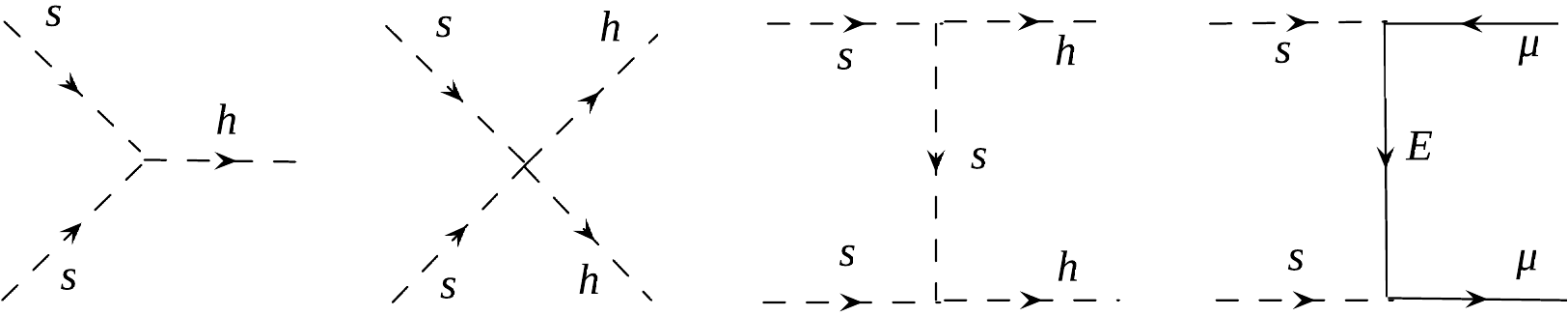}
\caption{Starting from the left, the first 3 diagrams show the well-known scalar portal interactions that can potentially lead to the correct 
dark matter relic abundance. The last diagram on the right depicts the $t$-channel ``bulk'' mechanism via a new heavy fermion $E$.}
\label{fig:hportal}
\end{figure}

The scalar potential given in \refeq{scalpot_real} features well known dark matter properties 
(see\cite{McDonald:1993ex,Bento:2000ah,Burgess:2000yq,Davoudiasl:2004be,Patt:2006fw,Barger:2007im} for early
papers exploring the Higgs portal). The WIMP is here the scalar $s$, which efficiently annihilates in the early Universe
through the interaction vertices depicted in the first three diagrams of \reffig{fig:hportal}.
However, it is also well known that the portal coupling and dark matter mass are subject to strong bounds from direct detection searches, 
which, after the most recent bound from XENON-1T\cite{Aprile:2017iyp} are considered, 
exclude the mass range $m_{\textrm{DM}}=m_s\approx 10-800\gev$ if one imposes the 
relic density constraint. This is precisely the mass range where the new scalar field and VL leptons can positively contribute 
to the \gmtwo\ anomaly.

Interestingly, the presence of VL fermions opens up additional mechanisms for dark matter annihilation, 
as shown in the last diagram on the right of \reffig{fig:hportal}.
The limits from direct detection searches can be then easily evaded, thanks to the leptophilic nature of the interaction between the 
dark matter scalar and the SM. By borrowing a term often used in supersymmetry, we will hence refer to this mechanism as 
the \textit{bulk}. (See, e.g, Refs.\cite{Drees:1992am,Baer:1995nc} for a definition of the bulk in supersymmetry,\cite{Bai:2014osa} in the context of simplified models, and\cite{Fukushima:2014yia,Belanger:2015nma} in relation to \gmtwo.)

We will see in what follows that, once the new fermions are introduced in the theory to explain the \gmtwo\ anomaly, the bulk 
emerges in the vast majority of cases as the favored mechanism for the dark matter relic density. 
However, we leave in our numerical scans the portal parameters of the scalar potential free to float, 
as they are allowed by the symmetries and can be constrained to small values case by case by the phenomenology. 
Besides, this also allows us to investigate regions of the parameter space characterized by a mixed bulk/portal mechanism for dark matter.

For the nature of the VL fermions, we consider all possibilities allowed by the SM gauge symmetry:\medskip 

\textbf{Model 1: Fermion singlets.}

In the first class of models we add a pair of charged lepton SU(2) singlets,
\begin{equation}
E:(\mathbf{1},\mathbf{1},-1),\quad E':(\mathbf{1},\mathbf{1},1),\label{singqn}
\end{equation}
which are odd under $\mathbb{Z}_2$. 
We can write new muophilic Weyl terms in the Lagrangian,
\begin{equation}\label{mod1lag}
\mathcal{L}\supset -Y_S\,E\,e_R^{\ast} s -M_E E' E+\textrm{h.c.},
\end{equation}
in terms of a new Yukawa coupling to the second generation, $Y_S$, and a new VL mass, $M_E$.

One can also construct a new Dirac spinor, $\psi_{\widetilde{E}}=(E,E'^{\ast})^T$, 
which leads to the Dirac-type interaction
\begin{equation}
\mathcal{L}\supset -Y_S\, \bar{\psi}_{\widetilde{E}}P_R\psi_e\,s +\textrm{h.c.}\,,\label{Dirac1}
\end{equation}
with $P_R=(1+\gamma_5)/2$. 

From \refeq{Dirac1} one can immediately read off the scalar and pseudoscalar couplings of \refeq{scalarint}: 
$g_s=Y_S/2$ and $i g_p=Y_S/2$. 
Alternatively, one can use the chiral formalism defined before \refeq{chiralint} and get
$c_L=0$, $c_R=Y_S$, which can be then plugged in 
to calculate \deltagmtwomu. 
Since in this scenario the amplitude is not enhanced by a chirality-flip term one expects to fit the \gmtwo\ anomaly 
either in the presence of large $Y_S$ values, or with relatively small mass for the new scalar and fermion particles.

We subject Model~1, and all the models we define in the next paragraphs and sections, 
to several constraints from different experiments, which we describe in detail in \refsec{sec:constraints}. 
This will allow us to systematically discriminate which region of the parameter space is more likely to
give a signature in future \gmtwo\ experiments, 
within the individual model themselves and in relation to the others.

The constraints we consider in Model~1 are
\begin{itemize}
\item LHC 13 TeV bounds from searches for leptons and missing $E_T$\cite{ATLAS:2016uwq,CMS:2017fij}
\item LHC 13 TeV mono-jet search bounds\cite{Aaboud:2016tnv}
\item EWPO constraints from the $Z$ lineshape and asymmetry data at LEP
and measurements of the muon lifetime and $W$ mass\cite{Olive:2016xmw}
\item Where applicable (portal couplings), direct detection constraints from LUX\cite{Akerib:2016vxi} and XENON1T\cite{Aprile:2017iyp}.
\end{itemize}
\medskip

\textbf{Model 2: Fermion doublets.}

One can instead add to the Lagrangian a VL pair of SU(2) doublets,
\begin{equation}\label{vl_doublets}
L:(\mathbf{1},\mathbf{2},-1/2),\quad L':(\mathbf{1},\mathbf{\bar{2}},1/2),
\end{equation}
where we explicitly write the doublets as $L=(N_1,E_1)^T$ and $L'=(N_2,E_2)$.
The heavy Dirac lepton of charge $-1$ is given, similarly to Model~1, 
by $\psi_E=(E_1,E_2^\ast)^T$, and there is also a heavy Dirac neutrino, whose  
mass is at the tree level degenerate with the charged lepton's.

As before we write new muophilic Weyl terms,
\begin{equation}
\mathcal{L}\supset -Y_D\,L'l s-M_L L'L+\textrm{h.c.}
\label{Lagr1}
\end{equation}
or, explicitly,
\begin{equation}
\mathcal{L}\supset -Y_D\,E_2\,e_L\,s - Y_D N_2\,\nu_L\,s +\textrm{h.c.}\,,
\label{Lagr1b}
\end{equation}
which lead to the same type of Dirac interaction with muons as in \refeq{Dirac1}:
\begin{equation}
\mathcal{L}\supset -Y_D\, \bar{\psi}_{E}P_L\psi_e\,s +\textrm{h.c.}\,,
\end{equation}
where $P_L=(1-\gamma_5)/2$. 
As before, one can read off $g_s=Y_D/2$, $-ig_{p}=Y_D/2$, or $c_L=Y_D$, $c_R=0$.
The effect on the calculation of \gmtwo\ ends up being the same as in Model~1. 

\newpage
\textbf{Model~3. Fermion singlets and fermion doublets.}

The most straightforward way of introducing a chirality-flip term in the \gmtwo\ calculation is letting
the singlet and doublet fermions of Model~1 and Model~2 mix with each other through the SM Higgs vev.

One can thus write down
\begin{equation}
\mathcal{L}\supset -Y_S\,E\,e_R^{\ast} s-Y_D\,L'l s-\widetilde{Y}_1\,\phi^{\dag}L\,E'-\widetilde{Y}_2\,L'\phi\,E- 
M_E E'E-M_L L'L+\textrm{h.c.}\,,\label{doubsing}
\end{equation}
where, in agreement with the SM and $\mathbb{Z}_2$ symmetries, we have introduced two additional 
Yukawa couplings to the Higgs boson, $\widetilde{Y}_1$ and $\widetilde{Y}_2$\,.

There are now two, charged, heavy Dirac leptons, and their mass matrix 
in the basis $\{(E_2,E')\times(E_1,E)^T\}$ is given by 
\begin{equation}\label{physmas}
M_C= \left( \begin{array}{cc}
M_{L} & \frac{\widetilde{Y}_2\,v}{\sqrt{2}}  \\
\frac{\widetilde{Y}_1\,v}{\sqrt{2}} & M_{E}  \\
\end{array} \right),
\end{equation}
and is diagonalized in the usual way by two unitary matrices $U$ and $V$, 
with the convention that $\textrm{diag}(m_{E_1^p},m_{E_2^p})=U^{\ast}M_C V^{\dag}$.

For each heavy lepton, $E_{i=1,2}^p$, 
the product $c_L^i c_R^{i\,\ast}$ depends on the mixing parameters: 
\bea\label{mixing}
c_L^i c_R^{i\,\ast}&=& -Y_D V_{i1}\,Y_S U_{i2}\nonumber\\
 &\approx& Y_D Y_S\,\frac{\max(M_L,M_E)\,M_{C\{ij\}}+\min(M_L,M_E)\,M_{C\{ji\}}}{m_{E^p_j}^2-m_{E^p_i}^2}\big|_{j\neq i}\,,
\eea
where the second line of \refeq{mixing} is given in terms of $M_{C\{12\}}=\widetilde{Y}_2v/\sqrt{2}$ and $M_{C\{21\}}=\widetilde{Y}_1v/\sqrt{2}$. Note that $c_L^i c_R^{i\,\ast}$ vanishes when \textit{both} 
$\widetilde{Y}_1$ and $\widetilde{Y}_2$ are zero, thus reproducing the limit of Models~1 and 2.

We subject Model~3 to the constraints described above for Model~1. Since the explicit mixing terms of \refeq{doubsing} 
break chiral symmetry, we expect the constraints from EWPOs to bite significantly into the parameter space that produces 
an enhancement in \deltagmtwomu.
Note also that by introducing an additional source of chiral-symmetry breaking this terms 
potentially induce large loop corrections to the muon mass.
However, as was pointed out in, e.g.,\cite{Fukushima:2014yia} for the equivalent supersymmetric case, 
these first order corrections do not depend on the cut-off scale and can be simply treated as a mild source of fine tuning, which should 
be minimal after the constraints from EWPOs are taken into account.
   
\subsection{Models with a complex neutral scalar singlet\label{sec:compscal}}

We extend the models of \refsec{sec:realsca} by replacing the real scalar field of \refeq{scalarqn} with a complex scalar. 
We define $S=(s+ia)/\sqrt{2}$\,, expressed in terms of 2 real degrees of freedom.

The $\mathbb{Z}_2$ symmetric scalar potential now reads, 
\begin{eqnarray}
V&=&-\mu^2 \phi^{\dag}\phi+\lambda/2\;  (\phi^{\dag}\phi)^2+\mu_S^2/2\;|S|^2+\lambda_S/2\; |S|^4+ \lambda_{12}|S|^2\phi^{\dag}\phi\nonumber\\
 & &+(\mu_S'^2/2\;S^2+\lambda_S'/2\;S^4+\lambda_{12}'S^2\phi^{\dag}\phi+\lambda_{22}'/2\;S^2|S|^2+\textrm{h.c.})\,,
\end{eqnarray}
where we have introduced new portal couplings, $\lam'_{12}$ and $\lam'_{22}$, which again we leave free to float but 
will not play an important role in the dark matter discussion.

Tree-level vacuum stability requires in this case $\lambda>0$, $\tilde{\lambda}_s>0$, $\tilde{\lambda}_a>0$, $\lambda_S>-1/6\sqrt{\tilde{\lambda}_a\tilde{\lambda}_s}$, $\lambda_{12}-2\lambda_{12}'>-\sqrt{\lambda\tilde{\lambda}_a}$, and $\lambda_{12}+2\lambda_{12}'>-\sqrt{\lambda\tilde{\lambda}_s}$,
where $\tilde{\lambda}_s=\lambda_S+2\lambda_S'+2\lambda_{22}'$, $\tilde{\lambda}_a=\lambda_S+2\lambda_S'-2\lambda_{22}'$.
The $\mathbb{Z}_2$ symmetry is preserved if $\mu_S^2-2\mu_S'^2>0$ and $\mu_S^2+2\mu_S'^2>0$.

The masses of dark sector scalars are given at the tree level by 
\begin{eqnarray}
m_s^2&=&\frac{1}{2}(\mu^2_S+\lambda_{12}v^2+2\mu_S'^2+2\lambda_{12}'v^2)\nonumber\\
m_a^2&=&\frac{1}{2}(\mu^2_S+\lambda_{12}v^2-2\mu_S'^2-2\lambda_{12}'v^2)\,.
\end{eqnarray}
Thus, this scenario is endowed with two possible dark matter candidates, a scalar and a pseudoscalar WIMP
(see, e.g.,\cite{Barger:2008jx,Gonderinger:2012rd} for early studies of the complex scalar/Higgs portal).

By mirroring the models of \refsec{sec:realsca}, we introduce \textbf{Model~1a},
\textbf{Model~2a}, and  \textbf{Model~3a}, obtained by performing in the Lagrangians of 
Eqs.~(\ref{mod1lag}), (\ref{Lagr1}), and (\ref{doubsing}) the substitutions $s\rightarrow S$ and $s\rightarrow S^{\ast}$.
For example, Model~1a is characterized by the Lagrangian
\begin{equation}
\mathcal{L}\supset -Y_S\,E\,e_R^{\ast} S -Y_{S^{\ast}}\,E\,e_R^{\ast} S^{\ast} -M_E E' E+\textrm{h.c.},
\end{equation}
from which one reads the chiral couplings of the muon to the scalar and pseudoscalar fields:
$c_R^{(s)}=(Y_S+Y_{S^{\ast}})/\sqrt{2}$\,, which can become significant even within the perturbativity bound for the individual Yukawa couplings, and $c_R^{(a)\ast}=i(Y_S-Y_{S^{\ast}})/\sqrt{2}$, which instead, being expressed as a difference, is in general a bit smaller. Besides, $c_L^{(s)}=c_L^{(a)}=0$ like in the model with a real scalar.

Similar arguments apply to the corresponding extensions of Models~2 and 3 of \refsec{sec:realsca}. 
In particular, in Model 3a, we mirror the singlet/doublet fermion mixing of Model~3 by adding 2 additional Yukawa couplings,
which have the effect of broadly opening up the parameter space, and making it more difficult to constrain.

We conclude this subsection by pointing out that we do not treat in this work models with charged scalar singlet fields 
and neutral fermions only, as in their simplest implementation they provide a negative contribution to \deltagmtwomu, see \refeq{chargescal}.  
We shall see in subsequent sections, however, that in the presence of neutral fermion mixing the second term of \refeq{chargescal}
can provide a positive contribution to \gmtwo.

\subsection{Models with a scalar doublet\label{sec:scadoub}}

Finally, we consider models with an inert scalar doublet in place of the inert scalar singlet of 
\refsec{sec:realsca} and \refsec{sec:compscal}.
We introduce
\begin{equation}\label{doubqn}
\Phi:(\mathbf{1},\mathbf{2},1/2),
\end{equation} 
where $\Phi=(S^+,S^0/\sqrt{2})^T$ in terms of a charged scalar field $S^{\pm}$ and a neutral complex $S^0=s+ia$.

At the tree level the scalar potential reads, 
\begin{eqnarray}
V&=&-\mu^2\,\phi^{\dag}\phi+\lambda/2\; (\phi^{\dag}\phi)^2+\mu_{\Phi}^2/2\, \Phi^{\dag}\Phi+\lambda_{\Phi}/2\; ( \Phi^{\dag}\Phi)^2\nonumber\\
 & &+\lambda_3\,\Phi^{\dag}\Phi\,\phi^{\dag}\phi+\lambda_4\,(\Phi^{\dag}\phi)(\phi^{\dag}\Phi)+(\lambda_5\,(\Phi^{\dag}\phi)^2+\textrm{h.c.})\,.
\end{eqnarray}
Vacuum stability requires $\lambda>0$, $\lambda_{\Phi}>0$, $\lambda_{3}>-\sqrt{\lambda\lambda_{\Phi}}$, and  $\lambda_{3}+\lambda_{4}\pm2\lambda_{5}>-\sqrt{\lambda\lambda_{\Phi}}$, as well as $\mu_{\Phi}^2>0$.

At the tree level the masses of dark sector scalars are given by 
\begin{eqnarray}\label{doublet_mass}
m_s^2&=&\frac{1}{2}\mu^2_{\Phi}+\frac{1}{2}v^2(\lambda_3+\lambda_{4}+2\lambda_{5})\nonumber\\
m_a^2&=&\frac{1}{2}\mu^2_{\Phi}+\frac{1}{2}v^2(\lambda_3+\lambda_{4}-2\lambda_{5})\nonumber\\
m_{S^{\pm}}^2&=&\frac{1}{2}(\mu^2_{\Phi}+v^2\lambda_3)\,.
\end{eqnarray}
As a consequence, the dark matter can be either a scalar or pseudoscalar, but additional constraints on the parameter space arise 
from the condition that the charged scalar is not the lightest one, and it evades the LEP limit, $m_{S^{\pm}}>100\gev$. 
The portal interactions and dark matter of this class of models have been investigated extensively
(early studies include\cite{Barbieri:2006dq,Ma:2006km,Ma:2006fn,LopezHonorez:2006gr,Majumdar:2006nt})
but, once more, in most of the cases described here the relevant mechanism for the relic density will be provided by the bulk.

The fermion fields can in this case be grouped in 4 categories according to their SU(2) representation:
singlet, doublet, triplet, and adjoint triplet. As before, we also consider the possibility of doublet/singlet mixing and doublet/triplet mixing through the Higgs vev.
\medskip

\textbf{Model~4. Fermion singlets.}

In terms of the singlets defined in \refeq{singqn} the Lagrangian reads
\begin{equation}
\mathcal{L}\supset -Y_S\,\Phi^{\dag}l\,E' -M_E E E'+\textrm{h.c.},
\end{equation}
where we have introduced, with some redundancy in the notation, a new Yukawa coupling, $Y_S$, and VL mass, $M_E$.
Writing explicitly the components of the SM doublet, one gets
\be\label{scal_doub_fer_sing}
\mathcal{L}\supset -Y_S\left(\nu_L S^-E'+e_L \frac{S^{0\ast}E'}{\sqrt{2}}\right)+\textrm{h.c.}\,,
\ee
which gives $c_L^{(s)}=Y_S/\sqrt{2}$, $c_L^{(a)}=-iY_S/\sqrt{2}$, and
$c_R^{(s,a)}=0$.
Note that even though the chiral structure of \refeq{scal_doub_fer_sing} resembles the scalar singlet case of Model 1a, 
the presence of a non-zero coupling of VL fermions to muon neutrinos will have an important impact on the LHC phenomenology.
\medskip

\textbf{Model~5. Fermion doublets.}

For the VL doublets introduced in \refeq{vl_doublets}, the Lagrangian reads
\begin{equation}\label{doubdoub}
\mathcal{L}\supset -Y_D\,\Phi^{\dag}L e^{\ast}_R -M_L L' L+\textrm{h.c.}\,,
\end{equation}
in terms of a new Yukawa coupling, $Y_D$, and VL mass, $M_L$. 

Beside a positive contribution to \gmtwo, similar to \refeq{scal_doub_fer_sing} and \refeq{Lagr1b}, there is 
a negative contribution due to the charged scalar field,
\begin{equation}\label{sca_doub_fer_doub}
\mathcal{L}\supset -Y_D \left(N_1 S^- e^{\ast}_R+\frac{E_1 S^{0\ast}}{\sqrt{2}} e^{\ast}_R\right)+\textrm{h.c.}
\end{equation}
This feature suggests that boosting \gmtwo\ to the experimentally measured value may be more challenging in this case, see \refeq{chargescal}.
\medskip

\textbf{Model~6. Fermion singlets and fermion doublets.}

The fermion doublets of Model~5 can mix with the singlets of Model~4 through the Higgs boson vev.
The Lagrangian includes the terms
\be
\mathcal{L}\supset -Y_S\,\Phi^{\dag}\,l\,E'-Y_D\,\Phi^{\dag}L e^{\ast}_R
-\widetilde{Y}_1\,\phi^{\dag}L\,E'-\widetilde{Y}_2\,L'\phi\,E+\textrm{h.c.}\,,
\ee
with two additional Yukawa couplings, in a fashion similar to Model~3 in \refsec{sec:realsca}. 

Note, however, that even if the mixing provides a source of chiral symmetry breaking similar to the one giving a boost to 
\gmtwo\ in Model~3, we do not expect an enhancement in this model. 
The reason is that there are two neutral real scalar fields, $s$ and $a$, 
whose couplings to the muon have opposite parity but the same size.
One can read off from Eqs.~(\ref{scal_doub_fer_sing}) and (\ref{sca_doub_fer_doub}) that $c_{L}^{(a)}=-i c_{L}^{(s)}$ and 
$c_{R}^{(a)\ast}=-i c_{R}^{(s)\ast}$, so that $\Re(c_{L}^{(a)}c_{R}^{(a)\ast})=-\Re(c_{L}^{(s)}c_{R}^{(s)\ast})$ and 
the last term in \refeq{chiralint} is identically zero. 
As a consequence, Model~6 does not result much more interesting from a 
phenomenological point of view than the individual models comprising it, and we do not consider it further. 
\medskip

\textbf{Model~7. Fermion triplets.}

In this case one extends the SM by adding fermion SU(2) triplets with the following quantum numbers,
\begin{equation}
\Psi_T:(\mathbf{1},\mathbf{3},-1)\,,\quad \Psi'_T:(\mathbf{1},\mathbf{\bar{3}},1)\,,
\end{equation}
which can be parameterized in terms of a neutral, charged, and doubly charged component as
\[  \Psi_T'=
  \left( {\begin{array}{cc}
\frac{\Psi^+}{\sqrt{2}} & \Psi^{++} \\
\Psi^{0} & -\frac{\Psi^+}{\sqrt{2}}
 \end{array} } \right),
\]
and equivalent decomposition applies to $\Psi_T$.

The Lagrangian reads
\begin{equation}
\mathcal{L}\supset -Y_T\,\Phi^{\dag}\,\Psi'_T\,l -M_T\,\textrm{Tr}(\Psi_T'\,\Psi_T) +\textrm{h.c.}\,,
\end{equation}
which is expanded into
\begin{equation}\label{trip}
\mathcal{L}\supset -Y_T\,\left(-\frac{\Psi^+ S^{0\ast}}{2}+\Psi^{++} S^-\right)e_L
-Y_T\left(\frac{\Psi^{0} S^{0\ast}}{\sqrt{2}}+\frac{\Psi^+S^-}{\sqrt{2}}\right)\nu_L+\textrm{h.c.}
\end{equation}
Equation~(\ref{trip}) includes the well-known doubly-charged fermion/charged scalar coupling, which will generate the 
large positive contribution to \gmtwo\ given in \refeq{doubl_char}.

Note that in this case too, the quantum numbers allow for doublet/triplet fermion mixing through the Higgs vev. 
As in the case just discussed above, however, the couplings of $s$ and $a$ to the muon have opposite parity and equal size, providing, again, an identical cancellation of the chirality-flip term.

\newpage
\textbf{Model 8. Fermion adjoint triplet.}

If the hypercharge of the VL fermion is zero, one obtains an adjoint SU(2) triplet:
\begin{equation}
\Psi_A:(\mathbf{1},\mathbf{3},0)
\end{equation}
where the triplet's matrix form is
\[  \Psi_A=
  \left( {\begin{array}{cc}
 \frac{\Psi^0}{\sqrt{2}} & \Psi^+ \\
 \Psi^- & -\frac{\Psi^0}{\sqrt{2}}
 \end{array} } \right),
\]
in terms of a charged fermion and a neutral Majorana field.

The Lagrangian in this scenario reads
\begin{equation}
\mathcal{L}\supset -Y_A\,(i\sigma_2\Phi)^T \Psi_A l -M_A \textrm{Tr}(\Psi_A\Psi_A)+\textrm{h.c.}
\end{equation}
This gives 
\begin{equation}
\mathcal{L}\supset -Y_A \left(\frac{\Psi^0 S^{+}}{\sqrt{2}}+\frac{\Psi^{+} S^0}{\sqrt{2}}\right)e_L
-Y_A  \left(\frac{\Psi^0 S^0}{2}-\Psi^- S^+\right)\nu_L+\textrm{h.c.}
\end{equation}
Like in Model~5, there is here a negative contribution to \gmtwo\ arising from the coupling of the muon with the charged scalar.
\medskip

\textbf{Model 9. Fermion adjoint triplet and fermion doublets.}

Positive contributions to \gmtwo\ arise when the adjoint fermion triplet mixes with the doublet through the SM Higgs vev.
As before, we add to the Lagrangian
\be\label{adj_mix}
\mathcal{L}\supset -\widetilde{Y}_1\,(i\sigma_2\phi)^T\Psi_A L -\widetilde{Y}_2\,(i\sigma_2\phi^{\ast})^T\Psi_A L' +\textrm{h.c.}
\ee

Note that, besides the mixing between two heavy charged fermions, \refeq{adj_mix} leads to the additional 
presence of two heavy neutral fermions mixing with each other. 
The chirality-flip contribution to \gmtwo\ is not suppressed in this case as, on the one hand,   
$c_{L}^{(a)}=i c_{L}^{(s)}$ and 
$c_{R}^{(a)\ast}=-i c_{R}^{(s)\ast}$, and on the other there is an additional positive-value loop involving the heavy
mixing ``neutrinos'', see the last term in \refeq{chargescal}.  
\bigskip

We conclude this section by pointing out that we do not treat cases with scalar SU(2) triplets in this work. 
The reason is twofold. On the one hand, it was pointed out in Ref.\cite{Freitas:2014pua} that in some cases 
(scalar triplet/fermion doublets, scalar adjoint triplet/fermion doublets, and scalar triplet/fermions adjoint-triplet and singlet) 
the 1-loop contribution to \gmtwo\ is negative.
On the other hand, even for the cases where a positive contribution exists (scalar adjoint triplet/fermions triplet and singlet),
the correct dark matter 
relic density can only be obtained with the scalar mass in the range of 5.5\tev\cite{Araki:2011hm}, which is obviously too 
high to accommodate the \gmtwo\ anomaly.

\section{Experimental constraints\label{sec:constraints}}

We review in this section the experimental constraints that can affect the allowed parameter space of the 
BSM models introduced in \refsec{sec:models}.

\subsection{Electroweak precision observables}

We subject all our models to electroweak precision constraints. Since VL fermions do not have tree-level 
axial-vector couplings, their contribution to EWPOs is expected to be small. However, in the models with mixing between 
fermions of different representations, and in models with scalar multiplets whose components are not mass-degenerate, 
loop-induced effects can be significant. 

In this work we compare to the experimental data two observables. We calculate the $Z\mu\bar{\mu}$ effective coupling 
and confront the result with precision fits for $g_A$ and $g_V$ from the $Z$ lineshape and asymmetry data 
at LEP and SLC\cite{ALEPH:2005ab}, as reported by the PDG\cite{Olive:2016xmw}. We also confront the corrections to the 
$W$ mass with its measured value.

To calculate loop corrections to the $Z\mu\bar{\mu}$ couplings and $W$ mass we follow the formalism of\cite{Hagiwara:1994pw}, 
which was adopted in, e.g.,\cite{Cho:1999km} for a precision analysis of supersymmetry. 
The conventions for Passarino-Veltman functions are also taken from\cite{Hagiwara:1994pw,Cho:1999km} and we 
use \texttt{LoopTools}\cite{Hahn:1998yk} for their calculation. 
The impact of precision observables in models with VL fermions for the \gmtwo\ anomaly 
has been also recently investigated in\cite{Kanemitsu:2012dc}.
\medskip

\textbf{1. Oblique parameters.}

Oblique parameters $S$, $T$ and $U$\cite{Peskin:1991sw} capture the BSM contributions to the gauge bosons' vacuum polarization.
$S$ is related to the difference between the number of the left- and right-handed weak fermion doublets, 
thus providing a measure of the breaking of the axial part of SU(2). 
This means that in the models with degenerate VL fermions the contribution to $S$ vanishes.

The oblique parameter $T$ is related to the difference between the $Z$ and $W$ bosons' self-energies, 
thus providing a measure of the breaking of the vector part of SU(2). 
As a result, it is sensitive to the mass splitting between the components of an electroweak doublet\cite{Olive:2016xmw},
\be
\Delta T=\frac{1}{32\pi^2 v^2\alpha}\sum \Delta m^2,
\ee
where $\alpha$ is the fine-structure constant, the sum runs over all non-degenerate doublets, and
\be
\Delta m^2=m_1^2+m_2^2-\frac{2m_1^2m_2^2}{m_1^2-m_2^2}\ln\frac{m_1^2}{m_2^2}.
\ee 

In cases where VL fermions mix, as in Model~3, 3a, and 9, 
their contribution to the parameters $S$ and $T$ can be parameterized in terms of the mixing matrices and of the physical masses. 
When calculating these effects, we use formulas derived in Ref.\cite{Joglekar:2012vc}.
A contribution to the oblique parameter $U$ is generally much smaller and can be neglected.
\medskip

\textbf{2. $\boldsymbol{Z\mu\bar{\mu}}$ vertex corrections.}

\begin{figure}[t]
\centering
\includegraphics[width=0.90\textwidth]{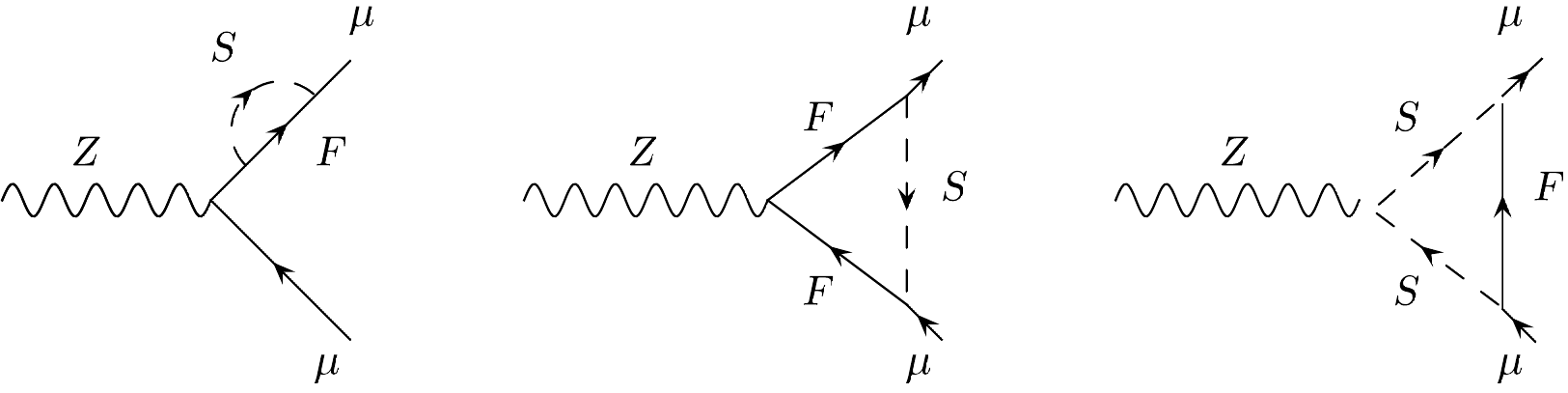}
\caption{One-loop BSM contributions to the $Z\mu\bar{\mu}$ vertex in extensions of the SM with a generic new fermion $F$ and a scalar $S$. }
\label{fig:zmumu}
\end{figure}

One-loop corrections to the coupling of the $Z$ boson with the left-handed and right-handed muon arise from diagrams 
like the ones depicted in \reffig{fig:zmumu}.
They are given by
\be
\Delta g_{L,R}^{\mu}=\frac{1}{\sqrt{4\sqrt{2}G_FM_Z^2}}\left[g_{L,R}^{\textrm{SM}}\,\Sigma'_{L,R}(0)-\Gamma_{L,R}(M_Z^2)\right],
\ee
where the $\Sigma'$ terms are the derivatives of the self-energy functions of the external fermion legs at zero momentum,
$\Gamma_{L,R}(M_Z^2)$ arise from triangular vertex corrections, and 
$g_{L,R}^{\textrm{SM}}=(-\hat{s}_{\textrm{W}}^2 Y+\hat{c}_{\textrm{W}}^2 T_3)\,g/c_{\textrm{W}}$.

In terms of Passarino-Veltman functions the $\Sigma'$ terms read
\be
\Sigma'_{L,R}(0)=\frac{1}{16\pi^2}\left[\sum_{F_i,S_j}|c_{L,R}^{ij}|^2\left(B_0+B_1)\right(0;m_{S_j},m_{F_i})\right],
\ee
where $c_{L,R}^{ij}$, the couplings of the generic new fermion $F_i$ and scalar $S_j$ to the muon, are given for our models 
in \refsec{sec:models}.

The contributions from triangle diagrams are
\begin{multline}\label{Z_gam}
\Gamma_{L,R}(M_Z^2)=-\frac{1}{16\pi^2}\sum_{F_i,S_j,F_k,S_l}\left\{ c_{L,R}^{ij\ast}c_{L,R}^{kj}\left[g_{L,R}^{Zik}\, m_{F_i}m_{F_k} C_0(p_1,p_2:m_{F_i},m_{S_j},m_{F_k})\right.\right.\\
\left.+g_{R,L}^{Zik}\,\left(-M_Z^2 C_{12}-M_Z^2 C_{23}-2C_{24}+\frac{1}{2}\right)(p_1,p_2:m_{F_i},m_{S_j},m_{F_k})\right]\\
\left.-c_{L,R}^{ij\ast}c_{L,R}^{il}\,g^{Zjl}\,2\,C_{24}(p_1,p_2:m_{S_j},m_{F_i},m_{S_l})\right\}\,,
\end{multline}
where the $g^{Zik}_{R,L}$ and $g^{Zjl}$ give the SM-like couplings of the $Z$ to the new fermions and scalars.

\begin{figure}[t]
\centering
\includegraphics[width=0.50\textwidth]{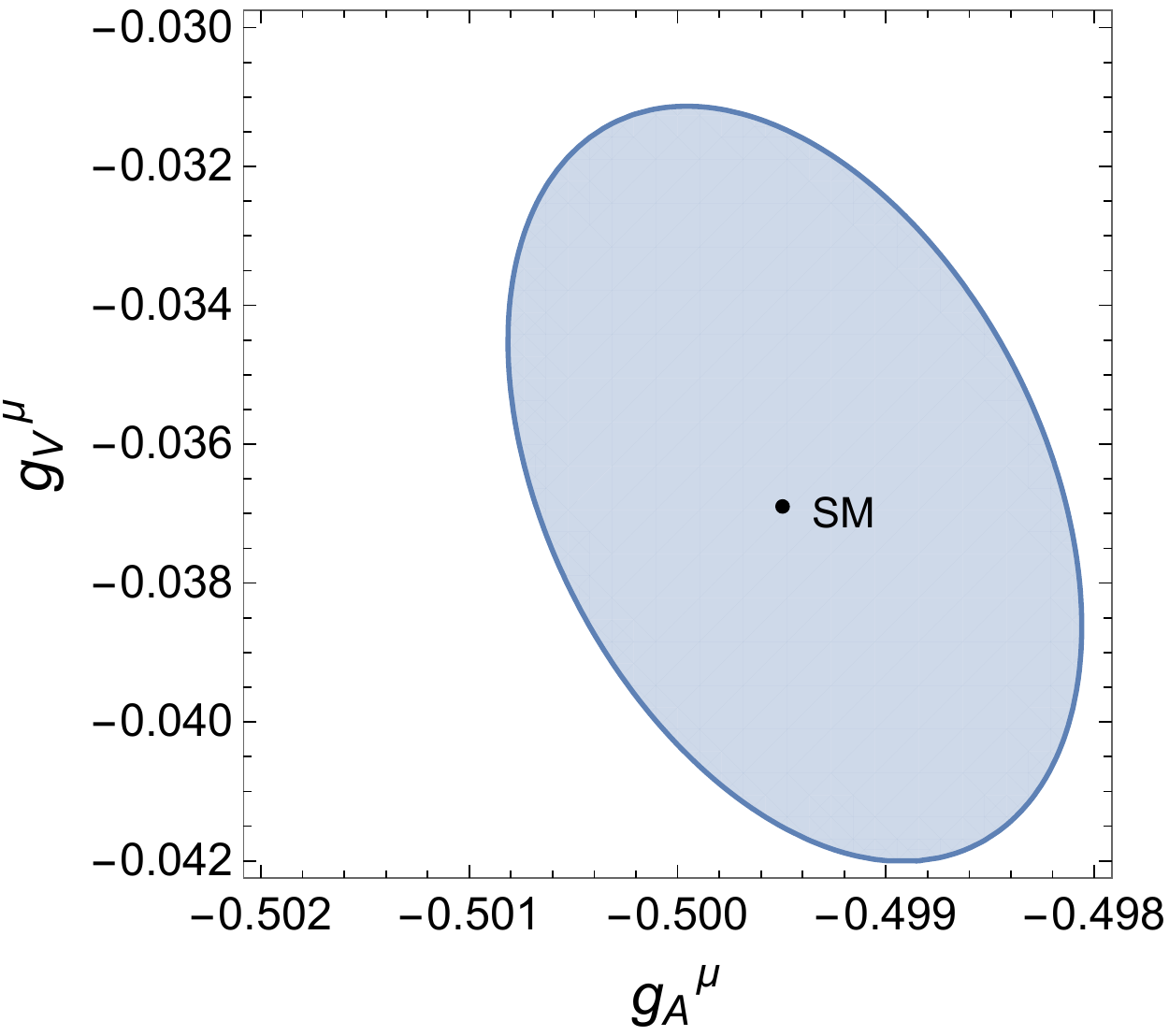}
\caption{An estimate of the 95\%~C.L. region in $g_A^{\mu}$, $g_V^{\mu}$ from a fit of $Z$ lineshape and asymmetry data 
at LEP and SLC\cite{ALEPH:2005ab}. The ellipse is obtained by rescaling the 39\%~C.L. region given in\cite{Olive:2016xmw}.}
\label{fig:Zline}
\end{figure}

After evaluating the observable couplings $g_V^{\mu}= g_L^{\mu}+g_R^{\mu}$ and $g_A^{\mu}=g_L^{\mu}-g_R^{\mu}$ from $g_L^{\mu}\equiv -0.2682+\Delta g_L^{\mu}$ and 
$g_R^{\mu}\equiv 0.2313+\Delta g_L^{\mu}$, we compare them to the 95\%~C.L. contour presented in \reffig{fig:Zline}, 
which we have approximately determined by rescaling up the 39\%~C.L. contour given in\cite{Olive:2016xmw}.

We also confront all our models with rough estimates of the reach in the possible future high-precision experiments 
GigaZ\cite{Erler:2000jg}, with an estimated improvement of a factor 20\cite{Baer:2013cma}
in the systematic uncertainty, and TLEP\cite{Gomez-Ceballos:2013zzn}, with an improvement of a factor 100.
The discovery potential of those experiments through precision measurements of $Z$-boson observables was 
also analyzed, e.g., in the context of leptoquarks in~Ref.\cite{ColuccioLeskow:2016dox}.

In principle, our models introduce loop contributions to the $h\rightarrow \mu^+\mu^-$ decay rate as well. 
However, these are supposed to be well within the present and foreseeable future uncertainties at the LHC, and for this reason we do not discuss them in this paper any longer. 
\medskip

\textbf{3. Constraints from the $\boldsymbol{W}$ mass.}

Additionally, we calculate the corrections to the $W$ mass in our models.
We follow, again, Refs.\cite{Hagiwara:1994pw,Cho:1999km}, which parameterize
\bea
M_W&=&M_W^{\textrm{SM}}+\Delta M_W\\
\Delta M_W&=&-0.288\Delta S+0.418 \Delta T+0.337 \Delta U-0.126 \frac{\Delta\overline{\delta}_G}{\bar{\alpha}},
\eea 
in terms of the usual oblique parameters $S,\,T,\,U$. We use $M_W^{\textrm{SM}}=80.361\gev$ and $\bar{\alpha}^{-1}=127.95$.
We neglect $\Delta U$, which is small, and we calculate the corrections to the $S$ and $T$ parameters using analytic formulas of 
Ref.\cite{Joglekar:2012vc} for the fermions, and of Ref.\cite{Barbieri:2006dq} for the scalars.

We calculate $\Delta\overline{\delta}_G=2 \delta^{v}$, the correction to the muon lifetime, as
\be\label{muondec}
\delta^{v}=\frac{\sqrt{2}}{\hat{g}}\,\Gamma^{W\mu\nu_{\mu}}(0)-\frac{1}{2}\left[\Sigma'_{\mu_L}(0)+\Sigma'_{\nu_{\mu}}(0)\right]\,,
\ee
where $\hat{g}$ is the $\overline{MS}$ value of the weak coupling constant and, once more, 
$\Sigma'$ and $\Gamma^{W\mu\nu_{\mu}}$ parameterize corrections to the external legs and triangle diagrams
modifying the $W\mu\nu_{\mu}$ vertex. 

When expressed in terms of the Passarino-Veltman functions they are given by 
\be
\Sigma'_{\mu_L,\nu_{\mu}}(0)=\frac{1}{16 \pi^2}\sum_{F_i,S_j} |c^{ij}_L|^2 (B_0+B_1) (0; m_{S_j}, m_{F_i})
\ee
and
\begin{multline}\label{mulife_vert}
\Gamma^{W\mu\nu_{\mu}}(0)=\frac{1}{16 \pi^2}\sum_{F_i,S_j,F_k,S_l}\left\{-c_L^{ij\ast}c_L^{kj}\left[g_L^{Wik} m_{F_i}m_{F_k}C_0\left(0;m_{F_i},m_{S_j},m_{F_k}\right)\right.\right.\\
\left.\left.+g_R^{Wik}\left(-2 C_{24}+\frac{1}{2}\right)\right]\left(0;m_{F_i},m_{S_j},m_{F_k}\right)
+c_L^{ij\ast}c_L^{il}g^{Wjl}\,2C_{24}\left(0;m_{S_j},m_{F_i},m_{S_l}\right)\right\}\,,
\end{multline}
where the symbols have equivalent meaning as in \refeq{Z_gam}.

\subsection{Collider constraints}

In all our models, we apply a default hard cut on the mass of new charged particles, $m_{E^{\pm},S^{\pm}}>100\gev$ to roughly 
take into account LEP~II limits. 

Moreover, since the VL fermions are charged under the SM electroweak gauge symmetry group, they can be pair-produced at the 
LHC in Drell-Yan processes, $pp\to Z,\gamma,W^{\pm}\to F\,\bar{F}$, and subsequently undergo Yukawa-driven decays into a DM scalar and a muon, $F\to S\, \mu$, thus leading to a characteristic 2 leptons plus missing energy (MET) signature.
Such a topology has been investigated by both ATLAS and CMS in the context of supersymmetry in the searches dedicated to sleptons, charginos and neutralinos.

In this work, we employ two different 2-lepton searches based on 13\tev\ data: the ATLAS search\cite{ATLAS:2016uwq}, based on hard leptons in the finals state, with an integrated luminosity of 13.3\invfb; and the CMS search\cite{CMS:2017fij}, based on soft leptons, 
with an integrated luminosity of 35.9\invfb. We numerically recast these two experimental analyses for 
the models introduced in \refsec{sec:models}, following the procedure described in detail in\cite{Kowalska:2015zja,Kowalska:2016ent} 
and references therein.
The main kinematical variable used in the ATLAS search to discriminate between the signal and the SM background is the 
stransverse mass $m_{T2}$\cite{Barr:2003rg}, with the end point correlated to the mass splitting between fermion and dark matter scalar, $\delta m$. 
As a result, the sensitivity of the search weakens when the mass splitting decreases, dropping to zero when $\delta m\approx 70\gev$. 
Conversely, the CMS analysis targets low-momentum leptonic final states and therefore can test the compressed spectra region, 
where the mass difference between fermion and dark matter is smaller than $\sim40\gev$.

Singlet scalar dark matter can be pair-produced at the LHC through the off-shell Higgs boson, $pp\to h^*\to S\,S$. 
The cross-section is in this case directly proportional to the size of the portal couplings and is not expected to be significant, given the discussed stringent bounds from direct detection experiments. For doublet scalar dark matter, Drell-Yan production with electroweak-size cross section is also possible, $pp\to Z,\gamma,W^{\pm}\to \Phi^{\dagger}\,\Phi$.
Such a signature can be probed by monojet searches, which tag an energetic jet from initial-state radiation recoiling against the produced dark matter. To capture this possibility, we recast the ATLAS 13\tev\ analysis\cite{Aaboud:2016tnv} with 3.2\invfb\ of data. 
Besides the present exclusion bounds, we also calculate the sensitivity of ATLAS searches\cite{ATLAS:2016uwq,Aaboud:2016tnv} 
with the assumed luminosity of $300\invfb$ at the LHC 14\tev\ run.

If $m_{DM}<m_h/2$, the Higgs boson can invisibly decay into dark matter with branching ratio proportional to the portal coupling(s). For completeness, we apply the CMS upper bound on the corresponding branching ratio, $\textrm{BR}(h\to\textrm{invisible})<0.24$ 
at the 95\% confidence level\cite{Khachatryan:2016whc}.

In the models characterized by the mixing of fermions with different SU(2) quantum numbers through the Higgs boson vev 
(Models~3, 3a, 9) we also apply $2\sigma$ 
constraints from the branching ratio $\textrm{BR}(h\rightarrow \gamma\gamma)$ at the LHC, 
which has been measured, e.g, by ATLAS\cite{ATLAS-CONF-2017-045}:
$R_{\gamma\gamma}\equiv\textrm{BR}(h\rightarrow \gamma\gamma)/\textrm{BR}(h\rightarrow \gamma\gamma)_{\textrm{SM}}=0.99\pm0.14$.
We calculate the $R_{\gamma\gamma}$ ratio at one loop following, e.g.,\cite{Kearney:2012zi}:
\be
R_{\gamma\gamma}\approx \left|1-0.109\sum_{i=1,2}\frac{2\,C_{hii}\,v}{m_{E^p_i}}\,A_{1/2}\left(\frac{4 m_{E^p_i}^2}{m_h^2}\right)\right|^2,
\ee
in terms of the physical masses defined in \refeq{physmas}, and the tree-level couplings of the Higgs boson $h$ to the heavy fermions,
$C_{hii}=(\widetilde{Y}_1 V_{i1}U_{i2}+\widetilde{Y}_2 V_{i2} U_{i1})/\sqrt{2}$. The loop function $A_{1/2}(x)$ can be found in\cite{Kearney:2012zi}.

Finally, we recall that we apply to all our models the relic density constraint from Planck\cite{Ade:2015xua}, $\abund=0.1188\pm 0.0010$, 
to which we add in quadrature a $\sim10\%$ theoretical uncertainty, and the XENON1T 90\%~C.L. upper bound on the spin-independent 
cross section \sigsip\cite{Aprile:2017iyp} as a hard cut. 

\section{Numerical analysis\label{sec:results}}

We present now the results of the numerical analysis of the models introduced in \refsec{sec:models}. 
We will start with a brief description of the numerical tools utilized in the study. 

Each of the considered models has been generated with \sarah\ v.4.9.3\cite{Staub:2013tta} and the corresponding \spheno\cite{Porod:2003um,Porod:2011nf} modules have been produced to calculate mass spectra and decay branching ratios. 
Flavor observables, including \deltagmtwomu, have been calculated with the \fkit\ package\cite{Porod:2014xia} of \sarah-\spheno.
Model files for \chep\cite{Belyaev:2012qa} were also generated and passed to \micromegas\ v.4.3.1\cite{Belanger:2013oya} to calculate dark matter related observables.

In order to efficiently scan the multidimensional parameter space, all the packages were interfaced to 
\multinest\ v.3.10\cite{Feroz:2008xx} for sampling. We emploied a Gaussian likelihood function to find the regions favored by 
the dark matter relic density and \deltagmtwomu. 

The parameters of the models were scanned in the following ranges:
\bea
0.001 \leq &\rm{Yukawa\;\; couplings} &\leq \sqrt{4\pi},\nonumber\\
-1 \leq &\rm{portal\;\; couplings}& \leq 1,\nonumber\\
100\; \rm{GeV} \leq & M_{L,E,T,A} & \leq 10000\; \rm{GeV},\nonumber\\
(10\gev)^2 \leq & \mu_s^2, \mu_S^2, \mu_{\Phi}^2 & \leq (5000\gev)^2, \nonumber\\
-0.5\,\mu_S^2 \leq & \mu_S'^2 & \leq 0.5\,\mu_S^2.
\eea

The LHC limits from the ATLAS 2-lepton and monojet searches have been implemented using the recast procedure described in detail in\cite{Kowalska:2015zja,Kowalska:2016ent} and adapted to handle non-SUSY scenarios. To this end, {\tt UFO} files have been generated with \sarah\ and passed to \mad\cite{Alwall:2014hca}, where a set of new BSM processes with a corresponding output for \pythia\cite{Sjostrand:2007gs} were created. Finally, the hadronization products were passed to the fast detector simulator \delphes\cite{deFavereau:2013fsa}.

We perform our numerical analysis at the tree level. 
In doing so, we are relying on some underlying assumptions that may not be always warranted, 
especially in theories including several scalar fields like the ones investigated here.

Scalar masses are affected by significant loop corrections, in particular those that depend on the BSM Yukawa couplings which, 
as we shall see, in general have to be sizable to accommodate the experimentally measured value of \deltagmtwomu. 
\sarah\ v.4.9.3 allows one to calculate 1-loop corrections to all scalar masses, and define the input parameter at a high 
renormalization scale of choice.
However, it has been pointed out\cite{Braathen:2017izn} that the results so implemented maintain a significant residual 
dependence on the renormalization scale. One should therefore make use of the full 2-loop calculation, 
which has only become available very recently\cite{Braathen:2017izn}.    

Thus, in this work we limit ourselves to the assumption generally adopted in the literature when dealing with non-supersymmetric 
BSM scalar fields, i.e., that it is possible to absorb corrections to the scalar masses into the counterterms of the free parameters.   
Note that effects of 1-loop corrections to the scalar mass in relation to the relic density in inert scalar models have been 
analyzed, e.g., in\cite{Hambye:2007vf,Goudelis:2013uca}. In Ref.\cite{Goudelis:2013uca}, in particular, it was shown that as long as the 
input parameters are defined not far above the EWSB scale, say up to 10\tev\ or so, the parameter space regions in agreement with the relic 
density are not altered drastically with respect to the tree level.  

\subsection{Real scalar with a singlet or doublet VL fermion\label{sec:result1}}

We begin our discussion with \textbf{Model~1}, characterized by the addition of a real scalar particle and VL fermion singlet fields to the spectrum of the SM. 
The $\mathbb{Z}_2$-odd scalar plays the role of the dark matter particle in this case, and we will refer to it with 
$m_{\textrm{DM}}\equiv m_s$ interchangeably.

\begin{figure}[t]
\centering
\subfloat[]{%
\includegraphics[width=0.47\textwidth]{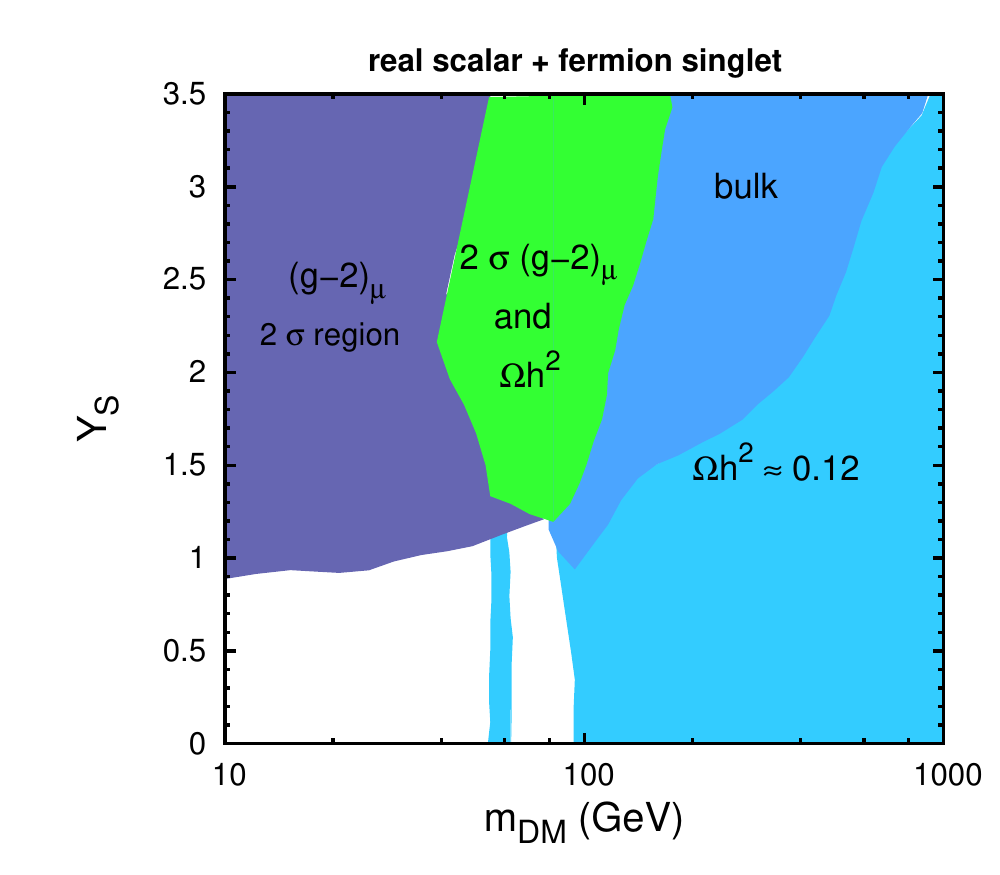}
}%
\\
\subfloat[]{%
\includegraphics[width=0.47\textwidth]{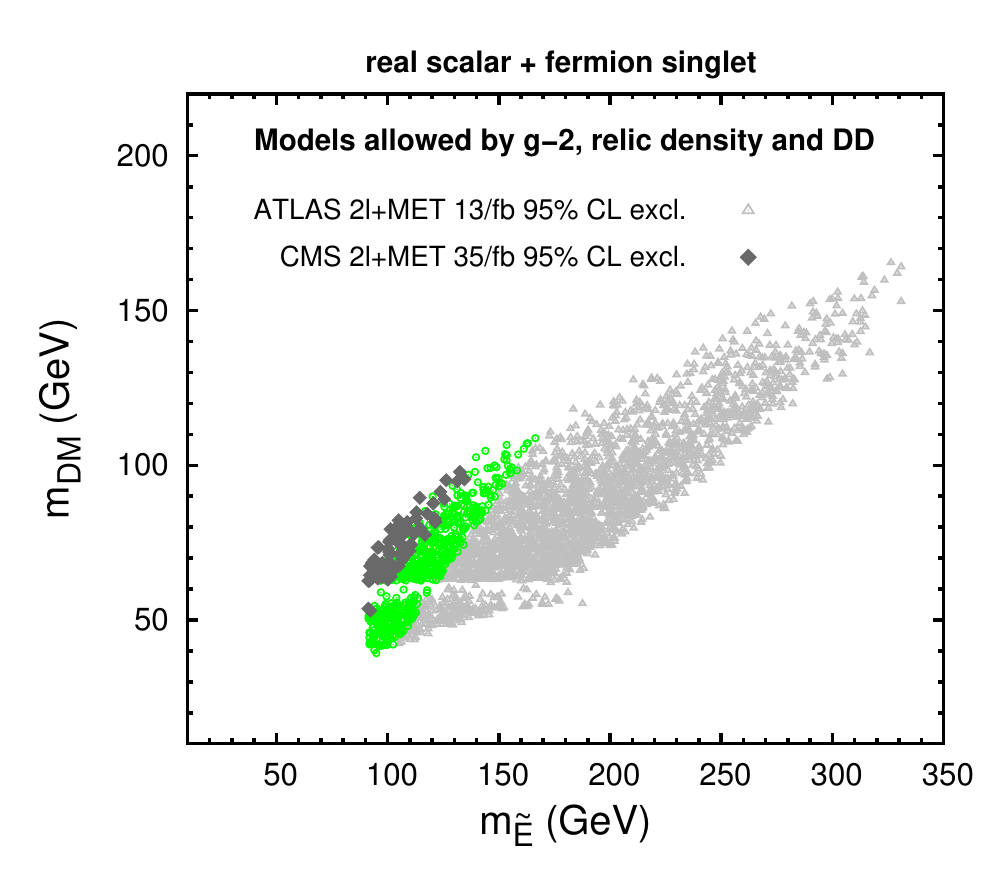}
}%
\hspace{0.02\textwidth}
\subfloat[]{%
\includegraphics[width=0.47\textwidth]{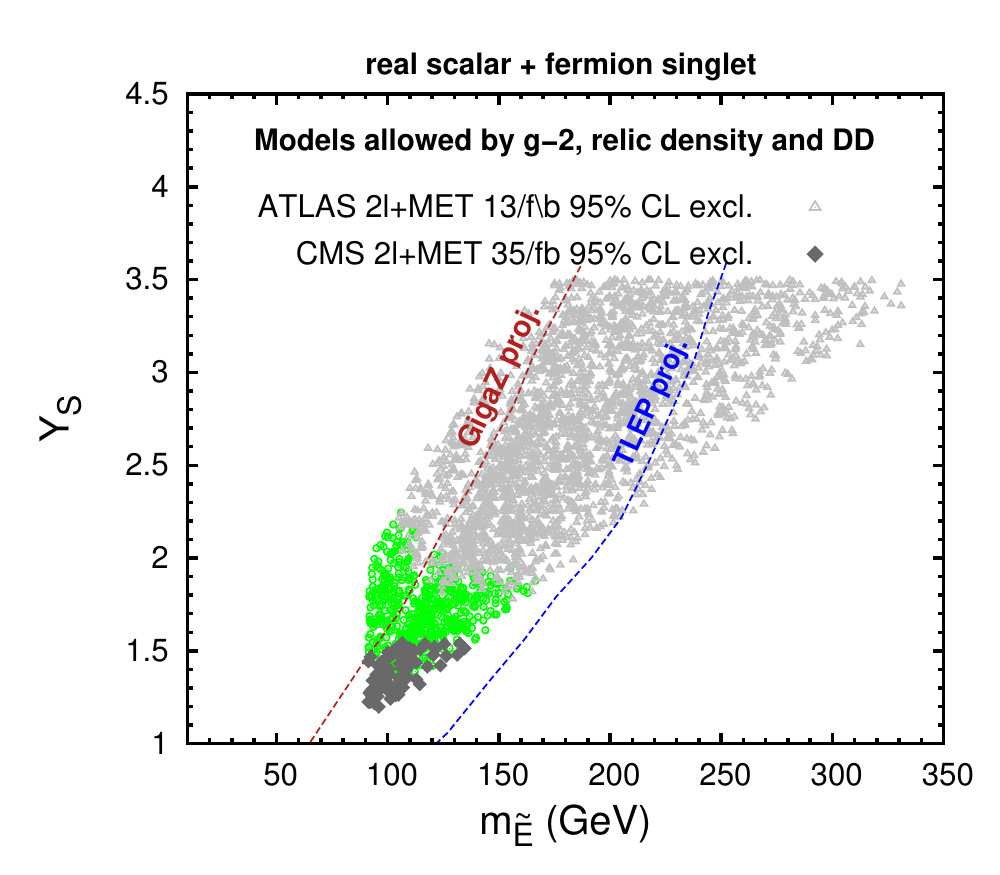}
}%
\caption{(a) The ($m_{\textrm{DM}}$, $Y_S$) plane for Model~1 (real scalar field and VL fermion singlet). 
In cyan, the parameter space favored at $2 \sigma$ by the relic density is shown, while the one favored 
by the \gmtwo\ measurement is shown in dark blue. Green region corresponds to those values of model parameters where both constraints are satisfied simultaneously.
(b) The parameter space common to the relic density and \gmtwo\ in the ($m_{\widetilde{E}}$, $m_{\textrm{DM}}$) plane. Gray triangles show the parameter space excluded by the ATLAS 2 hard leptons search\cite{ATLAS:2016uwq}, whereas dim gray diamonds show 
that excluded by the CMS 2 soft leptons search\cite{CMS:2017fij}.
(c) Same as (b) but in the ($m_{\widetilde{E}}$, $Y_S$) plane. The projected reach of precision measurements at GigaZ\cite{Erler:2000jg} 
and TLEP\cite{Gomez-Ceballos:2013zzn} is also shown.}
\label{fig:real_sing}
\end{figure}

In \reffig{fig:real_sing}(a) we present a plot of the model's parameter space in the plane of the new coupling to the muon, $c_R=Y_S$,
versus the dark matter mass. The parameter space allowed at $2 \sigma$ (including a $\sim10\%$ theory error)
by the relic density is shown in cyan, and we highlight with a darker shade the region in which $\abund\approx 0.12$ is due with good approximation exclusively to the bulk. The \gmtwo\ constraint is shown in dark blue and 
we do not impose at this stage any LHC or precision constraints.

Higgs-portal dark matter plays a small role, almost exclusively limited to the region above $0.8-1\tev$, in which the recent 
bounds from XENON1T can be evaded. Note that the relic abundance imposes a lower bound on the mass of the scalar particle,
$m_s=m_{\textrm{DM}}\gsim 40-50\gev$, as the bulk mechanism loses its efficiency when the spread between
$m_s$ and $m_{\widetilde{E}}$ is significant (recall that $m_{\widetilde{E}}\gsim100\gev$ by LEP bounds).
As we shall see below, this lower bound on the dark matter mass is model-dependent and can be evaded in other scenarios.

The parameter space allowed at $2\sigma$ by the combination of relic density and \gmtwo\ is 
shown in green. The $2\sigma$ region from the BNL measurement places an upper bound on the mass of the dark matter scalar, $m_s\lesssim 170-180\gev$, beyond which one is forced to resort to non-perturbative values for the new Yukawa coupling $Y_S$, 
independently of the size of $m_{\widetilde{E}}>m_s$.

In \reffig{fig:real_sing}(b) we show the points of the allowed parameter space -- the green region of \reffig{fig:real_sing}(a) -- in the 
($m_{\widetilde{E}}$, $m_{\textrm{DM}}$) plane, best suited for interpreting the LHC constraints. We also apply here 
the constraints from precision observables, which have no visible effect in this case due to the VL nature of the new fermion.  
Both 2-lepton + missing $E_T$ bounds are applied and the excluded points are shown with gray triangles and dim gray diamonds.
The ATLAS bound is quite aggressive, and excludes most of the parameter space in the picture, 
with the exception of a limited region in which $m_{\widetilde{E}}$ and $m_s$ become increasingly close to each other, 
as the $m_{T2}$ variable loses its discriminating power for compressed and semi-compressed spectra. 
We have also calculated the reach of the ATLAS search with 300\invfb, but the latter shows little if any impact 
on the surviving region, due to the intrinsic ineffectiveness of $m_{T2}$ for these spectra. 
On the other hand, the remaining parameter space is extensively tested by the CMS low-momentum search, 
which allows one to exclude BSM fermions with masses lower than $140\gev$ for a dark matter scalar heavier than 
$\sim 60\gev$. 
Note, incidentally, that neither the monojet search nor Higgs invisible decays 
can provide a complementary way of testing the surviving region due to the smallness of the portal Higgs coupling.

The same points are shown in \reffig{fig:real_sing}(c), in the plane of the new Yukawa coupling, $Y_S$, versus the VL fermion mass. 
The CMS soft-lepton search excludes in this case essentially the  whole parameter space with $Y_S\lesssim 1.5$. 
Because of the overall large Yukawa values, future precision experiments like GigaZ or TLEP, with a projected 
improvement by a factor 20 or more over LEP, have the potential to probe the surviving region. 
We show with dashed lines the projected reach of these experiments in precision measurements of $Z$-physics observables. 
Note also that, quite possibly, $e^+ e^-$ colliders like the ILC or TLEP may be able to cover the surviving region directly by 
pair production of the charged particles $\widetilde{E}^{\pm}$. 
A precise estimate of the direct reach of electron/positron colliders in this case requires a numerical simulation of the event generation and detector response, which is beyond the scope of this paper.
   
One might wonder at this point if a large BSM Yukawa coupling, which is required to fit \deltagmtwomu\,, can
remain perturbative up to the Planck scale. We have checked that this is often not the case.
For example, in Model~1, $Y_S\approx 1.5$ induces a breakdown of perturbativity at about $10^5\gev$.
This, however, does not necessarily pose a problem here, as we are dealing with simplified models that, by constuction, are not 
meant to be UV complete. 
It is not hard to imagine that in a more general framework the asymptotic behavior of $Y_S$ could be modified in a way that makes the Landau Pole below the Planck scale disappear.

\begin{figure}[t]
\centering
\includegraphics[width=0.55\textwidth]{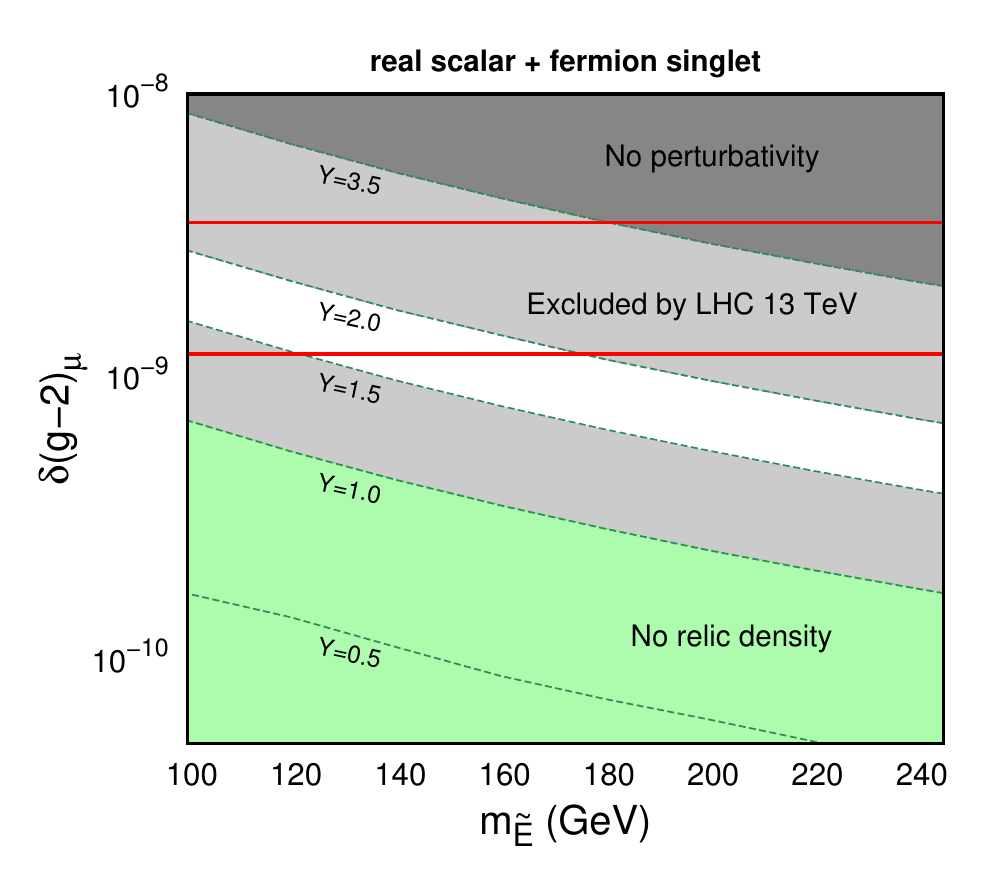}
\caption{The measurement of \deltagmtwomu\ as a function of the VL fermion mass and the new Yukawa coupling in Model~1. 
The scalar dark matter mass is fixed here at $m_s=m_{\textrm{DM}}=80\gev$, but the plot is not very sensitive to its value. 
The horizontal red solid lines show the $2\sigma$ region of the BNL experiment.}
\label{fig:g2_real_s}
\end{figure}

We summarize the case of Model~1 in \reffig{fig:g2_real_s}, where we present predictions for this model based 
on an eventual measurement of \gmtwo\ at Fermilab. The different bounds discussed above are applied.
The surviving parameter space in Model~1 is thus confined to a very narrow strip.

\textbf{Model~2}, characterized by the coupling of the muon to a real scalar singlet and a fermion doublet, 
does not show significant differences from Model~1 at the tree level. This could have been anticipated by a simple inspection of the
two Lagrangians, \refeq{mod1lag} and \refeq{Lagr1}. The dark matter particle is the same in both cases and 
the contributions to the \gmtwo\ calculation are also the same, 
as one can simply switch the role of $c_L$ and $c_R$ in \refeq{chiralint} (and one of them is always zero). 
The bounds from EWPOs are easily satisfied in both models, as both lack an explicit source of chiral-symmetry violation beyond the VL mass. 

The LHC multi-lepton bound is in principle stronger for the fermion doublet case in the region of large 
$m_{E}$ and small $m_{\textrm{DM}}$, due to the possibility of producing the doublet through the $W$ boson, which enhances the cross section. However, we do not expect any difference from what is shown in \reffig{fig:real_sing}(b),
as the limit is strong enough to exclude this region in Model~1 as well.
Thus, no difference can be observed between Model~1 and Model~2, and the reader can refer to 
Figs.~\ref{fig:real_sing} and \ref{fig:g2_real_s} for Model~2 as well.

\subsection{Real scalar with mixing singlet and doublet VL fermions}

As was discussed in \refsec{sec:realsca}, when the singlet and doublet fermion are both included in the theory, 
they can mix through the Higgs boson field vev. This introduces an additional explicit source of chiral-symmetry violation in the \gmtwo\ loop,
which can boost its value as now both $c_L$ and $c_R$ differ from zero in \refeq{chiralint}. 
For the same reason, however, this model, which we dubbed \textbf{Model~3}, 
is also subject to strong constraints from EWPOs.

\begin{figure}[t]
\centering
\subfloat[]{%
\includegraphics[width=0.47\textwidth]{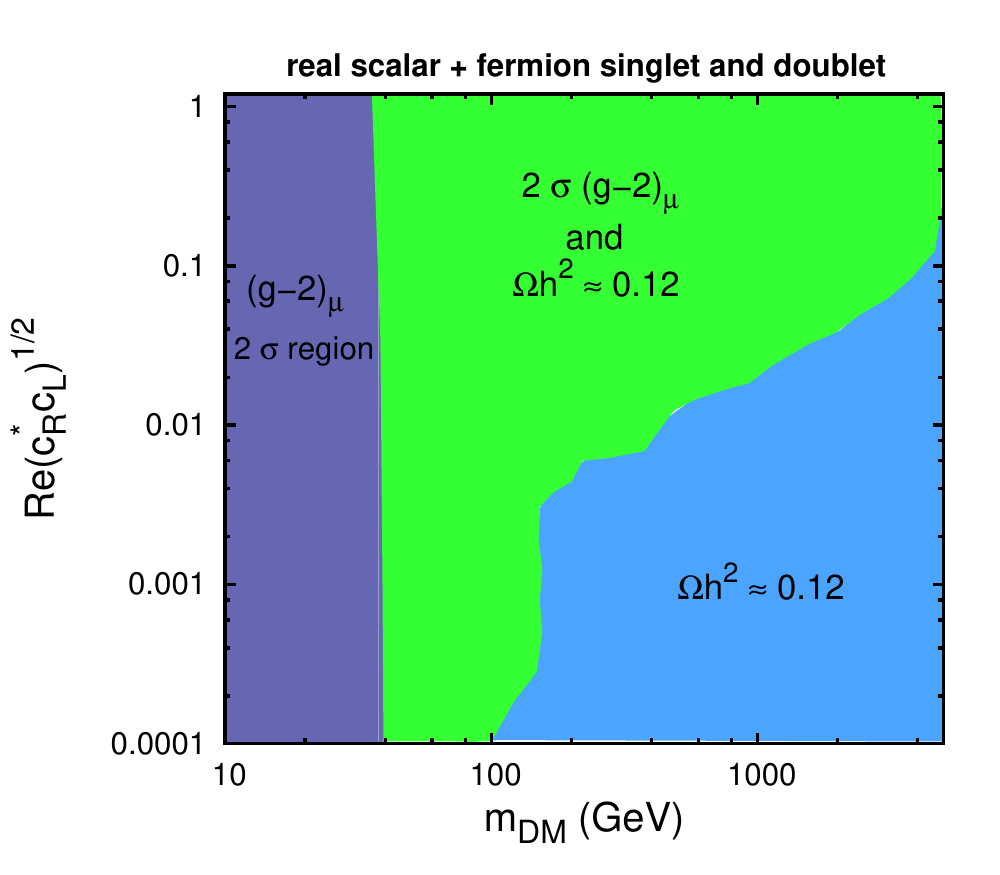}
}%
\hspace{0.02\textwidth}
\subfloat[]{%
\includegraphics[width=0.47\textwidth]{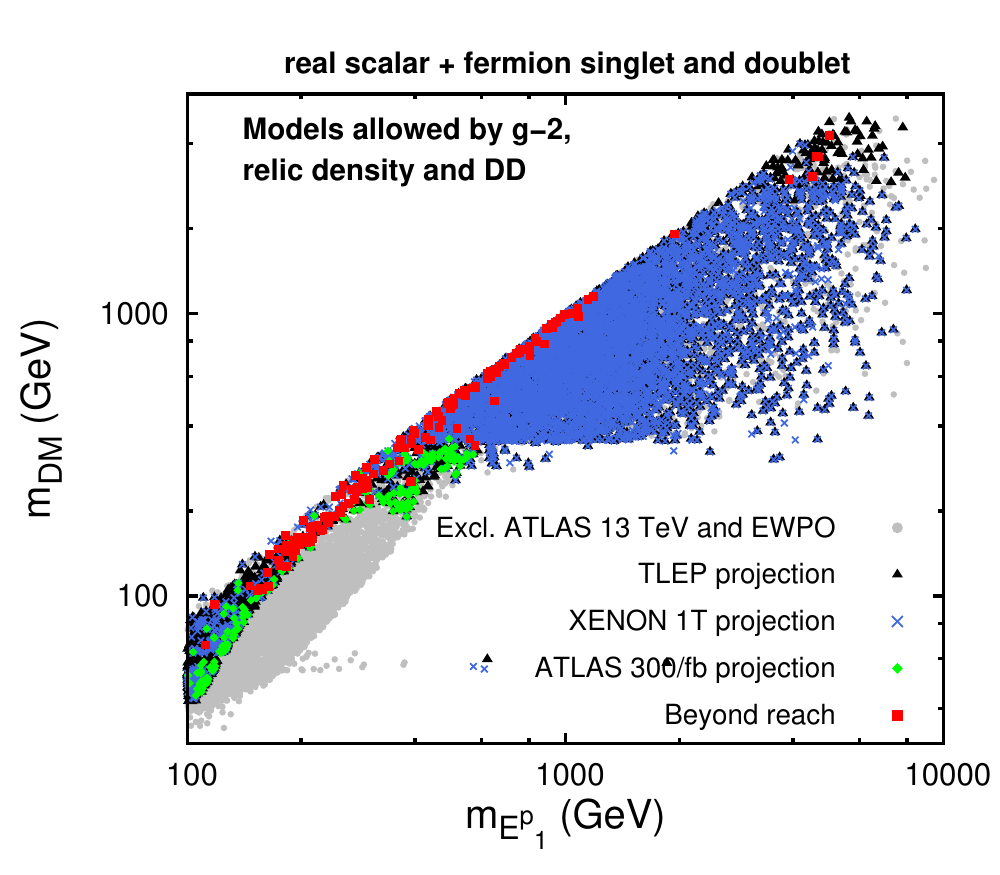}
}%
\caption{(a) The [$m_{\textrm{DM}}$, $\Re(c^1_L c^{1\ast}_R)^{1/2}$] plane for Model~3 (real scalar field and mixing singlet and doublet VL fermions). The color code is the same as in \reffig{fig:real_sing}(a). (b) The parameter space of Model~3 in agreement with the relic density and the \gmtwo\ anomaly at $2\sigma$ in the 
($m_{E_1^p}$, $m_{\textrm{DM}}$) plane. 
Gray dots are excluded by a combination of electroweak precision data, the ATLAS and CMS 2-lepton + MET searches 
at 13\tev, and the constraints from $\textrm{BR}(h\rightarrow \gamma\gamma)$ at ATLAS\cite{ATLAS-CONF-2017-045}. 
The projected reach of ATLAS 2-lepton searches with 300\invfb\ probes the points shown as green dots; 
new data from XENON-IT will test the points shown as royal-blue crosses;
and improved electroweak precision at TLEP will test the points marked by black triangles. 
The points shown as red squares are possibly outside of foreseeable reach.}
\label{fig:real_mixing}
\end{figure}

We present in \reffig{fig:real_mixing}(a) the parameter space of Model~3 in the $[m_{\textrm{DM}}$, $\Re(c^1_L c^{1\ast}_R)^{1/2}]$ 
plane, without for the moment applying the limits from EWPOs and the LHC. The color code is the same as in \reffig{fig:real_sing}(a). 
The \gmtwo\ anomaly can be accommodated for very large dark matter mass, 
well into the TeV range, thanks to the above-mentioned boost provided by the chirality-flip term involving $c_L c_R^{\ast}$.
For the same reason, the new Yukawa couplings are also allowed to span a broader range, 
with the sum of them that can be as small as $\sim 0.2$.
 
We show the points of the parameter space favored by \abund\ and \gmtwo\ in the ($m_{E_1^p}$, $m_{\textrm{DM}}$) plane 
in \reffig{fig:real_mixing}(b). After applying the remaining constraints, we mark with gray dots the points excluded by a combination of the ATLAS and CMS 2-lepton + MET searches (mostly points on the bottom left of the picture), ATLAS $R_{\gamma\gamma}$ constraint (a handful of points that are also excluded by the 2-lepton searches), and EWPOs (mostly points on the top right). 
Electroweak precision bounds are able to exclude solutions with large fermion mass since large singlet/doublet mixing is required to 
fit the \gmtwo\ constraint at $2\sigma$. 

More specifically, the BSM contributions to the $Z\mu\bar{\mu}$ vertex imply $\Re(c^1_L c^{1\ast}_R)^{1/2}\lesssim 0.01$, 
unless the ratio $Y_S/Y_D\sim\mathcal{O}(1)$. In that case the limit weakens to $\Re(c^1_L c^{1\ast}_R)^{1/2}\lesssim 0.1$. 
On the other hand, the corrections to the $W$ mass, which arise from BSM contributions to the muon lifetime and the oblique parameters, 
test two distinctive regions of the parameter space: the small mass regime, with $m_s<100\gev$ and small fermion mixing, where the corrections to muon decay are dominant; and the large mass regime, in which the large mixing induces large splitting of the doublet fermion masses, 
which subsequently increases the parameter $T$.

The points that are not excluded by LEP or the recent LHC constraints are in reach of 
the high-luminosity LHC or future experiments sensitive to corrections to the $Z\mu\bar{\mu}$ effective coupling.
The points in reach of the LHC with 300\invfb\ are shown as green diamonds, whereas  black triangles mark the points in reach of future precision experiments. Note that points characterized by a non-negligible portal coupling, $\lam_{12}\gsim 0.1$, will be tested in complementarity with the next release from XENON-1T data, and we show these points in \reffig{fig:real_mixing}(b) as royal-blue crosses.
Finally, we highlight with red squares the points that appear to be beyond the reach of all of the 
projected measurements considered in this work. 

We point out that these points are almost all characterized by the scalar and lightest fermion mass 
being very close to each other. This is not enforced by any of the symmetries of the Lagrangian considered in \refsec{sec:realsca}, 
so that we can conclude that Model~3 requires a certain amount of fine tuning to evade all future bounds.

\subsection{Complex singlet scalar with VL fermions}

\begin{figure}[t]
\centering
\subfloat[]{%
\includegraphics[width=0.47\textwidth]{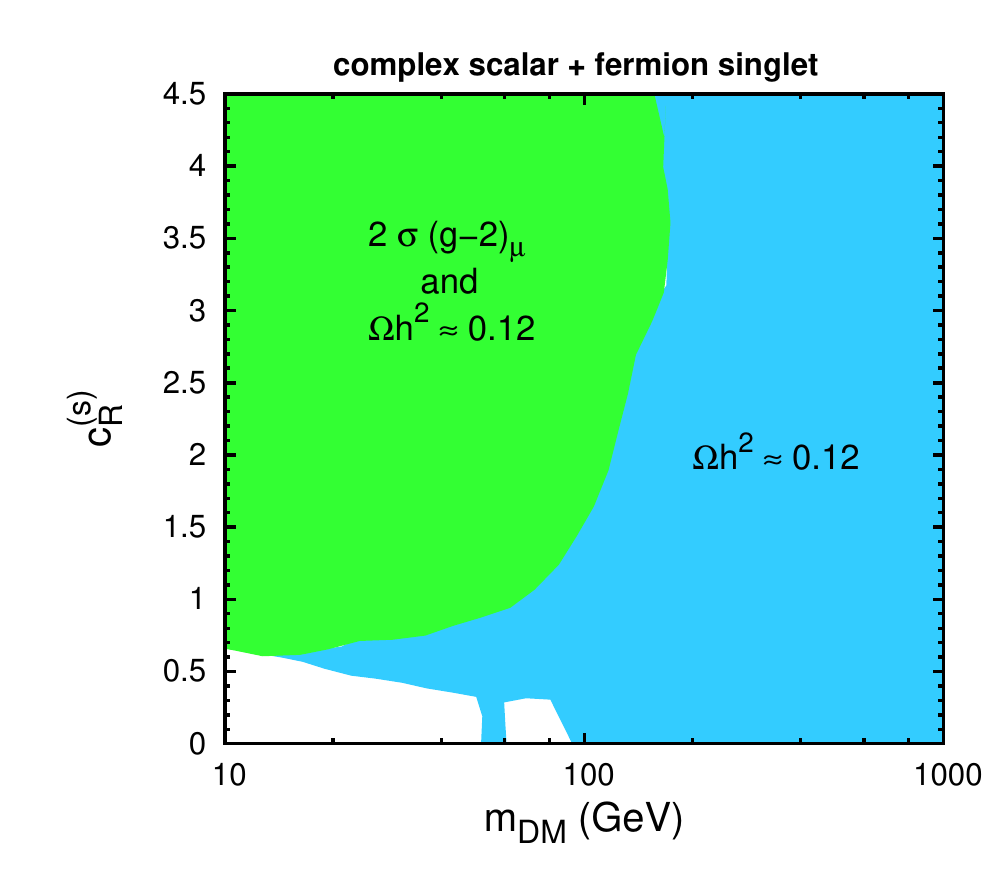}
}%
\hspace{0.02\textwidth}
\subfloat[]{%
\includegraphics[width=0.47\textwidth]{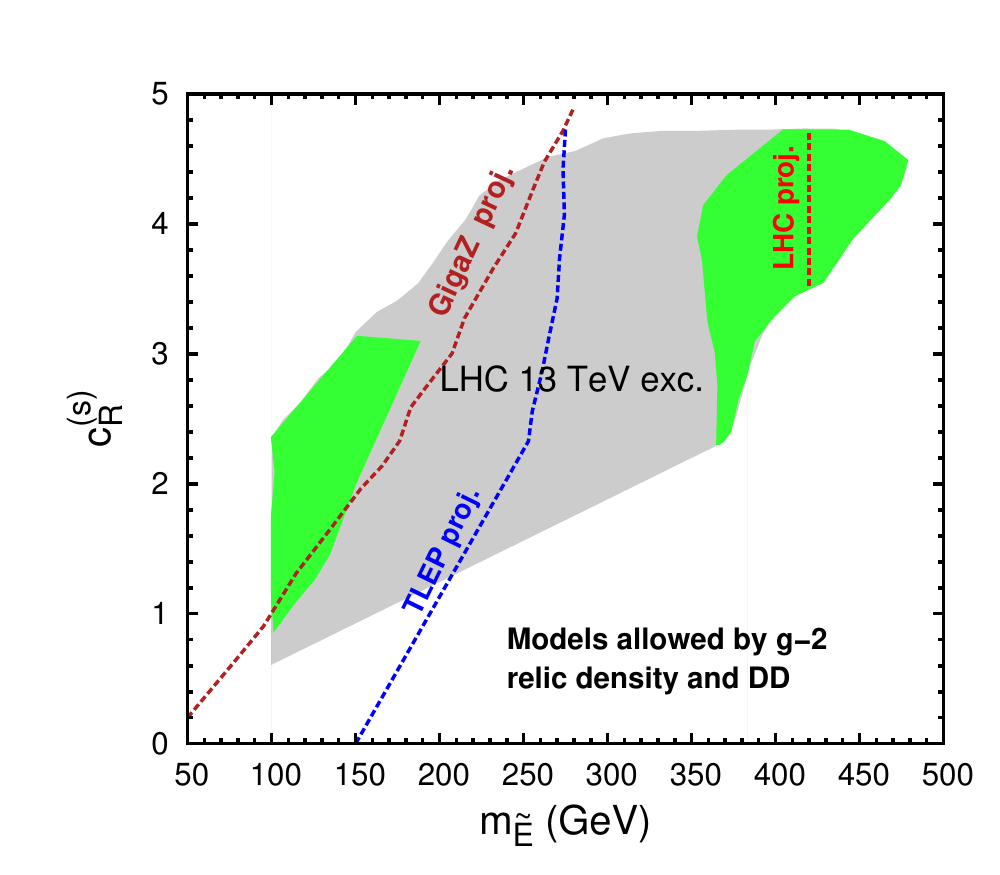}
}%
\caption{(a) The ($m_{\textrm{DM}}$, $c_R^{(s)}$) plane for Model~1a (complex scalar field and singlet VL fermions). The color code is the same as in \reffig{fig:real_sing}(a). (b) Region of Model~1a favored by a combination of the relic density and \gmtwo\ constraints in the plane ($m_{\widetilde{E}}$, $c_R^{(s)}$). The gray area is excluded by the LHC 2-lepton + MET searches. Projected reach at the LHC with 300\invfb, and future precision experiments at GigaZ and TLEP, are shown as dashed lines.}
\label{fig:compl_sing}
\end{figure}

In \reffig{fig:compl_sing}(a) we present the parameter space of \textbf{Model~1a} in the ($m_{\textrm{DM}}$, $c_R^{(s)}$) plane (recall that $c_R^{(s)}=(Y_S+Y_{S^{\ast}})/\sqrt{2}$). At this stage the LHC and precision constraints have not been yet applied, 
and the color code is the same as in \reffig{fig:real_sing}(a), with the only difference being that we do not explicitly highlight here the 
parameter space region belonging to the bulk.

The most striking difference with Model~1 is that in Model~1a one can fit the \gmtwo\ anomaly with lighter dark matter,
as light as our prior range allows, $m_s, m_a\approx 10\gev$. This is due to the presence of two possible dark matter particles in this mass range,
the scalar $s$ and the pseudoscalar $a$. When their masses are not far apart from one another, 
the relic density can be effectively diluted in the early Universe thanks to additional bulk processes like 
$a\,s\rightarrow \mu^+\mu^-$, or $a\,a\rightarrow \mu^+\mu^-$.  

When one applies the constraints from the LHC (precision bounds do not alter the picture much in this model) 
the available parameter space is much reduced.
We project the region favored by a combination of the relic density and \gmtwo\ constraints to the plane ($m_{\widetilde{E}}$, $c_R^{(s)}$). 
The result is presented in \reffig{fig:compl_sing}(b). The region in gray is excluded by the LHC 2-lepton searches. 
Note that, while both LHC 2-lepton searches are quite effecfive in excluding large swaths of parameter space in Model~1b, 
there is a larger region of the parameter space that survives the contraints than in Model~1, as the number of free parameters is here larger, and there are more ways to evade the bounds.  
We also show with dashed lines of different colors the projected reach of the ATLAS search with 
300\invfb, and of future precision experiments at GigaZ and TLEP.    

As dark matter particles, the scalars $s$ and $a$ behave symmetrically, with the only difference being that, by construction, 
the available range for the coupling of $s$ to the muon, $c_R^{(s)}=(Y_S+Y_{S^{\ast}})/\sqrt{2}$, is slightly larger than the allowed coupling
for $a$, $|c_R^{(a)}|=|Y_S-Y_{S^{\ast}}|/\sqrt{2}$.
In fact, the region with good dark matter in \reffig{fig:compl_sing}(a) features solutions of various types, 
$m_s\approx m_a$, $m_s\ll m_a$, $m_a\ll m_s$, and cases in between. 

\begin{figure}[t]
\centering
\subfloat[]{%
\includegraphics[width=0.47\textwidth]{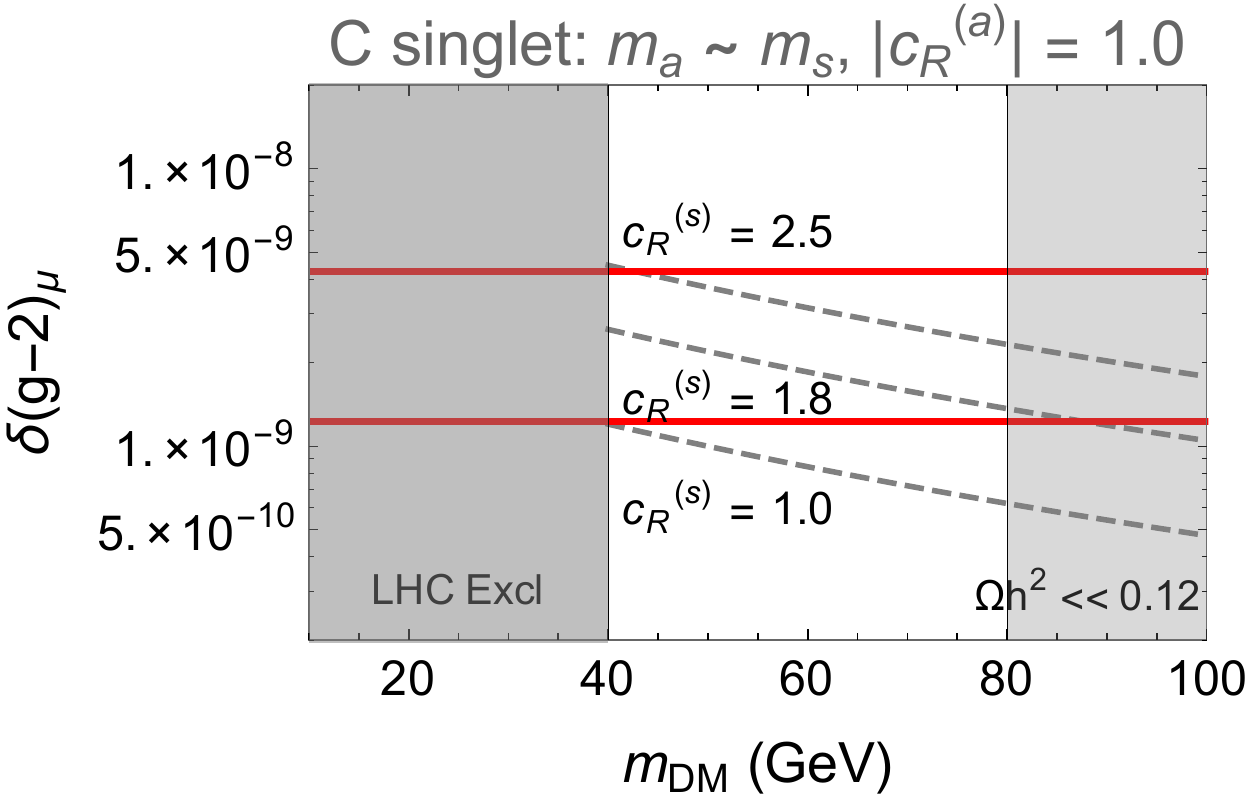}
}%
\hspace{0.02\textwidth}
\subfloat[]{%
\includegraphics[width=0.47\textwidth]{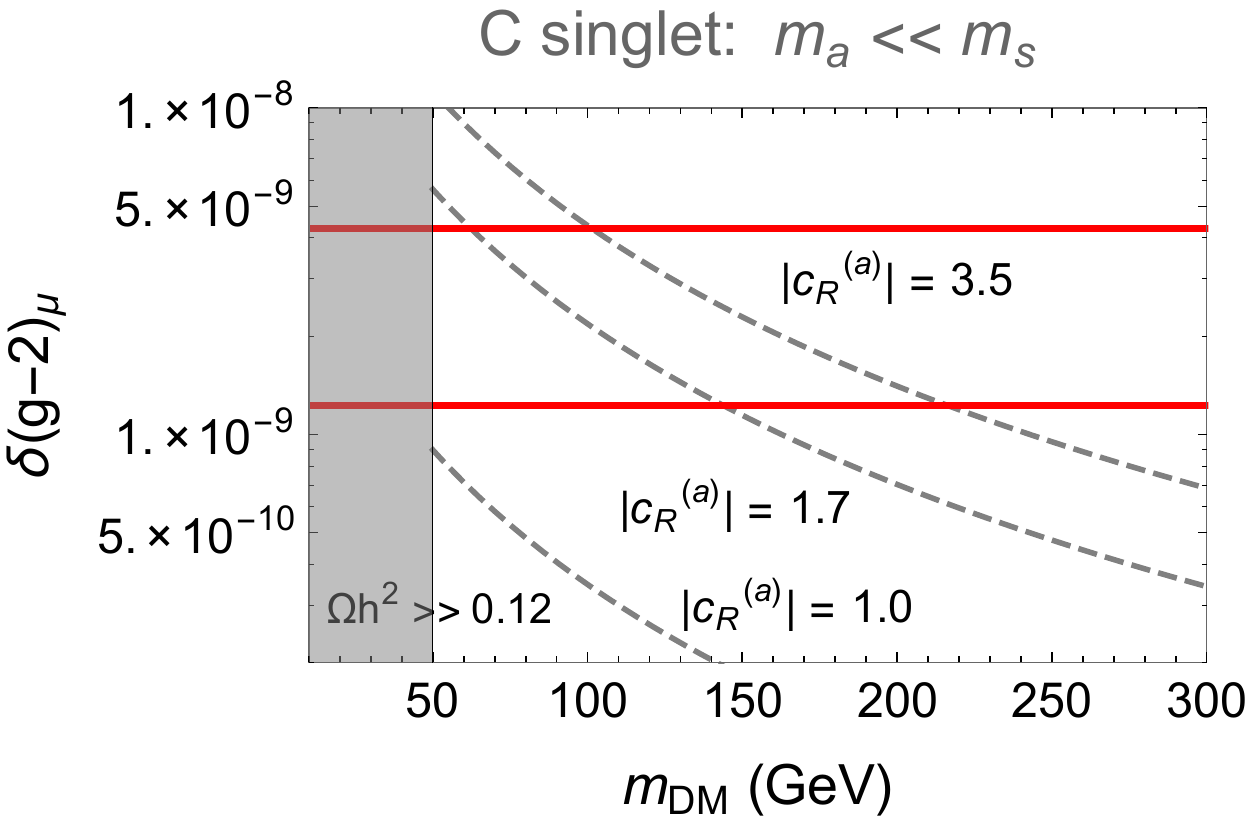}
}%
\caption{The computed \deltagmtwomu\ versus the dark matter mass for the parameter space allowed by dark matter 
in Model~1a (complex singlet scalar and singlet VL fermions). (a) The case of scalar masses being close to degenerate, $m_{\textrm{DM}}\equiv m_a\approx m_s$, as a function of the coupling to the muon, $c_R^{(s)}=(Y_S+Y_{S^{\ast}})/\sqrt{2}$. 
The new fermion mass is fixed slightly above 100\gev, 
and we set the pseudoscalar coupling to $|c_R^{(a)}|=|Y_S-Y_{S^{\ast}}|/\sqrt{2}=1$. 
The dark matter becomes under-abundant for $m_{\textrm{DM}}\gsim 80\gev$.
(b) The case $m_s\gg m_a$, for different values of the coupling of $a$ to the muon,  $|c_R^{(a)}|$. 
This case resembles the behavior of Model~1.}
\label{fig:compl_sing_1d}
\end{figure}

Because of a quite large number of free parameters, however, a measurement of \gmtwo\ is not sufficient in Model~1a to 
pinpoint specific features and correlations of some parameters with respect to others. 
Thus, we identify two limiting cases that we summarize in \reffig{fig:compl_sing_1d}. 

In \reffig{fig:compl_sing_1d}(a) we show the value of \deltagmtwomu\ versus the dark matter mass in the case where the scalar and pseudoscalar are close to being degenerate, $|m_s-m_a|\lesssim$ a few GeV. 
The parameter of greatest impact on the \gmtwo\ calculation is in this case the sum of the new Yukawa couplings, or $c_R^{(s)}$. 
We show in the figure the dependence of \deltagmtwomu\ on selected values of $c_R^{(s)}$, 
when $m_{\widetilde{E}}\approx 100\gev$ to avoid LHC bounds, and $|c_R^{(a)}|$ is set to 1.
One can see that, as $m_s\approx m_a$ approaches $m_{\widetilde{E}}$, the bulk becomes more and more efficient\cite{Bai:2014osa} 
until \abund\ drops below the lower bound when $m_{\textrm{DM}}\approx 80\gev$. 
At that point, in order to maintain the constraint from $\abund\approx 0.12$ in place, the scalar masses must become more separated.
Note also that for $m_{\widetilde{E}}\approx 100\gev$, a dark matter mass below $\sim 40\gev$ is excluded by the ATLAS 2-lepton search,
independently of the value of $c_R^{(s)}$.

In \reffig{fig:compl_sing_1d}(b) we show the case $m_{\textrm{DM}}=m_a\ll m_s$. 
We impose in the plot $m_{\widetilde{E}}\approx m_a+50\gev$, as under this condition one can evade the LHC constraints. 
Here the scalar particle $s$ effectively decouples from the \gmtwo\ calculation, thus reproducing the limit of Model~1, 
with the difference that the coupling of the scalar $a$ to the muon is expressed in terms of $|c_R^{(a)}|=|Y_S-Y_{S^{\ast}}|/\sqrt{2}$.  

Similarly to the equivalent cases in \refsec{sec:result1}, if we consider \textbf{Model~2a},
in which the complex scalar couples to a doublet VL fermion, the differences with Model~1a are not substantial. 
However, because of the enlarged production cross section for doublet fermions, the LHC excludes in this case the region 
shown on the top right in \reffig{fig:compl_sing}(b), characterized by charged leptons between 400 and 500\gev. 

\textbf{Model~3a}, finally, presents a pattern that in the [$m_{\textrm{DM}}$, $\Re(c_L^{(s,a)1}c_R^{(s,a)1\ast})^{1/2}$] plane
would look similar to the one depicted in \reffig{fig:real_mixing}(a), with large regions of the parameter space allowed 
by the constraints before the bounds from precision and the LHC are applied. 
However, because there are now several Yukawa couplings and two scalar masses that can interplay, it appears that the 
large $m_{\textrm{DM}}$ region requires slightly lower values for the mixing angles than in Model~3 and, as a consequence,  
in Model~3a the parameter space cannot be easily constrained.
In fact, the ($m_{E_1^p}$, $m_{\textrm{DM}}$)-plane plot equivalent to \reffig{fig:real_mixing}(b) 
is not very informative in Model~3a, with no clear correlation emerging in the masses of the surviving points. 
For this reason we do not show this figure here, and we are forced to conclude that a complex scalar singlet with mixing 
doublet/singlet VL fermion can accommodate \abund\ and the \gmtwo\ anomaly for large ranges of parameters that will possibly 
avoid most future constraints, as long as the singlet-doublet mixing is not very large.

\subsection{Doublet scalar with singlet, doublet, or triplet VL fermions\label{sec:doub_sca-results}}

In order to analyze the doublet scalar case, let us first briefly recall its dark matter properties. 
Inert doublet scalar dark matter has been analyzed in the literature in several papers 
(we refer to, e.g.,\cite{Goudelis:2013uca,Arhrib:2013ela} for recent studies).
When the relic density is driven by the Higgs portal couplings the characteristics are well known. 
There are two viable regions for dark matter in the parameter space: one for $m_{\textrm{DM}}\lesssim 100\gev$ 
and the other in the range from $\sim 700\gev$ to several~TeV. 
For the parameter space in between, the dark matter is under-abundant, 
as at about 85\gev\ the annihilation channel $s s(a)\rightarrow W^+W^-$ opens up 
(in a fashion similar to the higgsino case in supersymmetry).
For the models with no fermion mixing, the TeV-range region does not admit solutions for the \gmtwo\ anomaly, so that we will limit 
our analysis to scalar masses below $\sim 100\gev$.

\begin{figure}[t]
\centering
\subfloat[]{%
\includegraphics[width=0.47\textwidth]{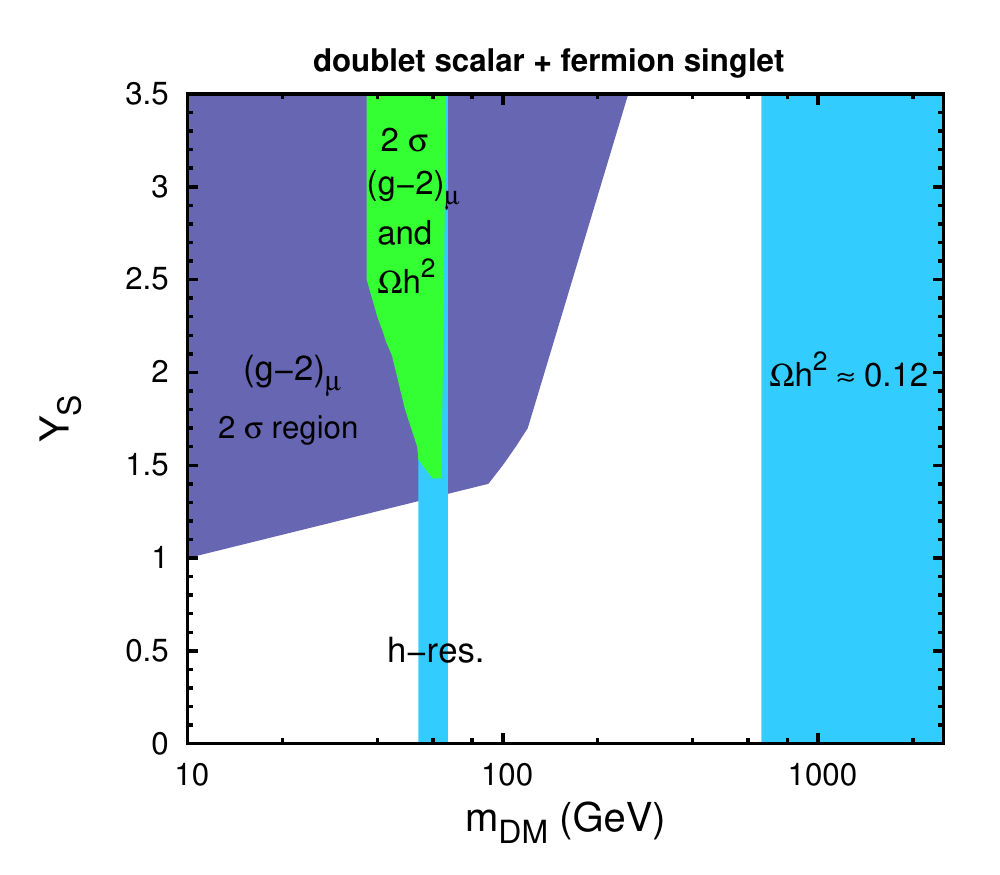}
}%
\hspace{0.02\textwidth}
\subfloat[]{%
\includegraphics[width=0.47\textwidth]{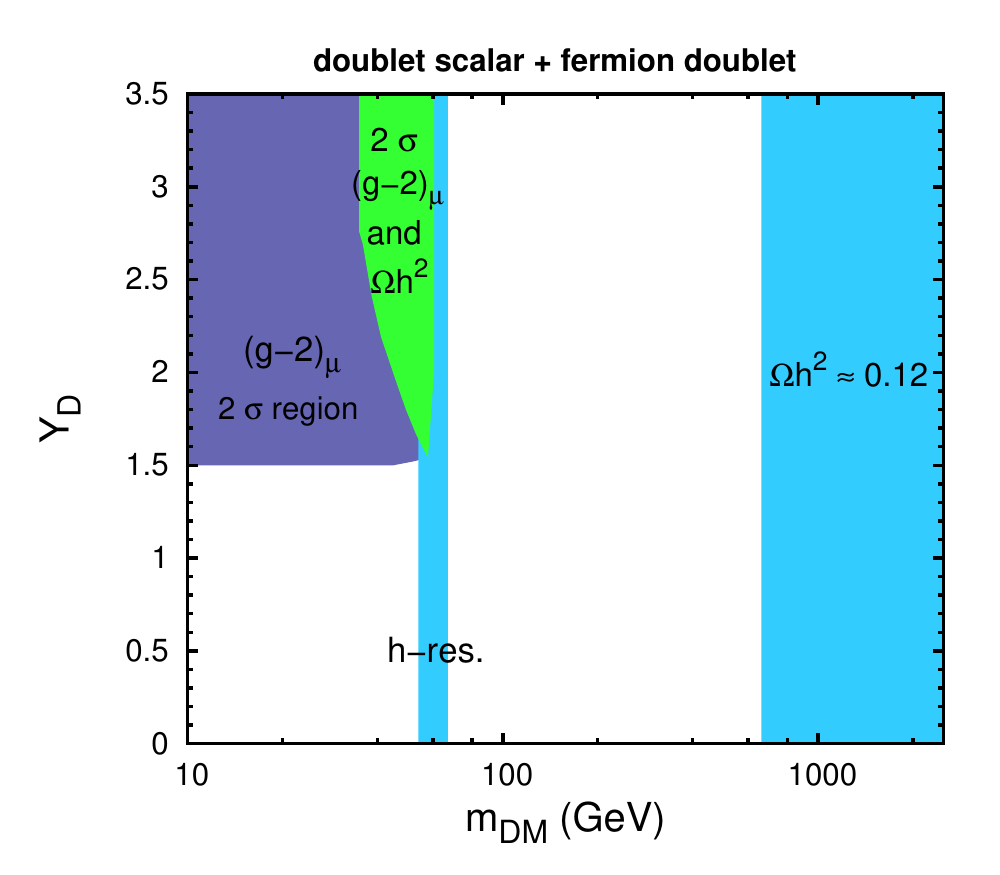}
}\\
\subfloat[]{%
\includegraphics[width=0.47\textwidth]{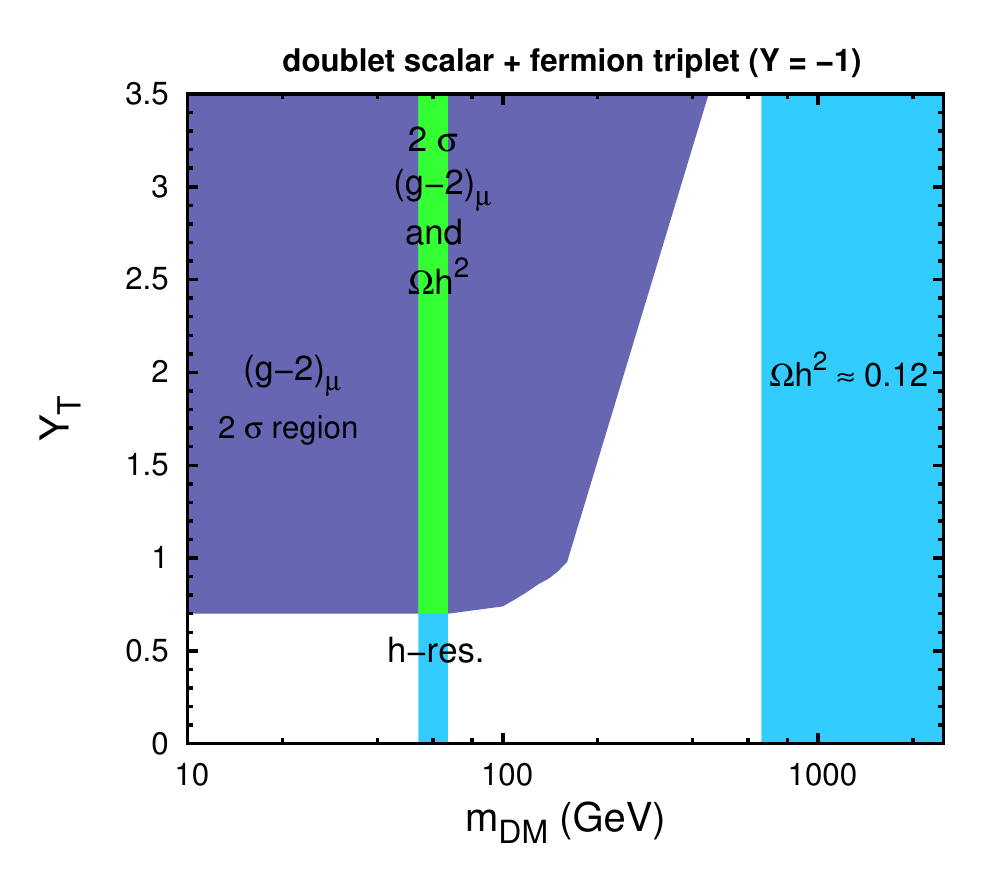}
}%
\hspace{0.02\textwidth}
\subfloat[]{%
\includegraphics[width=0.47\textwidth]{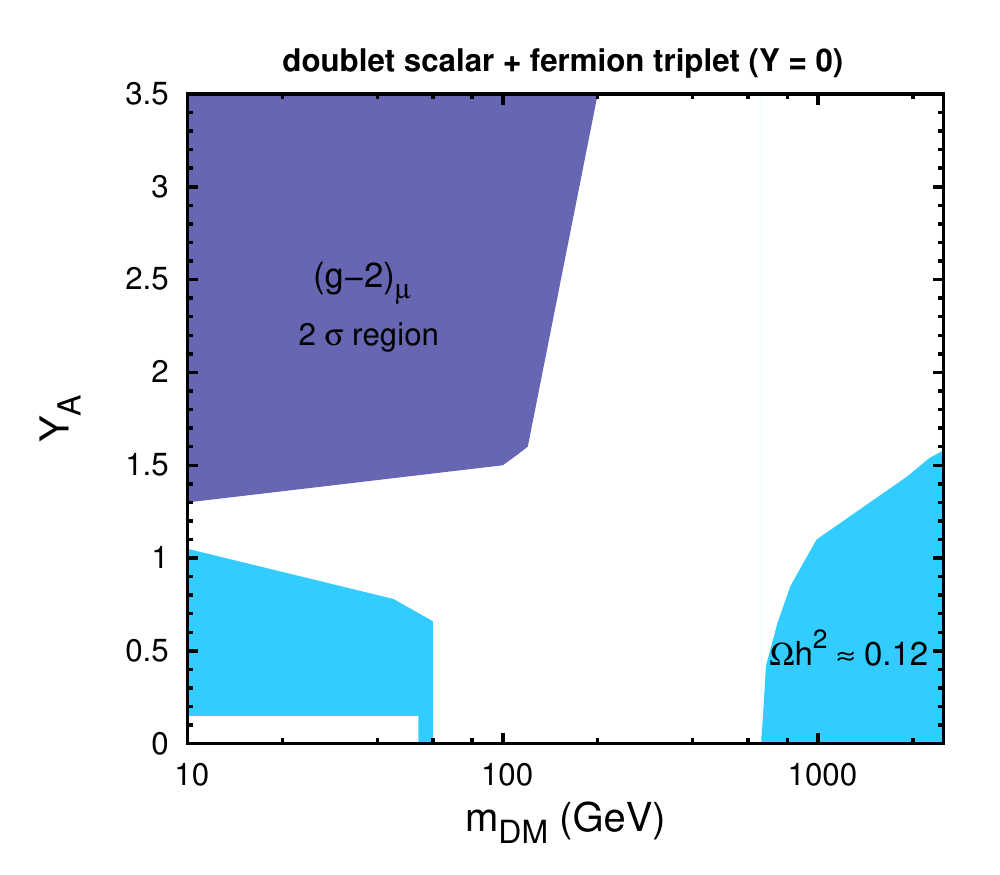}
}
\caption{Planes of the Yukawa coupling to muons versus the dark matter mass in models with a scalar doublet. 
The region where the \gmtwo\ anomaly can be accommodated at $2\sigma$ is shown in blue, the region where
the relic abundance is correct within $2\sigma$ is shown in cyan, and the combined parameter space is highlighted in green.
(a) Model~4 (doublet scalar and singlet VL fermions), (b) Model~5 (doublet scalar and doublet VL fermions), (c) Model~7 
(doublet scalar and triplet VL fermions), (d) Model~8 (doublet scalar and adjoint triplet VL fermions).}
\label{fig:doubl_mass}
\end{figure}

In \reffig{fig:doubl_mass}(a) we present the now familiar plane of the Yukawa coupling to muons versus the dark matter mass 
in \textbf{Model~4}, which features a doublet scalar and singlet VL fermions in the spectrum. As before, at this stage of the analysis 
we have not applied the bounds from EWPOs and the LHC.
The dark matter in the low-mass region is obtained mostly through the bulk, as values of the portal couplings that 
might produce $\abund\approx 0.12$ through interactions with the Higgs have long been excluded in direct detection experiments (with the exception of the Higgs resonance).

As in the previous cases the \gmtwo\ anomaly can be fitted in the bulk. 
Unlike in the complex singlet case, however, solutions with dark matter much lighter than 40\gev\ cannot be found here,
as LEP has excluded weakly-coupled charged particles below $\sim 100\gev$ and all scalars belong to the same doublet. 
Close inspection of \refeq{doublet_mass} reveals that, once the charged scalar satisfies that constraint, 
also the pseudoscalar becomes heavier than 100\gev\ (recall that $\lam_3+\lam_4+2\lam_5\ll 1$) 
and the dark matter sector approaches the limit of Model~1.

A similar plot for \textbf{Model~5}, which features a doublet scalar and doublet VL fermions, is shown in \reffig{fig:doubl_mass}(b).
After a quick look at \refeq{sca_doub_fer_doub} one can see that in this case there is a coupling of the muon to the 
charged scalar/neutral fermion loop. Thus, we expect \gmtwo\ to be generally smaller than in Model~4 in the equivalent parameter range, as there is a negative contribution damping its value. Note, in this regard, 
that the Higgs-resonance region does not present solutions to 
the \gmtwo\ anomaly in Model~5 since, in order to avoid excessively diluting \abund, either the Yukawa couplings should be 
there quite small, or the VL fermion mass larger than in other regions of the parameter space.
 
On the other hand, in \textbf{Model~7}, characterized by a doublet scalar and triplet VL fermions, simultaneous 
solutions to the \gmtwo\ anomaly and dark matter exist \textit{only} in the Higgs-resonance region. 
In fact, in general, due to the impact 
of the large doubly-charged contribution, one obtains \deltagmtwomu\ within $2 \sigma$ for fermion masses that must be 
quite large, $m_{\Psi^{\pm}},m_{\Psi^{\pm\pm}}\gg 100\gev$. As a consequence, the bulk annihilation cross section is not large enough to yield $\abund\approx 0.12$. However, in the Higgs-resonance region, the relic abundance is obtained via portal couplings, so that 
a solution that can accommodate the \gmtwo\ anomaly can be easily found for a wide range of Yukawa coupling and fermion mass values.

Finally, we illustrate the case of \textbf{Model~8}, characterized by a doublet scalar and adjoint triplet VL fermions,
in \reffig{fig:doubl_mass}(d). 
As one can see from the plot, there is no parameter space here that can accommodate the measured value of 
\gmtwo\ and the relic density at the same time. As a matter of fact, in Model~8 the bulk annihilation 
channel $s s(a)\rightarrow \nu_{\mu}\nu_{\mu}$ is so efficient that it effectively places an upper bound on the 
new Yukawa coupling, $Y_A\lesssim 0.8-1$, if one wants to satisfy $\abund\approx 0.12$. 
As a consequence, \deltagmtwomu\ is never large enough in the region with 
$\abund\approx 0.12$. 

We briefly come back here to the issue of a breakdown of perturbativity of the Yukawa couplings at some high scale, which is less severe in the models with a scalar doublet. For instance, values of $Y_T$ as large as 1.3 do not generate a Landau Pole below the Planck scale in the fermion triplet case, indicating that such a scenario can be considered as UV-complete. Conversely, for fermion singlets and doublets the lowest $Y_{S/D}$ consistent with the 
\gmtwo\ anomaly requires some extension of the corresponding simplified model at the scale of $\sim 10^{9}\gev$.

\begin{figure}[t]
\centering
\includegraphics[width=0.55\textwidth]{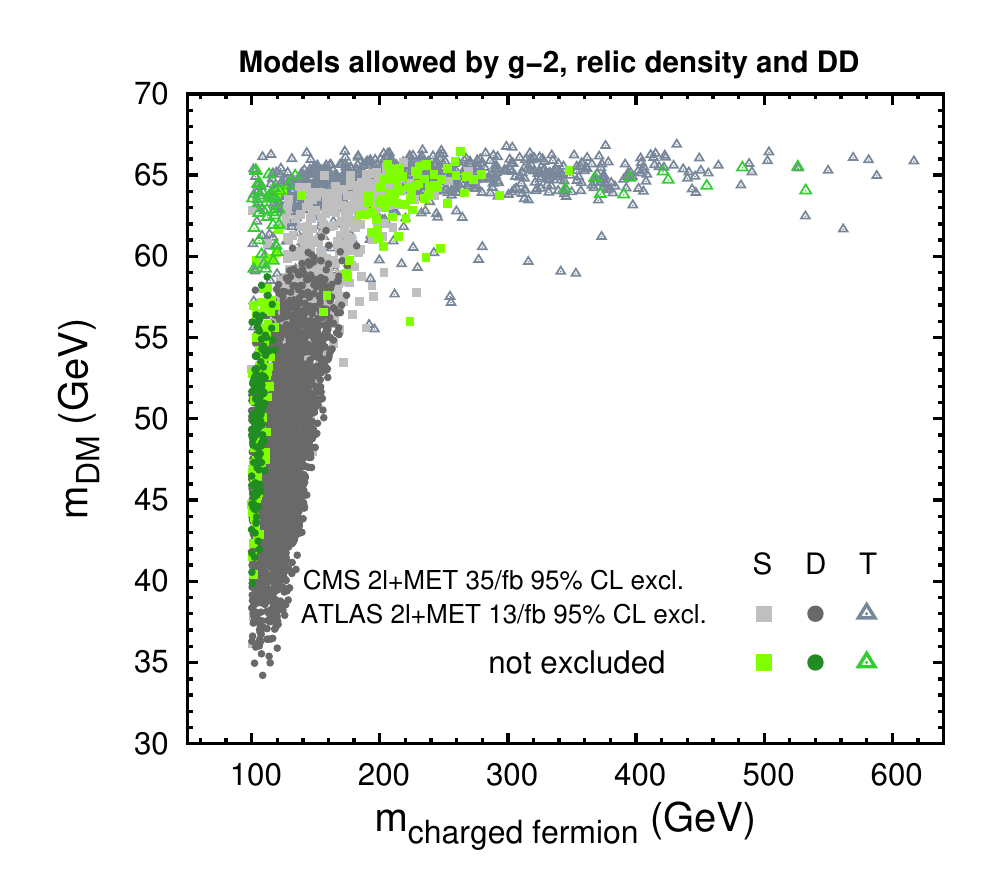}
\caption{The impact of LHC 13\tev\  on the parameter space in agreement with dark matter and the \gmtwo\ anomaly in Models~4 (squares), 5 (circles), and 7 (triangles). The points allowed are shown in three different shades of green, while those excluded in three different shades of gray.}
\label{fig:doubs_LHC}
\end{figure}

The impact of LHC 13\tev\ and electroweak precision constraints on the parameter space in agreement with dark matter and the 
\gmtwo\ anomaly for Models~4-8 is shown in \reffig{fig:doubs_LHC}. In all scenarios, 
as long as the charged fermion is situated around 100\gev\ a compressed region can be observed, 
which survives all of the constraints. It can only be minimally tested by the CMS soft-lepton analysis, 
as the mass difference between the charged fermion and dark matter is in general too large for the search to be effective.
Outside of the compressed region, across the Higgs resonance, the mass of the charged VL fermions 
becomes larger to suppress bulk annihilation. In Model~4 the LHC 2-lepton search places a lower bound on the fermions mass 
of about 160\gev. It is much weaker than the corresponding bound for Model~1, as in Model~4 the decay 
$E'\to S^+\bar{\nu}_{\mu}$ is possible, which cuts drastically the efficiency of the search due to a reduced branching ratio 
to the 2-lepton final state.

As before, we have also calculated the effect of the future LHC reach with 300\invfb\ on the models. 
Most of the currently allowed regions with fermion mass larger than $\sim 100\gev$ will be strongly 
reduced at the end of the current LHC run, although it does not appear that any of them can be probed in its absolute 
entirety even with large luminosity.  

\begin{figure}[t]
\centering
\includegraphics[width=0.60\textwidth]{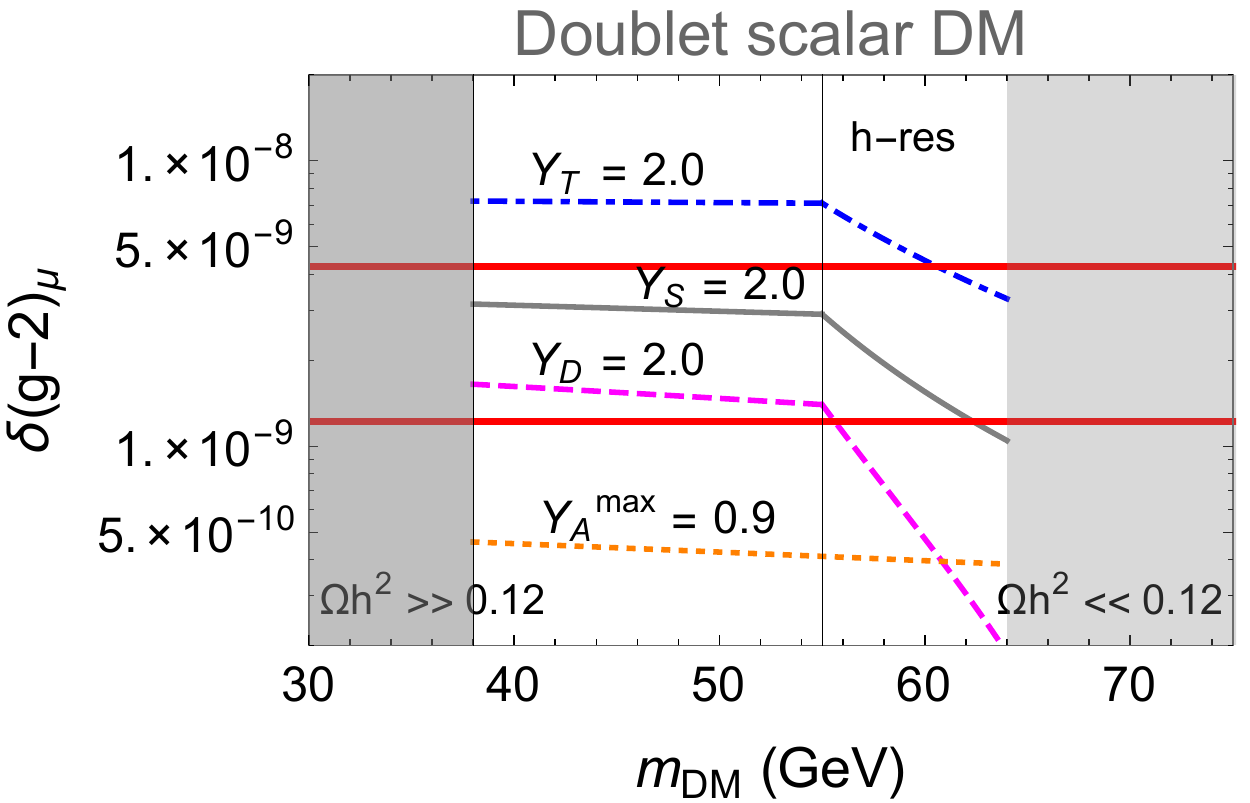}
\caption{The computed \deltagmtwomu\ as a function of the dark matter mass $m_{\textrm{DM}}=m_s$ 
and Yukawa coupling to the muon in the models with a doublet scalar, discussed in \refsec{sec:doub_sca-results}. 
The charged fermion mass is fixed at about 100\gev\ outside of the Higgs resonance to evade the LHC bounds, but it is progressively 
pushed to larger values on the resonance region. The pseudoscalar and charged scalar mass are fixed case by case to the typical values observed in the scans.
Gray solid line: singlet VL fermions; magenta dashed line: doublet VL fermions; blue dot-dashed line: triplet VL fermions; orange dotted line: adjoint triplet VL fermions.}
\label{fig:g2_doubs}
\end{figure}

We summarize the findings of this subsection in \reffig{fig:g2_doubs}, where we present the value of \deltagmtwomu\ 
versus the scalar dark matter mass in every model for a fixed value of the new Yukawa coupling, $Y=2$. 
The only exception is Model~8 for which, as we saw in \reffig{fig:doubl_mass}(d), 
there exist an upper bound on the Yukawa coupling due to the relic abundance. 
To avoid the bounds from the LHC, in \reffig{fig:g2_doubs} we have set the charged fermion mass just above 100\gev, 
while the pseudoscalar and charged scalar masses are fixed case by case to the typical values observed in the scans.
As the mass of the scalar $s$ approaches the Higgs resonance, however, we move the fermion mass up to larger values, to mimic the behavior of the scans. As a consequence, \deltagmtwomu\ drops.

Note that in Model~5 (magenta dashed) $Y_D\approx 2$ barely allows the model to sit inside the $2\sigma$ region for \gmtwo. 
As we discussed above, since in the Higgs-resonance either the allowed Yukawa coupling for $\abund\approx 0.12$ is much smaller than 2, 
or the fermion mass must be larger than 100\gev, it follows that in the Higgs-resonance there is no common 
parameter space for \gmtwo\ and dark matter.
Conversely in Model~7 (blue dot-dashed), for $Y_T\approx 2$ the model sits generally 
above the $2\sigma$ region for \gmtwo. Only the significant drop in the Yukawa coupling necessary to get $\abund\approx 0.12$ 
in the Higgs resonance can bring its value back into the allowed region.

We finally comment on the numerical results we get for \textbf{Model~9}, characterized 
by the mixing of the adjoint triplet fermions and the doublet fermions. As we mentioned when we introduced \refeq{adj_mix}, 
several substantial contributions to \gmtwo\ arise in this model, 
as the neutral scalar/charged fermion and pseudoscalar/charged fermion contributions sum up, and one also has the positive contribution from the mixing heavy neutrinos and charged scalar. As a consequence, solutions to the \gmtwo\ anomaly 
can be potentially found in the large dark-matter mass region, $m_{\textrm{DM}}\gsim 800\gev$. 
However, since the Yukawa coupling is always quite limited in size, 
the solutions we found in our scans are all characterized by large mixing angles, and inevitably fail to pass the constraints from EWPOs. 
Thus, up to the possibility of having missed some fine-tuned corners of the parameter space, we conclude that there is no simultaneous solution for \gmtwo\ and dark matter in Model~9.  

\section{Summary and conclusions\label{sec:summary}}

In this paper we have drawn some predictions for future measurements of \gmtwo\ under the hypothesis that 
the anomaly measured at BNL will be confirmed and that the same underlying BSM physics is responsible for 
the relic abundance of dark matter in the Universe. 

To investigate these scenarios, we have   
constructed a set of renormalizable, SU(2)$\times$U(1) invariant extensions of the SM, 
each comprising inert $\mathbb{Z}_2$-odd scalar fields and one or more VL pair of colorless fermions 
that communicate to the SM muons through Yukawa-type interactions.
Our new sectors are classified according to their transformation properties under the SM gauge group: real singlet, 
complex singlet, and doublet scalars, with all possible types of VL fermions allowed by the gauge symmetry. All 
models have been systematically confronted with a variety of experimental constraints: 
LEP mass bounds, direct LHC searches, electroweak precision observables, and direct searches for dark matter.

In general, the presence of a muon portal introduces a well known $t$-channel bulk mechanism for the dark matter relic density that 
extends the widely studied Higgs portal and allows one to evade strong bounds from direct detection experiments. 
In the case of a real singlet scalar dark matter particle, we find that before applying the LHC bounds 
both relic density and \deltagmtwomu\ can be accommodated 
for $m_{\textrm{DM}}\approx 40-160\gev$, provided VL fermions are lighter than 350\gev\ and the Yukawa coupling exceeds $\sim1.2$. The same pattern is observed for singlet and doublet fermions, since in both cases only the charged component contributes to the anomalous magnetic moment of the muon. When LHC bounds are applied, the parameter space is in large part excluded by the ATLAS 2-lepton + missing $E_T$ search, except for the region where the mass difference between the dark matter scalar and fermion drops below $\sim 70\gev$.
This latter region is extensively tested by the CMS 2 soft leptons + missing $E_T$ search, 
which excludes all the points with Yukawa coupling smaller than $\sim1.5$.
Interestingly, future high-precision experiments like GigaZ and TLEP have the potential to probe the remaining untested region entirely, 
as the new sector generates small but non-zero corrections to the $Z\mu\bar{\mu}$ vertex.

If the singlet scalar is complex, the region where \gmtwo\ and relic density are satisfied extends down to around 10\gev\ before the LHC bounds are applied, as additional annihilation channels are open due to the presence of a pseudoscalar particle in the spectrum. 
The lower bound on the allowed Yukawa coupling is reduced to $\sim 0.5$, since there are two Yukawa couplings summing up. 
For the same reason, fermions up to $\sim 500\gev$ are allowed since there are two contributions to \gmtwo. 

For both the real and complex singlet scalar, we have additionally allowed the singlet and doublet VL fermions to mix 
through interactions with the SM Higgs, thus introducing a source of chiral-symmetry violation that, 
by being proportional to the mass of the heavy leptons, can boost \deltagmtwomu. 
The anomaly can thus be accommodated for masses up to $\sim 3\tev$. The same effect, however, 
generates large contributions to the EWPOs that, for the large fermion mixing, exclude part of the parameter space. 
In particular, we find that in the real scalar case with mixing VL fermions the viable part of the parameter space 
can be probed almost entirely by a combination of the LHC, dark matter direct detection, and future electroweak precision experiments.

The parameter space is more constrained by the relic density in cases with an SU(2) scalar doublet. 
Typical masses range from $m_{\textrm{DM}}\approx 40\gev$ to the Higgs-resonance, $m_{\textrm{DM}}\approx m_h/2$.
Most of the parameter space is almost entirely tested by LHC 2-lepton searches, except for the region of compressed spectra, when the new fermions and scalars are almost degenerate.
Specific properties arise in the case with SU(2) triplet fermions where, 
thanks to the well known presence of a doubly charged lepton in the spectrum, which can enhance the calculation of \deltagmtwomu, the parameter space is confined to Higgs resonance.

Overall, our study shows that scenarios with one type of BSM scalar and fermions, which in general can accommodate \gmtwo\ for a 
relatively large range of masses and Yukawa couplings, become strongly constrained  when the relic density is added to the set of assumptions. That makes them very predictive in case a positive measurement is confirmed at Fermilab or J-PARC. 
That is in general not true, however, for scenarios with singlet-doublet mixing, since the number of free parameters is large enough to accommodate both \gmtwo\ and \abund\ and evade other experimental constraints. 
For these scenarios, increased precision measurements of 
$Z$-physics observables in future electron-positron colliders will be able to disentangle some of the degeneracies.

\bigskip
\noindent \textbf{Acknowledgments}
\medskip

\noindent We would like to thank Luc Darm\' e for valuable input on the interface of \texttt{SARAH} with \texttt{MultiNest},
Florian Staub for helpful correspondence regarding \texttt{SARAH}, Joachim Brod and Wei-Chih Huang for discussions.
KK is supported in part by the DFG Research Unit FOR 1873 ``Quark Flavour Physics and Effective Field
Theories''. 
The work of EMS is supported by the Alexander von Humboldt Foundation.
The use of the CIS computer cluster at the National Centre for Nuclear Research in Warsaw is gratefully acknowledged. 

\bibliographystyle{JHEP}
\bibliography{KE3}

\providecommand{\href}[2]{#2}\begingroup\raggedright\begin{thebibliography}{10}

\bibitem{Bennett:2006fi}
{\bf Muon g-2} Collaboration, G.~W. Bennett et~al., {\it {Final Report of the
  Muon E821 Anomalous Magnetic Moment Measurement at BNL}},  {\em Phys. Rev.}
  {\bf D73} (2006) 072003, [\href{http://arxiv.org/abs/hep-ex/0602035}{{\tt
  hep-ex/0602035}}].

\bibitem{Davier:2016iru}
M.~Davier, {\it {Update of the Hadronic Vacuum Polarisation Contribution to the
  muon g-2}},  {\em Nucl. Part. Phys. Proc.} {\bf 287-288} (2017) 70--75,
  [\href{http://arxiv.org/abs/1612.02743}{{\tt arXiv:1612.02743}}].

\bibitem{Jegerlehner:2017lbd}
F.~Jegerlehner, {\it {Muon g-2 Theory: the Hadronic Part}},
  \href{http://arxiv.org/abs/1705.00263}{{\tt arXiv:1705.00263}}.

\bibitem{Grange:2015fou}
{\bf Muon g-2} Collaboration, J.~Grange et~al., {\it {Muon (g-2) Technical
  Design Report}},  \href{http://arxiv.org/abs/1501.06858}{{\tt
  arXiv:1501.06858}}.

\bibitem{Chapelain:2017syu}
{\bf Muon g-2} Collaboration, A.~Chapelain, {\it {The Muon g-2 experiment at
  Fermilab}},  {\em EPJ Web Conf.} {\bf 137} (2017) 08001,
  [\href{http://arxiv.org/abs/1701.02807}{{\tt arXiv:1701.02807}}].

\bibitem{Ishida:2009zz}
{\bf Muon g-2} Collaboration, K.~Ishida, {\it {Ultra slow muon source for new
  muon g-2 experiment}},  {\em AIP Conf. Proc.} {\bf 1222} (2010) 396--399.

\bibitem{Mibe:2010zz}
{\bf J-PARC g-2} Collaboration, T.~Mibe, {\it {New g-2 experiment at J-PARC}},
  {\em Chin. Phys.} {\bf C34} (2010) 745--748.

\bibitem{Iinuma:2011zz}
{\bf J-PARC muon g-2/EDM} Collaboration, H.~Iinuma, {\it {New approach to the
  muon g-2 and EDM experiment at J-PARC}},  {\em J. Phys. Conf. Ser.} {\bf 295}
  (2011) 012032.

\bibitem{Saito:2012zz}
{\bf J-PARC g-'2/EDM} Collaboration, N.~Saito, {\it {A novel precision
  measurement of muon g-2 and EDM at J-PARC}},  {\em AIP Conf. Proc.} {\bf
  1467} (2012) 45--56.

\bibitem{Otani:2015jra}
{\bf E34} Collaboration, M.~Otani, {\it {Status of the Muon g-2/EDM Experiment
  at J-PARC (E34)}},  {\em JPS Conf. Proc.} {\bf 8} (2015) 025008.

\bibitem{Jegerlehner:2009ry}
F.~Jegerlehner and A.~Nyffeler, {\it {The Muon g-2}},  {\em Phys. Rept.} {\bf
  477} (2009) 1--110, [\href{http://arxiv.org/abs/0902.3360}{{\tt
  arXiv:0902.3360}}].

\bibitem{Leveille:1977rc}
J.~P. Leveille, {\it {The Second Order Weak Correction to (G-2) of the Muon in
  Arbitrary Gauge Models}},  {\em Nucl. Phys.} {\bf B137} (1978) 63--76.

\bibitem{Moore:1984eg}
S.~R. Moore, K.~Whisnant, and B.-L. Young, {\it {Second Order Corrections to
  the Muon Anomalous Magnetic Moment in Alternative Electroweak Models}},  {\em
  Phys. Rev.} {\bf D31} (1985) 105.

\bibitem{Fayet}
 {P. Fayet, in \textit{Unification of the Fundamental Particle Interactions},
  edited by S. Ferrara, J. Ellis, and P. van Nieuwenhuizen (Plenum, New York,
  1980), p. 587.}

\bibitem{Grifols:1982vx}
J.~A. Grifols and A.~Mendez, {\it {Constraints on Supersymmetric Particle
  Masses From ($g-2$) $\mu$}},  {\em Phys. Rev.} {\bf D26} (1982) 1809.

\bibitem{Ellis:1982by}
J.~R. Ellis, J.~S. Hagelin, and D.~V. Nanopoulos, {\it {Spin 0 Leptons and the
  Anomalous Magnetic Moment of the Muon}},  {\em Phys. Lett.} {\bf B116} (1982)
  283--286.

\bibitem{Barbieri:1982aj}
R.~Barbieri and L.~Maiani, {\it {The Muon Anomalous Magnetic Moment in Broken
  Supersymmetric Theories}},  {\em Phys. Lett.} {\bf B117} (1982) 203--207.

\bibitem{Romao:1984pn}
J.~C. Romao, A.~Barroso, M.~C. Bento, and G.~C. Branco, {\it {Flavor Violation
  in Supersymmetric Theories}},  {\em Nucl. Phys.} {\bf B250} (1985) 295--311.

\bibitem{Kosower:1983yw}
D.~A. Kosower, L.~M. Krauss, and N.~Sakai, {\it {Low-Energy Supergravity and
  the Anomalous Magnetic Moment of the Muon}},  {\em Phys. Lett.} {\bf B133}
  (1983) 305--310.

\bibitem{Yuan:1984ww}
T.~C. Yuan, R.~L. Arnowitt, A.~H. Chamseddine, and P.~Nath, {\it
  {Supersymmetric Electroweak Effects on G-2 (mu)}},  {\em Z. Phys.} {\bf C26}
  (1984) 407.

\bibitem{Vendramin:1988rd}
I.~Vendramin, {\it {Constraints on Supersymmetric Parameters from Muon Magnetic
  Moment}},  {\em Nuovo Cim.} {\bf A101} (1989) 731.

\bibitem{Moroi:1995yh}
T.~Moroi, {\it {The Muon anomalous magnetic dipole moment in the minimal
  supersymmetric standard model}},  {\em Phys. Rev.} {\bf D53} (1996)
  6565--6575, [\href{http://arxiv.org/abs/hep-ph/9512396}{{\tt
  hep-ph/9512396}}]. [Erratum: Phys. Rev.D56,4424(1997)].

\bibitem{Cho:2000sf}
G.-C. Cho, K.~Hagiwara, and M.~Hayakawa, {\it {Muon g-2 and precision
  electroweak physics in the MSSM}},  {\em Phys. Lett.} {\bf B478} (2000)
  231--238, [\href{http://arxiv.org/abs/hep-ph/0001229}{{\tt hep-ph/0001229}}].

\bibitem{Martin:2001st}
S.~P. Martin and J.~D. Wells, {\it {Muon anomalous magnetic dipole moment in
  supersymmetric theories}},  {\em Phys. Rev.} {\bf D64} (2001) 035003,
  [\href{http://arxiv.org/abs/hep-ph/0103067}{{\tt hep-ph/0103067}}].

\bibitem{Czarnecki:2001pv}
A.~Czarnecki and W.~J. Marciano, {\it {The Muon anomalous magnetic moment: A
  Harbinger for 'new physics'}},  {\em Phys. Rev.} {\bf D64} (2001) 013014,
  [\href{http://arxiv.org/abs/hep-ph/0102122}{{\tt hep-ph/0102122}}].

\bibitem{Lynch:2001zs}
K.~R. Lynch, {\it {A Note on one loop electroweak contributions to g-2: A
  Companion to BUHEP-01-16}},  \href{http://arxiv.org/abs/hep-ph/0108081}{{\tt
  hep-ph/0108081}}.

\bibitem{Freitas:2014pua}
A.~Freitas, J.~Lykken, S.~Kell, and S.~Westhoff, {\it {Testing the Muon g-2
  Anomaly at the LHC}},  {\em JHEP} {\bf 05} (2014) 145,
  [\href{http://arxiv.org/abs/1402.7065}{{\tt arXiv:1402.7065}}]. [Erratum:
  JHEP09,155(2014)].

\bibitem{Queiroz:2014zfa}
F.~S. Queiroz and W.~Shepherd, {\it {New Physics Contributions to the Muon
  Anomalous Magnetic Moment: A Numerical Code}},  {\em Phys. Rev.} {\bf D89}
  (2014), no.~9 095024, [\href{http://arxiv.org/abs/1403.2309}{{\tt
  arXiv:1403.2309}}].

\bibitem{Lindner:2016bgg}
M.~Lindner, M.~Platscher, and F.~S. Queiroz, {\it {A Call for New Physics : The
  Muon Anomalous Magnetic Moment and Lepton Flavor Violation}},
  \href{http://arxiv.org/abs/1610.06587}{{\tt arXiv:1610.06587}}.

\bibitem{Belanger:2015nma}
G.~Belanger, C.~Delaunay, and S.~Westhoff, {\it {A Dark Matter Relic From Muon
  Anomalies}},  {\em Phys. Rev.} {\bf D92} (2015) 055021,
  [\href{http://arxiv.org/abs/1507.06660}{{\tt arXiv:1507.06660}}].

\bibitem{Agrawal:2014ufa}
P.~Agrawal, Z.~Chacko, and C.~B. Verhaaren, {\it {Leptophilic Dark Matter and
  the Anomalous Magnetic Moment of the Muon}},  {\em JHEP} {\bf 08} (2014) 147,
  [\href{http://arxiv.org/abs/1402.7369}{{\tt arXiv:1402.7369}}].

\bibitem{Aaij:2017vbb}
{\bf LHCb} Collaboration, R.~Aaij et~al., {\it {Test of lepton universality
  with $B^{0} \rightarrow K^{*0}\ell^{+}\ell^{-}$ decays}},  {\em JHEP} {\bf
  08} (2017) 055, [\href{http://arxiv.org/abs/1705.05802}{{\tt
  arXiv:1705.05802}}].

\bibitem{Kannike:2011ng}
K.~Kannike, M.~Raidal, D.~M. Straub, and A.~Strumia, {\it {Anthropic solution
  to the magnetic muon anomaly: the charged see-saw}},  {\em JHEP} {\bf 02}
  (2012) 106, [\href{http://arxiv.org/abs/1111.2551}{{\tt arXiv:1111.2551}}].
  [Erratum: JHEP10,136(2012)].

\bibitem{Kanemitsu:2012dc}
S.~Kanemitsu and K.~Tobe, {\it {New physics for muon anomalous magnetic moment
  and its electroweak precision analysis}},  {\em Phys. Rev.} {\bf D86} (2012)
  095025, [\href{http://arxiv.org/abs/1207.1313}{{\tt arXiv:1207.1313}}].

\bibitem{Dermisek:2013gta}
R.~Dermisek and A.~Raval, {\it {Explanation of the Muon g-2 Anomaly with
  Vectorlike Leptons and its Implications for Higgs Decays}},  {\em Phys. Rev.}
  {\bf D88} (2013) 013017, [\href{http://arxiv.org/abs/1305.3522}{{\tt
  arXiv:1305.3522}}].

\bibitem{Lewis:2017dme}
I.~M. Lewis and M.~Sullivan, {\it {Benchmarks for Double Higgs Production in
  the Singlet Extended Standard Model at the LHC}},  {\em Phys. Rev.} {\bf D96}
  (2017), no.~3 035037, [\href{http://arxiv.org/abs/1701.08774}{{\tt
  arXiv:1701.08774}}].

\bibitem{McDonald:1993ex}
J.~McDonald, {\it {Gauge singlet scalars as cold dark matter}},  {\em Phys.
  Rev.} {\bf D50} (1994) 3637--3649,
  [\href{http://arxiv.org/abs/hep-ph/0702143}{{\tt hep-ph/0702143}}].

\bibitem{Bento:2000ah}
M.~C. Bento, O.~Bertolami, R.~Rosenfeld, and L.~Teodoro, {\it {Selfinteracting
  dark matter and invisibly decaying Higgs}},  {\em Phys. Rev.} {\bf D62}
  (2000) 041302, [\href{http://arxiv.org/abs/astro-ph/0003350}{{\tt
  astro-ph/0003350}}].

\bibitem{Burgess:2000yq}
C.~P. Burgess, M.~Pospelov, and T.~ter Veldhuis, {\it {The Minimal model of
  nonbaryonic dark matter: A Singlet scalar}},  {\em Nucl. Phys.} {\bf B619}
  (2001) 709--728, [\href{http://arxiv.org/abs/hep-ph/0011335}{{\tt
  hep-ph/0011335}}].

\bibitem{Davoudiasl:2004be}
H.~Davoudiasl, R.~Kitano, T.~Li, and H.~Murayama, {\it {The New minimal
  standard model}},  {\em Phys. Lett.} {\bf B609} (2005) 117--123,
  [\href{http://arxiv.org/abs/hep-ph/0405097}{{\tt hep-ph/0405097}}].

\bibitem{Patt:2006fw}
B.~Patt and F.~Wilczek, {\it {Higgs field portal into hidden sectors}},
  \href{http://arxiv.org/abs/hep-ph/0605188}{{\tt hep-ph/0605188}}.

\bibitem{Barger:2007im}
V.~Barger, P.~Langacker, M.~McCaskey, M.~J. Ramsey-Musolf, and G.~Shaughnessy,
  {\it {LHC Phenomenology of an Extended Standard Model with a Real Scalar
  Singlet}},  {\em Phys. Rev.} {\bf D77} (2008) 035005,
  [\href{http://arxiv.org/abs/0706.4311}{{\tt arXiv:0706.4311}}].

\bibitem{Aprile:2017iyp}
{\bf XENON} Collaboration, E.~Aprile et~al., {\it {First Dark Matter Search
  Results from the XENON1T Experiment}},
  \href{http://arxiv.org/abs/1705.06655}{{\tt arXiv:1705.06655}}.

\bibitem{Drees:1992am}
M.~Drees and M.~M. Nojiri, {\it {The Neutralino relic density in minimal $N=1$
  supergravity}},  {\em Phys. Rev.} {\bf D47} (1993) 376--408,
  [\href{http://arxiv.org/abs/hep-ph/9207234}{{\tt hep-ph/9207234}}].

\bibitem{Baer:1995nc}
H.~Baer and M.~Brhlik, {\it {Cosmological relic density from minimal
  supergravity with implications for collider physics}},  {\em Phys. Rev.} {\bf
  D53} (1996) 597--605, [\href{http://arxiv.org/abs/hep-ph/9508321}{{\tt
  hep-ph/9508321}}].

\bibitem{Bai:2014osa}
Y.~Bai and J.~Berger, {\it {Lepton Portal Dark Matter}},  {\em JHEP} {\bf 08}
  (2014) 153, [\href{http://arxiv.org/abs/1402.6696}{{\tt arXiv:1402.6696}}].

\bibitem{Fukushima:2014yia}
K.~Fukushima, C.~Kelso, J.~Kumar, P.~Sandick, and T.~Yamamoto, {\it {MSSM dark
  matter and a light slepton sector: The incredible bulk}},  {\em Phys. Rev.}
  {\bf D90} (2014), no.~9 095007, [\href{http://arxiv.org/abs/1406.4903}{{\tt
  arXiv:1406.4903}}].

\bibitem{ATLAS:2016uwq}
{\bf ATLAS} Collaboration, {\it {Search for supersymmetry with two and three
  leptons and missing transverse momentum in the final state at 13 TeV with the
  ATLAS detector}}, .

\bibitem{CMS:2017fij}
{\bf CMS} Collaboration, C.~Collaboration, {\it {Search for new physics in
  events with two low momentum opposite-sign leptons and missing transverse
  energy at $\sqrt{s}=13~\mathrm{TeV}$}}, .

\bibitem{Aaboud:2016tnv}
{\bf ATLAS} Collaboration, M.~Aaboud et~al., {\it {Search for new phenomena in
  final states with an energetic jet and large missing transverse momentum in
  pp collisions at 13 TeV using the ATLAS detector}},  {\em Phys. Rev.} {\bf
  D94} (2016), no.~3 032005, [\href{http://arxiv.org/abs/1604.07773}{{\tt
  arXiv:1604.07773}}].

\bibitem{Olive:2016xmw}
{\bf Particle Data Group} Collaboration, C.~Patrignani et~al., {\it {Review of
  Particle Physics}},  {\em Chin. Phys.} {\bf C40} (2016), no.~10 100001.

\bibitem{Akerib:2016vxi}
{\bf LUX} Collaboration, D.~S. Akerib et~al., {\it {Results from a search for
  dark matter in the complete LUX exposure}},  {\em Phys. Rev. Lett.} {\bf 118}
  (2017), no.~2 021303, [\href{http://arxiv.org/abs/1608.07648}{{\tt
  arXiv:1608.07648}}].

\bibitem{Barger:2008jx}
V.~Barger, P.~Langacker, M.~McCaskey, M.~Ramsey-Musolf, and G.~Shaughnessy,
  {\it {Complex Singlet Extension of the Standard Model}},  {\em Phys. Rev.}
  {\bf D79} (2009) 015018, [\href{http://arxiv.org/abs/0811.0393}{{\tt
  arXiv:0811.0393}}].

\bibitem{Gonderinger:2012rd}
M.~Gonderinger, H.~Lim, and M.~J. Ramsey-Musolf, {\it {Complex Scalar Singlet
  Dark Matter: Vacuum Stability and Phenomenology}},  {\em Phys. Rev.} {\bf
  D86} (2012) 043511, [\href{http://arxiv.org/abs/1202.1316}{{\tt
  arXiv:1202.1316}}].

\bibitem{Barbieri:2006dq}
R.~Barbieri, L.~J. Hall, and V.~S. Rychkov, {\it {Improved naturalness with a
  heavy Higgs: An Alternative road to LHC physics}},  {\em Phys. Rev.} {\bf
  D74} (2006) 015007, [\href{http://arxiv.org/abs/hep-ph/0603188}{{\tt
  hep-ph/0603188}}].

\bibitem{Ma:2006km}
E.~Ma, {\it {Verifiable radiative seesaw mechanism of neutrino mass and dark
  matter}},  {\em Phys. Rev.} {\bf D73} (2006) 077301,
  [\href{http://arxiv.org/abs/hep-ph/0601225}{{\tt hep-ph/0601225}}].

\bibitem{Ma:2006fn}
E.~Ma, {\it {Common origin of neutrino mass, dark matter, and baryogenesis}},
  {\em Mod. Phys. Lett.} {\bf A21} (2006) 1777--1782,
  [\href{http://arxiv.org/abs/hep-ph/0605180}{{\tt hep-ph/0605180}}].

\bibitem{LopezHonorez:2006gr}
L.~Lopez~Honorez, E.~Nezri, J.~F. Oliver, and M.~H.~G. Tytgat, {\it {The Inert
  Doublet Model: An Archetype for Dark Matter}},  {\em JCAP} {\bf 0702} (2007)
  028, [\href{http://arxiv.org/abs/hep-ph/0612275}{{\tt hep-ph/0612275}}].

\bibitem{Majumdar:2006nt}
D.~Majumdar and A.~Ghosal, {\it {Dark Matter candidate in a Heavy Higgs Model -
  Direct Detection Rates}},  {\em Mod. Phys. Lett.} {\bf A23} (2008)
  2011--2022, [\href{http://arxiv.org/abs/hep-ph/0607067}{{\tt
  hep-ph/0607067}}].

\bibitem{Araki:2011hm}
T.~Araki, C.~Q. Geng, and K.~I. Nagao, {\it {Dark Matter in Inert Triplet
  Models}},  {\em Phys. Rev.} {\bf D83} (2011) 075014,
  [\href{http://arxiv.org/abs/1102.4906}{{\tt arXiv:1102.4906}}].

\bibitem{ALEPH:2005ab}
{\bf SLD Electroweak Group, DELPHI, ALEPH, SLD, SLD Heavy Flavour Group, OPAL,
  LEP Electroweak Working Group, L3} Collaboration, S.~Schael et~al., {\it
  {Precision electroweak measurements on the $Z$ resonance}},  {\em Phys.
  Rept.} {\bf 427} (2006) 257--454,
  [\href{http://arxiv.org/abs/hep-ex/0509008}{{\tt hep-ex/0509008}}].

\bibitem{Hagiwara:1994pw}
K.~Hagiwara, S.~Matsumoto, D.~Haidt, and C.~S. Kim, {\it {A Novel approach to
  confront electroweak data and theory}},  {\em Z. Phys.} {\bf C64} (1994)
  559--620, [\href{http://arxiv.org/abs/hep-ph/9409380}{{\tt hep-ph/9409380}}].
  [Erratum: Z. Phys.C68,352(1995)].

\bibitem{Cho:1999km}
G.-C. Cho and K.~Hagiwara, {\it {Supersymmetry versus precision experiments
  revisited}},  {\em Nucl. Phys.} {\bf B574} (2000) 623--674,
  [\href{http://arxiv.org/abs/hep-ph/9912260}{{\tt hep-ph/9912260}}].

\bibitem{Hahn:1998yk}
T.~Hahn and M.~Perez-Victoria, {\it {Automatized one loop calculations in
  four-dimensions and D-dimensions}},  {\em Comput. Phys. Commun.} {\bf 118}
  (1999) 153--165, [\href{http://arxiv.org/abs/hep-ph/9807565}{{\tt
  hep-ph/9807565}}].

\bibitem{Peskin:1991sw}
M.~E. Peskin and T.~Takeuchi, {\it {Estimation of oblique electroweak
  corrections}},  {\em Phys. Rev.} {\bf D46} (1992) 381--409.

\bibitem{Joglekar:2012vc}
A.~Joglekar, P.~Schwaller, and C.~E.~M. Wagner, {\it {Dark Matter and Enhanced
  Higgs to Di-photon Rate from Vector-like Leptons}},  {\em JHEP} {\bf 12}
  (2012) 064, [\href{http://arxiv.org/abs/1207.4235}{{\tt arXiv:1207.4235}}].

\bibitem{Erler:2000jg}
J.~Erler, S.~Heinemeyer, W.~Hollik, G.~Weiglein, and P.~M. Zerwas, {\it
  {Physics impact of GigaZ}},  {\em Phys. Lett.} {\bf B486} (2000) 125--133,
  [\href{http://arxiv.org/abs/hep-ph/0005024}{{\tt hep-ph/0005024}}].
  [,1389(2000)].

\bibitem{Baer:2013cma}
H.~Baer, T.~Barklow, K.~Fujii, Y.~Gao, A.~Hoang, S.~Kanemura, J.~List, H.~E.
  Logan, A.~Nomerotski, M.~Perelstein, et~al., {\it {The International Linear
  Collider Technical Design Report - Volume 2: Physics}},
  \href{http://arxiv.org/abs/1306.6352}{{\tt arXiv:1306.6352}}.

\bibitem{Gomez-Ceballos:2013zzn}
{\bf TLEP Design Study Working Group} Collaboration, M.~Bicer et~al., {\it
  {First Look at the Physics Case of TLEP}},  {\em JHEP} {\bf 01} (2014) 164,
  [\href{http://arxiv.org/abs/1308.6176}{{\tt arXiv:1308.6176}}].

\bibitem{ColuccioLeskow:2016dox}
E.~Coluccio~Leskow, G.~D'Ambrosio, A.~Crivellin, and D.~M�ller, {\it
  {$(g-2)\mu$, lepton flavor violation, and $Z$ decays with leptoquarks:
  Correlations and future prospects}},  {\em Phys. Rev.} {\bf D95} (2017),
  no.~5 055018, [\href{http://arxiv.org/abs/1612.06858}{{\tt
  arXiv:1612.06858}}].

\bibitem{Kowalska:2015zja}
K.~Kowalska, L.~Roszkowski, E.~M. Sessolo, and A.~J. Williams, {\it
  {GUT-inspired SUSY and the muon g-2 anomaly: prospects for LHC 14 TeV}},
  {\em JHEP} {\bf 06} (2015) 020, [\href{http://arxiv.org/abs/1503.08219}{{\tt
  arXiv:1503.08219}}].

\bibitem{Kowalska:2016ent}
K.~Kowalska, {\it {Phenomenological MSSM in light of new 13 TeV LHC data}},
  {\em Eur. Phys. J.} {\bf C76} (2016), no.~12 684,
  [\href{http://arxiv.org/abs/1608.02489}{{\tt arXiv:1608.02489}}].

\bibitem{Barr:2003rg}
A.~Barr, C.~Lester, and P.~Stephens, {\it {m(T2): The Truth behind the
  glamour}},  {\em J. Phys.} {\bf G29} (2003) 2343--2363,
  [\href{http://arxiv.org/abs/hep-ph/0304226}{{\tt hep-ph/0304226}}].

\bibitem{Khachatryan:2016whc}
{\bf CMS} Collaboration, V.~Khachatryan et~al., {\it {Searches for invisible
  decays of the Higgs boson in pp collisions at sqrt(s) = 7, 8, and 13 TeV}},
  {\em JHEP} {\bf 02} (2017) 135, [\href{http://arxiv.org/abs/1610.09218}{{\tt
  arXiv:1610.09218}}].

\bibitem{ATLAS-CONF-2017-045}
{\bf ATLAS Collaboration} Collaboration, {\it {Measurements of Higgs boson
  properties in the diphoton decay channel with 36.1 fb$^{-1}$ $pp$ collision
  data at the center-of-mass energy of 13 TeV with the ATLAS detector}},  Tech.
  Rep. ATLAS-CONF-2017-045, CERN, Geneva, Jul, 2017.

\bibitem{Kearney:2012zi}
J.~Kearney, A.~Pierce, and N.~Weiner, {\it {Vectorlike Fermions and Higgs
  Couplings}},  {\em Phys. Rev.} {\bf D86} (2012) 113005,
  [\href{http://arxiv.org/abs/1207.7062}{{\tt arXiv:1207.7062}}].

\bibitem{Ade:2015xua}
{\bf Planck} Collaboration, P.~A.~R. Ade et~al., {\it {Planck 2015 results.
  XIII. Cosmological parameters}},  {\em Astron. Astrophys.} {\bf 594} (2016)
  A13, [\href{http://arxiv.org/abs/1502.01589}{{\tt arXiv:1502.01589}}].

\bibitem{Staub:2013tta}
F.~Staub, {\it {SARAH 4 : A tool for (not only SUSY) model builders}},  {\em
  Comput. Phys. Commun.} {\bf 185} (2014) 1773--1790,
  [\href{http://arxiv.org/abs/1309.7223}{{\tt arXiv:1309.7223}}].

\bibitem{Porod:2003um}
W.~Porod, {\it {SPheno, a program for calculating supersymmetric spectra, SUSY
  particle decays and SUSY particle production at e+ e- colliders}},  {\em
  Comput. Phys. Commun.} {\bf 153} (2003) 275--315,
  [\href{http://arxiv.org/abs/hep-ph/0301101}{{\tt hep-ph/0301101}}].

\bibitem{Porod:2011nf}
W.~Porod and F.~Staub, {\it {SPheno 3.1: Extensions including flavour,
  CP-phases and models beyond the MSSM}},  {\em Comput. Phys. Commun.} {\bf
  183} (2012) 2458--2469, [\href{http://arxiv.org/abs/1104.1573}{{\tt
  arXiv:1104.1573}}].

\bibitem{Porod:2014xia}
W.~Porod, F.~Staub, and A.~Vicente, {\it {A Flavor Kit for BSM models}},  {\em
  Eur. Phys. J.} {\bf C74} (2014), no.~8 2992,
  [\href{http://arxiv.org/abs/1405.1434}{{\tt arXiv:1405.1434}}].

\bibitem{Belyaev:2012qa}
A.~Belyaev, N.~D. Christensen, and A.~Pukhov, {\it {CalcHEP 3.4 for collider
  physics within and beyond the Standard Model}},  {\em Comput. Phys. Commun.}
  {\bf 184} (2013) 1729--1769, [\href{http://arxiv.org/abs/1207.6082}{{\tt
  arXiv:1207.6082}}].

\bibitem{Belanger:2013oya}
G.~Belanger, F.~Boudjema, A.~Pukhov, and A.~Semenov, {\it {micrOMEGAs 3: A
  program for calculating dark matter observables}},  {\em Comput.Phys.Commun.}
  {\bf 185} (2014) 960--985, [\href{http://arxiv.org/abs/1305.0237}{{\tt
  arXiv:1305.0237}}].

\bibitem{Feroz:2008xx}
F.~Feroz, M.~Hobson, and M.~Bridges, {\it {MultiNest: an efficient and robust
  Bayesian inference tool for cosmology and particle physics}},  {\em
  Mon.Not.Roy.Astron.Soc.} {\bf 398} (2009) 1601--1614,
  [\href{http://arxiv.org/abs/0809.3437}{{\tt arXiv:0809.3437}}].

\bibitem{Alwall:2014hca}
J.~Alwall, R.~Frederix, S.~Frixione, V.~Hirschi, F.~Maltoni, O.~Mattelaer,
  H.~S. Shao, T.~Stelzer, P.~Torrielli, and M.~Zaro, {\it {The automated
  computation of tree-level and next-to-leading order differential cross
  sections, and their matching to parton shower simulations}},  {\em JHEP} {\bf
  07} (2014) 079, [\href{http://arxiv.org/abs/1405.0301}{{\tt
  arXiv:1405.0301}}].

\bibitem{Sjostrand:2007gs}
T.~Sjostrand, S.~Mrenna, and P.~Z. Skands, {\it {A Brief Introduction to PYTHIA
  8.1}},  {\em Comput.Phys.Commun.} {\bf 178} (2008) 852--867,
  [\href{http://arxiv.org/abs/0710.3820}{{\tt arXiv:0710.3820}}].

\bibitem{deFavereau:2013fsa}
{\bf DELPHES 3} Collaboration, J.~de~Favereau et~al., {\it {DELPHES 3, A
  modular framework for fast simulation of a generic collider experiment}},
  {\em JHEP} {\bf 1402} (2014) 057, [\href{http://arxiv.org/abs/1307.6346}{{\tt
  arXiv:1307.6346}}].

\bibitem{Braathen:2017izn}
J.~Braathen, M.~D. Goodsell, and F.~Staub, {\it {Supersymmetric and
  non-supersymmetric models without catastrophic Goldstone bosons}},
  \href{http://arxiv.org/abs/1706.05372}{{\tt arXiv:1706.05372}}.

\bibitem{Hambye:2007vf}
T.~Hambye and M.~H.~G. Tytgat, {\it {Electroweak symmetry breaking induced by
  dark matter}},  {\em Phys. Lett.} {\bf B659} (2008) 651--655,
  [\href{http://arxiv.org/abs/0707.0633}{{\tt arXiv:0707.0633}}].

\bibitem{Goudelis:2013uca}
A.~Goudelis, B.~Herrmann, and O.~Stal, {\it {Dark matter in the Inert Doublet
  Model after the discovery of a Higgs-like boson at the LHC}},  {\em JHEP}
  {\bf 09} (2013) 106, [\href{http://arxiv.org/abs/1303.3010}{{\tt
  arXiv:1303.3010}}].

\bibitem{Arhrib:2013ela}
A.~Arhrib, Y.-L.~S. Tsai, Q.~Yuan, and T.-C. Yuan, {\it {An Updated Analysis of
  Inert Higgs Doublet Model in light of the Recent Results from LUX, PLANCK,
  AMS-02 and LHC}},  {\em JCAP} {\bf 1406} (2014) 030,
  [\href{http://arxiv.org/abs/1310.0358}{{\tt arXiv:1310.0358}}].

\end{thebibliography}\endgroup

\end{document}